\documentclass[11pt,onecolumn,draftcls]{IEEEtran}
\pdfoutput=1

\usepackage{ifpdf}

\usepackage{cite}

\ifCLASSINFOpdf
 \usepackage[pdftex]{graphicx}
 \DeclareGraphicsExtensions{.pdf,.jpeg,.png}
\else
 \usepackage[dvips]{graphicx}
 \DeclareGraphicsExtensions{.eps}
\fi
\usepackage[cmex10]{amsmath}
\usepackage{algorithmic}

\usepackage{array}

\usepackage{mdwmath}
\usepackage{mdwtab}

\usepackage{eqparbox}

\usepackage[tight,footnotesize]{subfigure}

\usepackage{stfloats}
\usepackage{amssymb}
\usepackage{bm}
\usepackage{overpic}
\usepackage{color}

\newtheorem{theorem}{Theorem}
\newtheorem{lemma}{Lemma}

\newtheorem{definition}{Definition}

\newtheorem{assumption}{Assumption}
\renewcommand{\IEEEQED}{\IEEEQEDopen}

\begin{document}
%
\title{Performance Analysis of the Matrix Pair Beamformer with Matrix Mismatch}

       \author{
        Jianshu~Chen,~\IEEEmembership{Student Member,~IEEE},
        Jian~Wang,~
        Xiu-Ming~Shan,~\\
        Ning~Ge,~\IEEEmembership{Member,~IEEE}~
        and~Xiang-Gen~Xia,~\IEEEmembership{Fellow,~IEEE}\\
        EDICS: SAM-PERF
        \thanks{
                This work was supported by the National Natural Science Foundation of China 
                under contract No. 60972019 and No. 60928001.
            }
        \thanks{
                Jianshu Chen was with the Department of Electronic Engineering,
                Tsinghua University, Beijing, P. R. China, 100084. 
                He is with the 
                Department of Electrical Engineering, University of
                California, Los Angeles, CA 90095, USA. (e-mail: jshchen@ee.ucla.edu)
               }
        \thanks{
                Jian Wang, Xiu-Ming Shan and Ning Ge are with the Department of Electronic Engineering,
                Tsinghua University, Beijing, P. R. China, 100084. (e-mail: \{jian-wang, shanxm, gening\}
                @tsinghua.edu.cn)
               }
        \thanks{
               Xiang-Gen Xia is with the Department of Electrical and Computer Engineering,
University of Delaware, Newark, DE 19716, USA. (e-mail:
xxia@ee.udel.edu). His work was suppported in part by the Air Force Office of Scientific 
Research (AFOSR) under Grant No. FA9550-08-1-0219, the National Science Foundation (NSF) under Grant CCF-0964500, 
and the World Class University (WCU) Program 2008-000-20014-0, National
Research Foundation, Korea.
               }
       }

\maketitle

\begin{abstract}
Matrix pair beamformer (MPB) is a blind beamformer. It exploits the temporal structure of 
the signal of interest (SOI) and applies generalized eigen-decomposition to a covariance matrix pair. 
Unlike other blind algorithms, it only uses the second order statistics. A key assumption in 
the previous work is that the two matrices have the same interference statistics. However, 
this assumption may be invalid in the presence of multipath propagations or certain ``smart'' 
jammers, and we call it as matrix mismatch. This paper analyzes the performance of MPB with 
matrix mismatch. First, we propose a general framework that covers the existing schemes. 
Then, we derive its normalized output SINR. It reveals that the matrix mismatch leads to  
a threshold effect caused by ``steering vector competition''. Second, using matrix perturbation 
theory, we find that, if there are generalized eigenvalues that are infinite, 
the threshold will increase unboundedly with the interference power. This is highly probable when 
there are multiple periodical interferers. Finally, we present simulation results to verify our analysis.
\end{abstract}

\begin{IEEEkeywords}
Adaptive beamforming, Matrix pair beamformer, Generalized eigen-decomposition, Matrix mismatch
\end{IEEEkeywords}

 \ifCLASSOPTIONpeerreview
 \begin{center} \bfseries EDICS Category: 3-BBND \end{center}
 \fi
%
\IEEEpeerreviewmaketitle

\section{Introduction}

\IEEEPARstart{B}{eamforming} is a spatial filter, which combines the outputs of multiple sensor elements by an appropriately designed weight vector so as to pass a signal of interest (SOI) while rejecting interfering signals. Since the pioneering work of Howells\cite{Howells1965}, Applebaum\cite{applebaum1976adaptive} and Widrow\cite{widrow1967adaptive}, it has been intensively studied during the past decades and widely applied in radar, sonar and wireless 
communications \cite{widrow1967adaptive, widrow1975adaptive, gabriel1976adaptive, Compton1978, compton1976adaptive, haykin1980array, LAL1997, Godara1997, paulraj1997stp} etc. For a comprehensive review, we refer to \cite{Van1988, haykin1992adaptive, Krim1996, Trees2002} and the references therein.

There are various forms of implementations for a beamformer. Some use the direction of arrival (DOA) of the SOI and directly calculate the weight vector by sample matrix inversion (SMI) \cite{Reed1974}. Some employ a reference signal (e.g. training signal\cite{Tanaka2000} and decision feedback signal\cite{Compton1978}) to iteratively calculate the weight vector. And there are also blind beamformers which do not require the DOA or the reference signal\cite{godard1980self, treichler1983new, shynk1993performance, ohgane1991characteristics, ohgane1993implementation, Naguib1996, Suard1993, choi2002nab, Yang2006fbba, Song2001, Torrieri2004, torrieri2007maa, chen2008opa, chen2008opb}.

To achieve blind beamforming, we need to exploit the special properties of the SOI, such as constant modulus, non-Gaussianness, high order statistics etc. One important property is the temporal structure of the SOI. Such kind of blind beamformers are extensively studied for CDMA systems, where the inherent structure of the spreading codes can be exploited\cite{Naguib1996,Suard1993,choi2002nab,Yang2006fbba,Torrieri2004,torrieri2007maa,chen2008opa,chen2008opb}. 
The advantage of this scheme is that it only relies on the second order statistics of the covariance matrix pair.
Although the implementation details differ, the main idea of these approaches is to exploit a pair of array covariance matrix and hence will be referred to as matrix pair beamformer (MPB) in this paper. The MPB projects the discrete sequence in each antenna onto the space spanned by the SOI's signature vector (for a CDMA system, it is the spreading code vector) and another carefully designed base vector. Then two sets of array snapshots (i.e. signal snapshot and interference snapshot) are acquired to calculate a pair of covariance matrices. With the processing gain, the desired signal power in the signal snapshot is generally greater than that in the interference snapshot. \emph{A key and common assumption is that the interference statistics in the two snapshots are identical.} These two features enable the separation of the signal steering vector and the interference covariance matrix from the two matices. And the weight vector derived from the dominant eigenvector of the matrix pair will maximize the output signal to interference plus noise ratio (SINR). 

However, this key assumption that the two matrices share the same interference statistics is not valid in many cases, which we refer to as \emph{matrix mismatch}. Matrix mismatch may occur when there are interferers with certain periodical structure, such as multiple access interference (MAI) in CDMA systems, tones and some other ``smart jammers'' in radar systems and it may lead to 
the failure of a system. In practical applications, to avoid such an unexpected failure, it is  necessary to understand the detailed performance of a scheme
when the assumption/condition is not satisfied.  To our best knowledge, 
 little effort has been devoted to analyzing the effect of matrix mismatch on the performance of MPB. This paper aims to analyze the MPB's performance under matrix mismatch, and our contributions are:
    \begin{itemize}
        \item
            proposing a general framework to model various existing MPB schemes;
       	\item
            deriving  analytical expressions for the normalized output SINR as the performance measure;
        \item
            discovering a threshold effect for MPB, 
            i.e., due to matrix mismatch, the performance of MPB  degrades rapidly when the input                   	
            signal to noise ratio (SNR) is below a predicted threshold, and the main beam  points to the directions of the interferers;
        \item
            explaining how MPB works ``blindly'' by ``steering vector competition'';
        \item
            discussing various factors that have impact on the threshold, and showing that when there is an 
            generalized eigenvalue that is infinite which 
             is called \emph{the noise-free covariance matrix pair}, the threshold SNR  increases  
            unboundedly with the interference power;
        \item
            discussing several typical scenarios and showing that MPB is very vulnerable to multiple periodical interferers.
    \end{itemize}

The rest of the paper is organized as follows. Sec. II presents a general framework of MPB to cover and reinterpret the basic ideas in \cite{Naguib1996, Suard1993, choi2002nab, Yang2006fbba, Song2001, Torrieri2004, torrieri2007maa, chen2008opa, chen2008opb}, followed by a formulation of the matrix mismatch problem. In Sec. III, we present the expressions for MPB's weight vector and its normalized output SINR, which uncovers the inherent threshold effect caused by matrix mismatch. The discussion relies heavily on the approximation of the generalized eigenvalue, which is derived in Appendix \ref{Sec:Appendix:Proof_of_Thm_lambda_max} using Gerschgorin theorem\cite{stewart1990mpt, horn1990ma}. Sec. IV applies the Weyl-Lidskii type theorem in matrix perturbation theory\cite{stewart1990mpt} to analyze MPB's threshold SNR, and discusses it in two typical scenarios. Finally, Sec. V presents simulation results to verify our theoretical analysis and Sec. VI concludes the whole paper.


\section{Problem Formulation}
\label{Sec:ProblemFormulation}
\subsection{Signal Model}
\label{Sec:ProblemFormulation:SignalModel}

    \begin{figure*}[!t]
    \centering
    \includegraphics[width=0.74\textwidth]{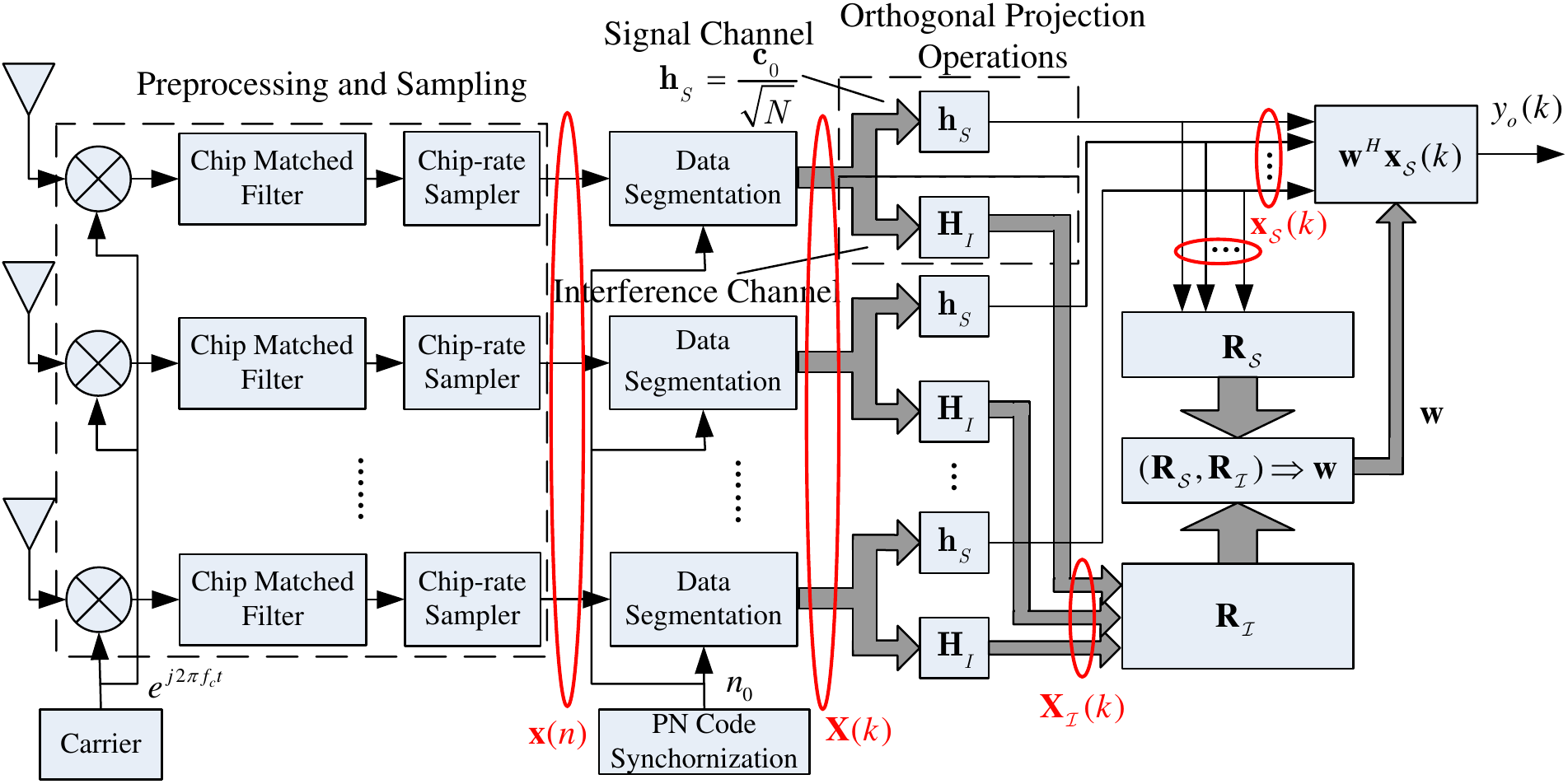}
    \caption{Block diagram of the matrix pair beamformer in a CDMA system.
    Data corresponding to the $k$ symbol are blocked and projected onto two spaces to
    produce signal channel snapshot $\mathbf{x}_\mathcal{S}(k)$ and interference channel
    snapshot $\mathbf{X}_\mathcal{I}(k)$, respectively.}
    \label{Fig:MatrixPairBeamformer_lthAntenna}
    \end{figure*}

Consider an antenna array of $L$ isotropic elements that receives
$D+1$ signals from a far field. After preprocessing (e.g.
mixing, filtering, etc.) and sampling, the output $L \times 1$ array
vector can be written as
    \begin{equation}
    \label{Equ:Signal_model:x} \mathbf{x}(n)= \sum_{i=0}^{D} \sqrt{P_i}
    s_i(n) \mathbf{a}_i + \mathbf{v}(n),
    \end{equation}
where $s_i(n)$ is the discrete sequence of the $i$th signal with
normalized power; $P_i$ is its power; $\mathbf{a}_i$ is
the steering vector, which depends on the DOA and the array geometry; $\mathbf{v}(n)$ is
the additive white Gaussian noise (AWGN) vector with zero mean and
covariance matrix $\sigma^2 \mathbf{I}$. Assume $s_0(n)$ is the
SOI, and $s_1(n),s_2(n),\ldots,s_D(n)$ model all the
possible interferers like MAI and jammer etc. This paper considers the case
when $s_0(n)$ has some inherent
temporal structure expressed as
    \begin{equation}
    \label{Equ:Signal_model:s0}
    s_0(n)=\sum_{k=-\infty}^{+\infty}b_0(k)c_0[n - \tau(k)],
    \end{equation}
where $c_0(n)$ is known and supported on $0 \le n \le N-1$.
Thus, $s_0(n)$ can be viewed as a train of pulses, with each of them being a delayed 
and scaled replica of $c_0(n)$. And $\tau(k)$ and $b_0(k)$ are the corresponding 
delay and amplitude of the $k$th pulse. This model is common
in modern communication systems. 
For instance, in CDMA system, $c_0(n)$ is the spreading code of the desired user, $b_0(k)$ is its data bits
and $\tau(k)$ is the delay of the $k$th symbol. Without loss of generality, let us consider 
$b_0(k) \in \{ \pm 1\}$, $c_0(n) \in \{\pm 1\}$ 
and $\tau(k) = kN + n_0$. Then $N$ is the processing gain and $n_0$
is the propagation delay.

\subsection{A General Framework of the Matrix Pair Beamformer}
\label{Sec:ProblemFormulation:MPBeamformer}

We first propose a framework called \emph{matrix pair beamformer} (MPB) to
cover the common ideas in \cite{Naguib1996,Suard1993,choi2002nab,Yang2006fbba,Song2001,Torrieri2004,torrieri2007maa,chen2008opa,chen2008opb}. Our strategy is to use orthogonal projection operation to model their ways of estimating
the covariance matrix pair. By this framework, we will find a common threshold effect in these methods.

The steering vector $\mathbf{a}_i$ in
(\ref{Equ:Signal_model:x}) is a spatial signature of the $i$th
signal, which is different from other $\mathbf{a}_j$ 
so long as they arrive from
different directions. Beamformer is a spatial filter that exploits
such difference to pass the desired signal $s_0(n)$ while
suppressing $s_1(n)\ldots s_D(n)$ and $\mathbf{v}(n)$. A
statistically optimum beamformer \cite{Van1988,Trees2002} generally
requires at least, either explicitly or implicitly, the information
about the steering vector $\mathbf{a}_0$ and the
interference covariance matrix. The latter one may be replaced by
the data covariance matrix, so the remaining problem is how to
acquire $\mathbf{a}_0$. In DOA-based beamformer, it is
calculated by the DOA and array manifold information. As for
training-based method or decision directed method, it is inherent in
the correlation vector between the reference signal and the data
vector.

To work ``blindly'', i.e. without any
 explicit information of DOA, the methods in
\cite{Naguib1996,Suard1993,choi2002nab,Yang2006fbba,Song2001,Torrieri2004,torrieri2007maa,chen2008opa,chen2008opb}
exploit the SOI's temporal signature $c_0(n)$ to acquire
these spatial statistical information. Specifically, it is
implemented by two orthogonal projections and a generalized eigen-decomposition of 
a covariance matrix pair. Hence, we refer to them as matrix pair beamformer (MPB) in this literature. 
In Fig. \ref{Fig:MatrixPairBeamformer_lthAntenna}, we summarize the common structure of MPB.
With the data segmentation, the array outputs corresponding to
the $k$th symbol of the desired user can be expressed in the
following matrix form:
    \begin{align}
    \label{Equ:MPBeamformer:X}
    \mathbf{X}(k) \! &\triangleq \! 
    				\begin{bmatrix}
				 	\mathbf{x}(kN+n_0) & \cdots & \mathbf{x}(kN+n_0+N-1)
				\end{bmatrix}
                     =          \! \left[\sqrt{P_0} b_0(k)\right] \! \mathbf{a}_0\mathbf{c}_0^T
                                 \!\!+\!\! 
                                 \underbrace{
                                 				\mathbf{A}_I \pmb{\Theta}_I^{\frac{1}{2}} \mathbf{S}_I^T(k)
                                 				\!\!+\!\! \mathbf{V}(k)
						  }_{\mathbf{Z}(k)}						  
    \end{align}
where
    \begin{align}
    \mathbf{A}_I &\triangleq \big[ \; \mathbf{a}_1 \; \mathbf{a}_2 \; \cdots \; \mathbf{a}_D \; \big]
    \nonumber\\
    \mathbf{c}_0 &\triangleq \big[ \; c_0(0) \; c_0(1) \; \cdots \; c_0(N-1)
    \; \big]^T
    \nonumber\\
    \mathbf{S}_I(k) &\triangleq \big[ \;
    \mathbf{s}_1(k) \; \mathbf{s}_2(k) \; \cdots \; \mathbf{s}_D(k) \; \big]
    \nonumber\\
    \mathbf{s}_i(k) &\triangleq \big[ \;
    s_i(kN+n_0) \; \cdots \; s_i(kN+n_0+N-1) \; \big]^T
    \nonumber\\
    \mathbf{V}(k) &\triangleq \big[ \;
    \mathbf{v}(kN+n_0) \; \cdots \; \mathbf{v}(kN+n_0+N-1) \; \big].
    \nonumber\\
    \pmb{\Theta}_I &\triangleq \mathrm{diag}\{P_1,P_2,\cdots,P_D\}.    \nonumber
    \end{align}
Then, the $k$th data block in each antenna is projected onto two subspaces:
signal space $\mathcal{S}$ spanned by the SOI's temporal signature vector $\mathbf{h}_\mathcal{S}\!=\!\mathbf{c}_0/\sqrt{N}$,
and a specifically designed interference space $\mathcal{I}\!=\!\mathcal{R}\{\mathbf{H}_\mathcal{I}\}$, respectively. Without loss of generality, 
assume the columns of $\mathbf{H}_\mathcal{I} \!\!\in\!\! \mathbb{C}^{N\times r_\mathcal{I}}$ are orthonormal.
Then, the projections produce the signal snapshot 
$\mathbf{x}_\mathcal{S}(k)\!\!=\!\!\mathbf{X}(k)\mathbf{h}_\mathcal{S}^\ast$ and the interference snapshot
$\mathbf{X}_\mathcal{I}(k)\!\!=\!\!\mathbf{X}(k)\mathbf{H}_\mathcal{I}^\ast$. 
Define $\sigma_{\mathcal{S}_0}^2 \!\!\triangleq\!\! NP_0 $ and 
$\sigma_{\mathcal{I}_0}^2 \!\!\triangleq\!\! 
\frac{P_0}{r_\mathcal{I}} \|\mathbf{H}_\mathcal{I}^H\mathbf{c}_0\|^2$.
Then, the covariance matrices of $\mathbf{x}_\mathcal{S}(k)$ 
and $\mathbf{X}_\mathcal{I}(k)$ are
    \begin{align}
        \label{Equ:MPBeamformer:R_S}
        \mathbf{R}_\mathcal{S}          	&\triangleq        \textsf{E}\{\mathbf{x}_\mathcal{S}(k) 
        												\mathbf{x}_\mathcal{S}^H(k)\}
                                        			=                  	\sigma_{\mathcal{S}_0}^2
										\mathbf{a}_0 \mathbf{a}^H_0 
										+
										\mathbf{Q}_\mathcal{S}
									        \\
        	\label{Equ:MPBeamformer:R_I}
        \mathbf{R}_\mathcal{I}          	&\triangleq        \frac{1}{r_\mathcal{I}}
        										\textsf{E}\{\mathbf{X}_\mathcal{I}(k) 
												\mathbf{X}_\mathcal{I}^H(k)\}
                                        			=                  	\sigma_{\mathcal{I}_0}^2
										\mathbf{a}_0 \mathbf{a}^H_0 
										+ 
										\mathbf{Q}_\mathcal{I}
    \end{align}
where $\sigma_{\mathcal{S}_0}^2$ and $\sigma_{\mathcal{I}_0}^2$
are the SOI's powers in the \emph{signal channel} and \emph{interference channel} (c.f. Fig. \ref{Fig:MatrixPairBeamformer_lthAntenna}), respectively.
$\mathbf{Q}_\mathcal{S}$ and $\mathbf{Q}_\mathcal{I}$
are the covariance matrices of the interference-plus-noise in them, defined by
	\begin{align}
		\label{Equ:MPBeamformer:Q_S}
		\mathbf{Q}_\mathcal{S}	&=	\textsf{E}\{ \mathbf{Z}(k)\mathbf{h}_\mathcal{S}^\ast
												 \mathbf{h}_\mathcal{S}^T\mathbf{Z}^H(k) \}
							=	\mathbf{A}_I\bm{\Phi}_\mathcal{S}\mathbf{A}_I^H
								+ \sigma^2\mathbf{I},
								\\
		\label{Equ:MPBeamformer:Q_I}
		\mathbf{Q}_\mathcal{I}	&=	\textsf{E}\{ \mathbf{Z}(k)\mathbf{H}_\mathcal{S}^\ast 
												\mathbf{H}_\mathcal{S}^T\mathbf{Z}^H(k) \}
							=	\mathbf{A}_I\bm{\Phi}_\mathcal{I}\mathbf{A}_I^H
								+ \sigma^2\mathbf{I},
	\end{align}
where $\bm{\Phi}_\mathcal{S} \!=\! \sigma^2 \textsf{INR} \cdot \bm{\Phi}_{\mathcal{S}_0}$,
$\bm{\Phi}_\mathcal{I} \!=\! \sigma^2 \textsf{INR} \cdot \bm{\Phi}_{\mathcal{I}_0}$,
$\textsf{INR} \!=\! P_1/\sigma^2$, and
	\begin{align}	
		\label{Equ:MPBeamformer:Phi_S0_AND_Phi_I0}
		&\bm{\Phi}_{\mathcal{S}_0}=	\bm{\Omega}_I^{\frac{1}{2}}
								\textsf{E}	\left\{
											\mathbf{S}_I^T(k)
											\mathbf{h}_\mathcal{S}^\ast
											\mathbf{h}_\mathcal{S}^T
											\mathbf{S}_I^\ast(k)
										\right\}
								\bm{\Omega}_I^{\frac{1}{2}},
								\qquad
		\bm{\Phi}_{\mathcal{I}_0}	=	\frac{1}{r_\mathcal{I}}
								\bm{\Omega}_I^{\frac{1}{2}}
								\textsf{E}	\left\{
											\mathbf{S}_I^T(k)
											\mathbf{H}_\mathcal{I}^\ast
											\mathbf{H}_\mathcal{I}^T
											\mathbf{S}_I^\ast(k)
										\right\}
								\bm{\Omega}_I^{\frac{1}{2}},
	\end{align}
with $\bm{\Omega}_I \!=\! (\bm{\Theta}_I/\sigma^2)/\textsf{INR}$. We can see that
$\bm{\Omega}_I$ is independent of $\textsf{INR}$ and only depends on 
the relative strength of the interferers. So are $\bm{\Phi}_{\mathcal{S}_0}$
and $\bm{\Phi}_{\mathcal{I}_0}$, and we will use this conclusion in Sec. \ref{Sec:Threshold}.


The MPB uses the eigenvector corresponding to the largest generalized eigenvalue of the
matrix pair $(\mathbf{R}_\mathcal{S},\mathbf{R}_\mathcal{I})$ as the weight vector $\mathbf{w}$, 
which is the solution to the following equation
	\begin{align}
		\label{Equ:MPBeamformer:EigenEquation0}
		\mathbf{R}_\mathcal{S}\mathbf{w} 
    						= \lambda_\mathrm{max} \mathbf{R}_\mathcal{I}\mathbf{w},
	\end{align}
where $\lambda_\mathrm{max}$ is the largest generalized eigenvalue of 
$(\mathbf{R}_\mathcal{S},\mathbf{R}_\mathcal{I})$. $\mathbf{w}$
is applied to $\mathbf{x}_\mathcal{S}(k)$ to yield the output
	\begin{align}
		\label{Equ:MPBeamformer:y_0}
		y_o(k)	\!\!=\!\!		\mathbf{w}^H \mathbf{x}_\mathcal{S}(k)
				\!\!=\!\!		\underbrace{
										\left[ \sigma_{\mathcal{S}_0} b_0(k) \right] 
										\mathbf{w}^H \mathbf{a}_0
									   }_{y_S(k)}
							\!\!+\!\!
							\underbrace{
										\mathbf{w}^H \mathbf{Z}(k) \mathbf{h}_\mathcal{S}^\ast
									   }_{y_{I}(k)}
	\end{align} 
where $y_S(k)$ and $y_I(k)$ are the output signal and interference-plus-noise.
Eq. \eqref{Equ:MPBeamformer:EigenEquation0} is also equivalent to
    \begin{equation}
    \label{Equ:MPBeamformer:EigenEquation}
    \left(\mathbf{R}_\mathcal{S}-\mathbf{R}_\mathcal{I}\right)\mathbf{w} 
    						= (\lambda_\mathrm{max}-1) \mathbf{R}_\mathcal{I}\mathbf{w}.
    \end{equation}
Using (\ref{Equ:MPBeamformer:R_S}) and (\ref{Equ:MPBeamformer:R_I}),
we have
    \begin{equation}
    \label{Equ:MPBeamformer:R_S_minus_R_I}
    \mathbf{R}_\mathcal{S}-\mathbf{R}_\mathcal{I} 	
    			=	(\sigma_{\mathcal{S}_0}^2-\sigma_{\mathcal{I}_0}^2)
    				\mathbf{a}_0\mathbf{a}_0^H + \mathbf{Q}_\mathcal{S}-\mathbf{Q}_\mathcal{I}.
    \end{equation}
Since the columns of $\mathbf{H}_\mathcal{I}$ are orthonormal, i.e. $\mathbf{H}_\mathcal{I}^H\mathbf{H}_\mathcal{I}=\mathbf{I}$,
its spectral norm is one and we can have
    \begin{align}
    	\sigma_{\mathcal{I}_0}^2 		\triangleq	\frac{P_0}{r_\mathcal{I}} 
									\|\mathbf{H}_\mathcal{I}^H\mathbf{c}_0\|^2
							\le          	\frac{P_0}{r_\mathcal{I}} 
									\|\mathbf{H}_\mathcal{I}\|^2 \|\mathbf{c}_0\|^2
							=		 \frac{NP_0}{r_\mathcal{I}}
							\le		\sigma_{\mathcal{S}_0}^2.			\nonumber
    \end{align}
The above inequality holds strictly when choosing $\mathbf{H}_\mathcal{I} \neq \mathbf{h}_\mathcal{S}$ 
(either $r_\mathcal{I} \!>\! 1$ or  $\mathbf{H}_\mathcal{I} \!\neq\! \mathbf{h}_\mathcal{S}$ for $r_\mathcal{I} \!=\! 1$). 
Furthermore, it is commonly assumed
\cite{Naguib1996,Suard1993,choi2002nab,Yang2006fbba,Song2001,
Torrieri2004,torrieri2007maa,chen2008opa,chen2008opb}
that $\mathbf{Q}_\mathcal{S} \!=\! \mathbf{Q}_\mathcal{I}$. 
Then, $\mathbf{Q}_\mathcal{S} \!-\! \mathbf{Q}_\mathcal{I} \!=\! \mathbf{O}$ in 
(\ref{Equ:MPBeamformer:R_S_minus_R_I}),
and the dominant eigenvector of
$(\mathbf{R}_\mathcal{S}-\mathbf{R}_\mathcal{I},\mathbf{R}_\mathcal{I})
=\left((\sigma_{\mathcal{S}_0}^2-\sigma_{\mathcal{I}_0}^2)\mathbf{a}_0\mathbf{a}_0^H,
\mathbf{R}_\mathcal{I}\right)$ is
	\begin{align}
		\label{Equ:MPBeamformer:w_opt}
		\mathbf{w}_\mathrm{opt} 	= 	\mu \mathbf{R}_\mathcal{I}^{-1} \mathbf{a}_0
	        						=	\mu' \mathbf{Q}_\mathcal{S}^{-1} \mathbf{a}_0,
	\end{align}
where $\mu$ and $\mu'$ are scalars. 
Then $\mathbf{w}_\mathrm{opt}$ will maximize the output 
interference plus noise ratio (SINR) \cite{Trees2002}, and the optimal SINR is
	\begin{align}
		\label{Equ:MPBeamformer:SINR_max}
		\textsf{SINR}_{\textsf{opt}} 	
		=	\left.\frac{\textsf{E}\{|y_S(k)|^2\}}{\textsf{E}\{|y_I(k)|^2\}}\right|_{\mathbf{w}_\mathrm{opt}}
		=	\sigma_{\mathcal{S}_0}^2 \mathbf{a}_0^H \mathbf{Q}_\mathcal{S}^{-1} \mathbf{a}_0		
	\end{align}

All the methods in\cite{Naguib1996,Suard1993,choi2002nab,Yang2006fbba,Song2001,Torrieri2004,torrieri2007maa,chen2008opa,chen2008opb}
share the structures described above. They
only differ in the dominant eigenvector searching algorithm and the interference space $\mathcal{I}$
(i.e. $\mathbf{H}_\mathcal{I}$). In most existing approaches,
$\mathcal{I}$ is a one dimension space ($r_\mathcal{I}=1$).
The pre- and post-correlation (PAPC)
scheme\cite{Naguib1996,Suard1993,choi2002nab,Yang2006fbba,Song2001}
uses $\mathbf{R}_\mathcal{I} \!=\! \textsf{E}\{ \mathbf{x}(n) \mathbf{x}^H(n) \}$, thus it is
equivalent to choosing $\mathbf{H}_\mathcal{I}$ as
    \begin{equation}
    \label{Equ:MPBeamformer:h_I_PAPC}
    \mathbf{H}_\mathcal{I}
    =
    \big[ \; 0 \;\; \cdots \;\; 0 \;\; 1 \;\; 0 \;\; \cdots \;\; 0 \;\big]^T.
    \end{equation}
where only one component in $\mathbf{H}_\mathcal{I}$ is nonzero. The Maximin scheme in \cite{Torrieri2004} and
\cite{torrieri2007maa} employs a \emph{monitor filter} to isolate the
interference, which can be interpreted as
    \begin{equation}
    \label{Equ:MPBeamformer:h_I_FP}
    \mathbf{H}_\mathcal{I}
    =
    \mathbf{c}_0 \odot \left[ \; 1 \;\; e^{j2 \pi f_\mathrm{MF}} \;\; \cdots \;\;
    e^{j2\pi f_\mathrm{MF} (N-1)} \;\right]^T,
    \end{equation}
where $f_\mathrm{MF}\in(0,1]$ is the normalized center frequency of
monitor filter, and $\odot$ is the Hadamard product.

\subsection{Matrix Mismatch}
We see that the MPB relies heavily on the key assumption that
$\mathbf{Q}_\mathcal{S}=\mathbf{Q}_\mathcal{I}$, namely the interferers 
have the same second order statistics in the two channels. 
By \eqref{Equ:MPBeamformer:Q_S}--\eqref{Equ:MPBeamformer:Phi_S0_AND_Phi_I0},
we know it is valid when each interferer
is random enough in the temporal domain, say directional white noise (c.f.
Sec. \ref{Sec:TwoTypicalScenarios:DWN}). However, it is generally not satisfied,
especially when there are multiple deterministic periodical interferers, like tones and MAI
(c.f. Sec. \ref{Sec:TwoTypicalScenarios:DPI}). 
This is because when the interferers are deterministic and periodical, the expectations in 
\eqref{Equ:MPBeamformer:Phi_S0_AND_Phi_I0} can be eliminated. Then
$\mathbf{Q}_\mathcal{S}\!=\!\mathbf{Q}_\mathcal{I}$ requires 
$\mathbf{S}_I^T(k)\mathbf{h}_\mathcal{S}^\ast \!=\! \mathbf{S}_I^T(k)\mathbf{H}_\mathcal{I}^\ast$,
which is highly improbable when $\mathbf{h}_\mathcal{S} \!\neq\! \mathbf{H}_\mathcal{I}$.
We term this as \emph{``matrix mismatch''}.
To our best knowledge, very little effort has been devoted to analyze this problem.
Therefore, we will investigate the
performance of MPB in this more general case. Before we proceed, we
define the normalized output SINR to measure performance degradation
with respect to that of no matrix mismatch. 
	\begin{definition}
		The \emph{normalized output SINR} is defined as the actual output SINR of MPB 
		normalized by the optimal value,
		i.e.	
			\begin{align}
				\label{Equ:Matrix_Mismatch:G}
				\textsf{G}	\triangleq	\left.\frac{\textsf{E}\{|y_S(k)|^2\}}
									{\textsf{E}\{|y_I(k)|^2\}}\right|_\mathbf{w}
							\cdot
							\frac{1}{\textsf{SINR}_\textsf{{opt}}}
			\end{align}	
		where $\textsf{SINR}_\textsf{opt}$ is given by (\ref{Equ:MPBeamformer:SINR_max}), and
		$\mathbf{w}$ is the solution to (\ref{Equ:MPBeamformer:EigenEquation}) without the
		assumption of $\mathbf{Q}_\mathcal{S}=\mathbf{Q}_\mathcal{I}$.
	\end{definition}

$\textsf{G}$ generally depends on the input SNR and the interference powers. 
So a reasonable way to characterize the performance is to plot $\textsf{G}$ against the input 
$\textsf{SNR} \triangleq \sigma_{\mathcal{S}_0}^2/\sigma^2$, when fixing
the interference powers. In the following sections, we will base our analysis on $\textsf{G}(\textsf{SNR})$, 
which we will refer to as \emph{operating curve}. Moreover, we assume infinite sample size 
so that the finite sample effect is ignored.
%

\section{Performance Analysis of the Matrix Pair Beamformer with Matrix Mismatch}
\label{Sec:Performance_Analysis}

In this section, we will derive the operating curve of MPB, and discuss how it works blindly.

\subsection{Operating Curve of Matrix Pair Beamformer}
\label{Sec:Performance_Analysis:Operating_Curve}

We base our discussions on the following assumptions, and summarize the main result in
theorem \ref{thm:operating_curve}.
	\begin{assumption}
		\label{Asm:Orthogonal}
		The spacing of DOA between any two signals is large enough (greater than a mainlobe),
		so that $\{\mathbf{a}_i\}_{i=0}^D$ are linearly independent and
		the projection of $\mathbf{a}_0$ onto 
		$\mathrm{span}\{\!\mathbf{a}_i\}_{i=1}^D$
		is much less than $\|\mathbf{a}_0\|$.
	\end{assumption}
	\begin{assumption}
		\label{Asm:Norm}
		The steering vectors of all signals are normalized so that $\|\mathbf{a}_i\|^2=L$, ($i=0,1,\ldots,D$).
	\end{assumption}
	\begin{theorem}[operating curve]
		\label{thm:operating_curve}
		The normalized output SINR of MPB with matrix mismatch is
			\begin{align}
				\label{Equ:Performance:G_Eb}
				\textsf{G}(\textsf{SNR})	&=	\left\{
										\begin{array}{ll}
											\frac{
													\displaystyle P_I \!+\! 1
											       }
											       {
											       		\displaystyle
											       		P_I
													/
													\left[
														1
														\!-\!
														{\textsf{SNR}_{\textsf{T}0}}
														/
														{\textsf{SNR}}
													\right]^2\!+\!1
											       }
											       G_U,
											&
											\textsf{SNR}>\textsf{SNR}_{\textsf{T2}}
											\\
											{
												\displaystyle
											      	\left[
													\frac{1\!\!+\!\!K_0}
													       {
													       	1
														\!\!-\!\!
														{\textsf{SNR}}
														/
														{\textsf{SNR}_{\textsf{T}0}}
														\!\!+\!\!
														K_0
														\left(
															{L\beta}
															\textsf{SNR}/{N}
															\!\!+\!\!
															1
														\right)
													       }
												\right]^2	
												G_L,	
											       }
											&
											\textsf{SNR}<\textsf{SNR}_{\textsf{T1}}
										\end{array}
									\right.
			\end{align}
		where $G_U$ and $G_L$ are the normalized output SINR when 
		$\textsf{SNR}\!\!=\!\!+\infty$ and $\textsf{SNR}\!\!=\!\!0$ ($-\infty$dB), respectively. 
		$\beta\!\!\triangleq\!\!N\sigma_{\mathcal{I}_0}^2/\sigma_{\mathcal{S}_0}^2$ is
		the normalized power leakage ratio (PLR) in interference channel.
		$P_I \!\!\triangleq\!\! 1/G_U\!\!-\!\!1$ is the output interference to noise ratio, and 
		$\textsf{SNR}_{\textsf{T}0}$ is the empirical threshold $\textsf{SNR}$. Their expressions
		are
			\begin{align}
				\label{Equ:Performance:G_U}
				&G_U 		= 		\frac{
									\mathbf{a}_0^H
									\mathbf{Q}_\mathcal{I}^{-1}
									\mathbf{a}_0
								        }
								        {
								        	\mathbf{a}_0^H
									\mathbf{Q}_\mathcal{S}^{-1}
									\mathbf{a}_0 
								        }
								\cdot
								\frac{
									\mathbf{a}_0^H
									\mathbf{Q}_\mathcal{I}^{-1}
									\mathbf{a}_0
								        }
								        {
								        	 \mathbf{a}_0^H
									 \mathbf{Q}_\mathcal{I}^{-1}
									 \mathbf{Q}_\mathcal{S}
									 \mathbf{Q}_\mathcal{I}^{-1}
									 \mathbf{a}_0
								        }
								\\
				\label{Equ:Performance:SNR_T0}
				&\textsf{SNR}_\textsf{T0}	=	\frac{N}{L}\cdot 
										\frac{1}{[{(N-\beta)}/{(\gamma_1)^+}-\beta]^{+}}
										\\
				\label{Equ:Performance:K0}
				&K_0	=	\frac{\beta+\frac{N}{L}/\textsf{SNR}_{\textsf{T}0}}
						       {N-\beta}
						\left(\gamma_1-\frac{N-\beta}{\beta}\right)^+
			\end{align}
		where $(\cdot)^+ = \max\{\cdot,0\}$, and $\gamma_1$ is the largest nonzero
		generalized eigenvalue of $(\mathbf{Q}_\mathcal{S}\!\!-\!\!\mathbf{Q}_\mathcal{I},
		\mathbf{Q}_\mathcal{I})$.
		\footnote{If $(\mathbf{Q}_\mathcal{S}\!\!-\!\!\mathbf{Q}_\mathcal{I},\mathbf{Q}_\mathcal{I})$ 
		has less than $D$ nonzero eigenvalues, pad them
		with zeros up to $D$ and order them decreasingly.}
		 The $\textsf{SNR}$ at which $\textsf{G}(\textsf{SNR})$
		is close to $G_L$ and $G_U$ (within $3$dB), which are given by
			\begin{align}
				\label{Equ:Performance:SNR_T1_AND_SNR_T2}
				\textsf{SNR}_{\textsf{T}1}	=	\left(1-\sqrt{\frac{1}{2}}\right)
										\textsf{SNR}_{\textsf{T}0},
				\qquad
				\textsf{SNR}_{\textsf{T}2}	=	\left(1-\sqrt{\frac{P_I}{2P_I+1}}\right)^{-1}
										\!\!\!
										\textsf{SNR}_{\textsf{T}0},
			\end{align}
respectively.
	\end{theorem}

\begin{figure*}[t]
	\centering{
				\subfigure[]
				{
				\begin{overpic}[width=0.42\textwidth]{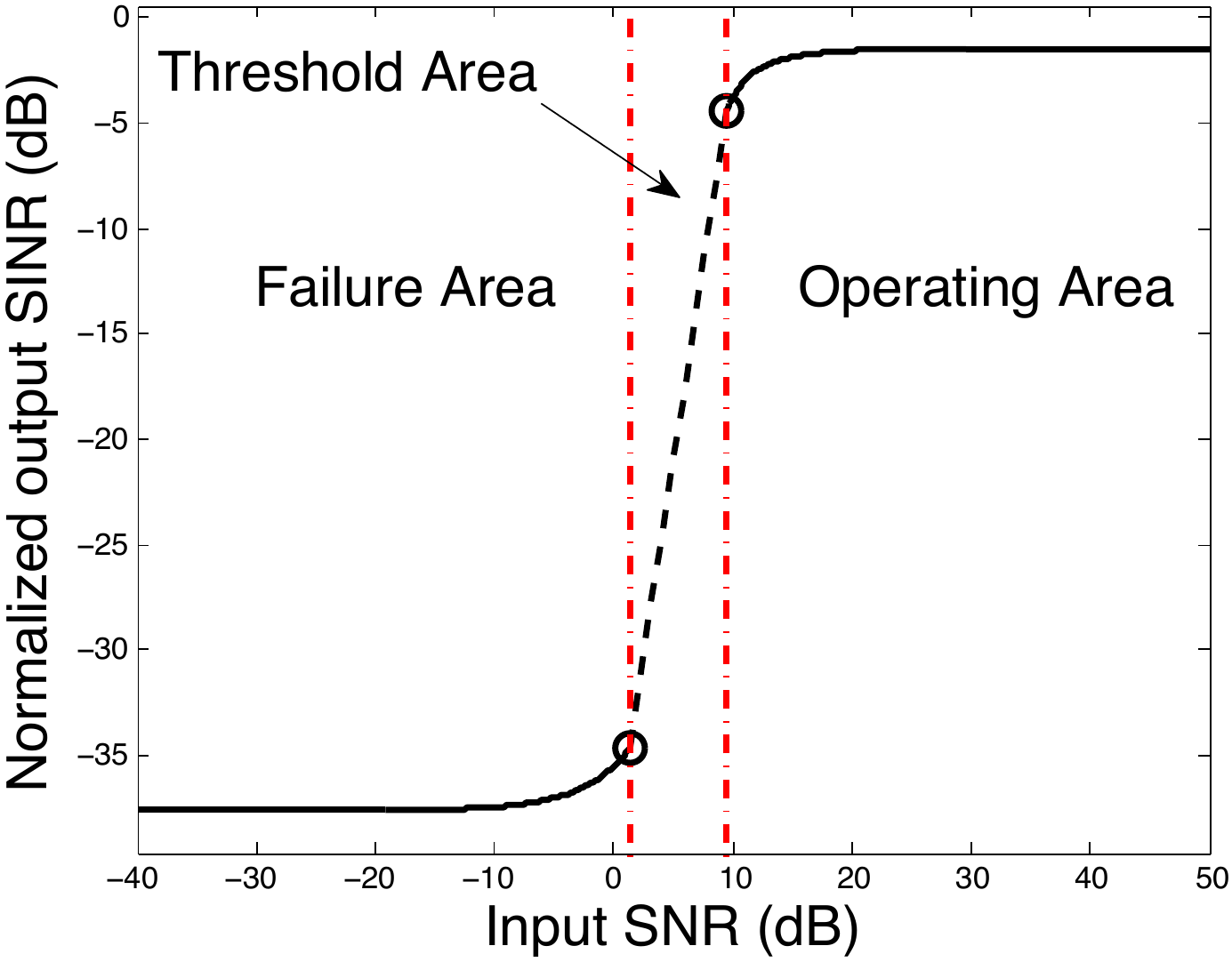}
				\put(36,25){\small{$\textsf{SNR}_{\textsf{T1}}$}}
				\put(61,25){\small{$\textsf{SNR}_{\textsf{T2}}$}}
				\end{overpic}
				\label{Fig:OperatingCurve:Regular}}
			}\\
	\centering{
				\subfigure[]
				{
				\begin{overpic}[width=0.42\textwidth]{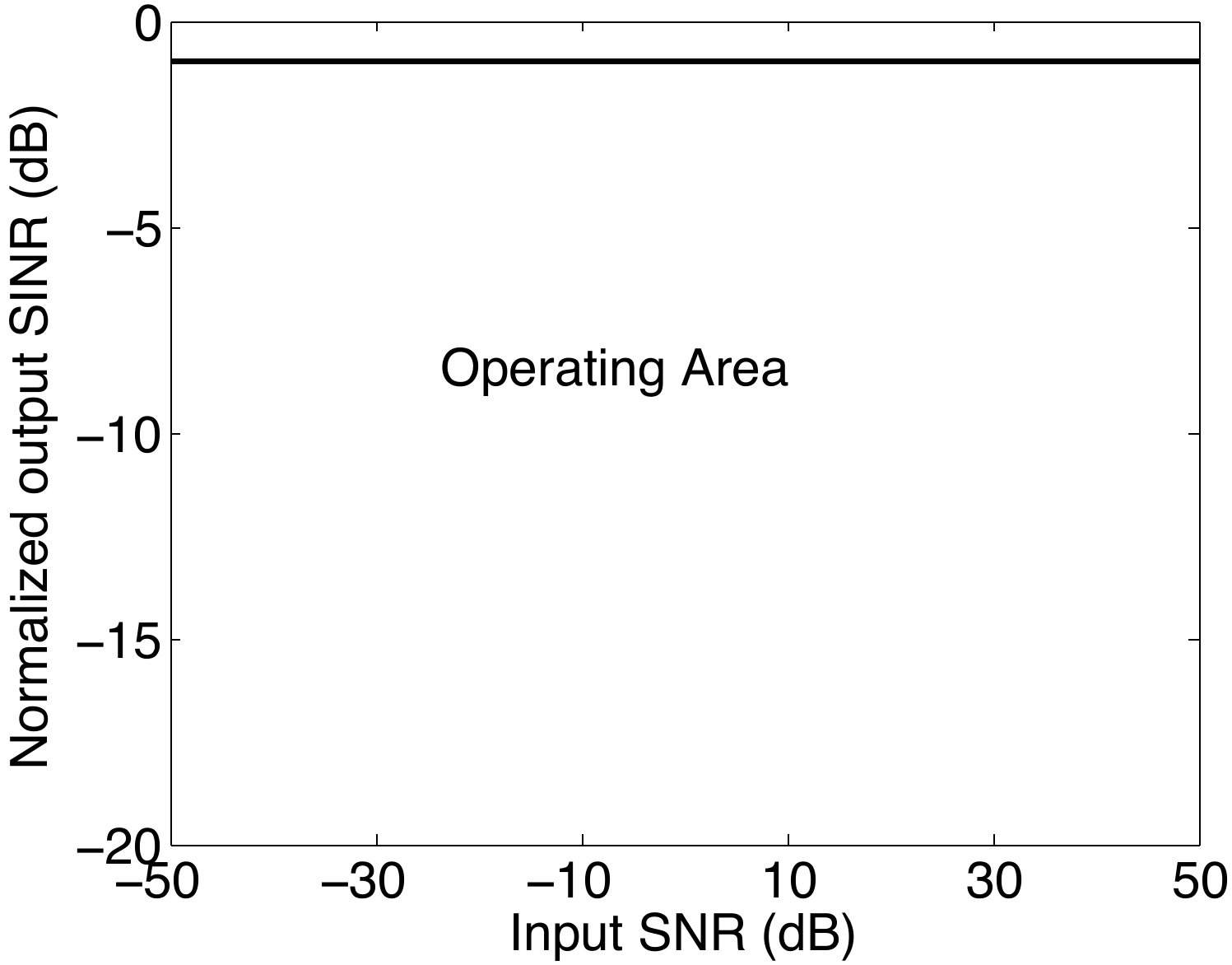}
				\put(20,25){\small{$\textsf{SNR}_{\textsf{T1}}=\textsf{SNR}_{\textsf{T2}}=-\infty$ dB}}
				\end{overpic}
				\label{Fig:OperatingCurve:Operating}}
				\subfigure[]
				{
				\begin{overpic}[width=0.42\textwidth]{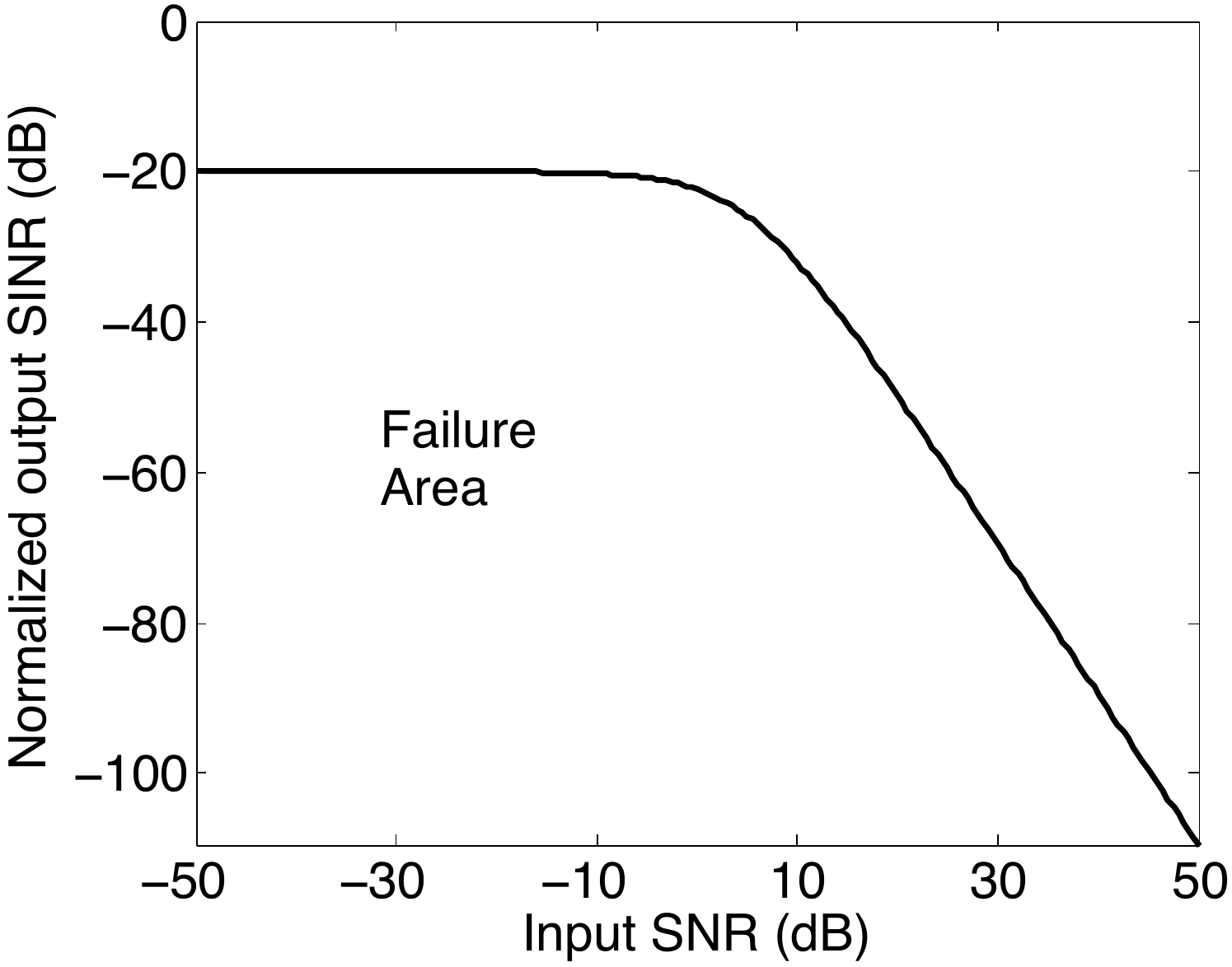}
				\put(20,25){\small{$\textsf{SNR}_{\textsf{T1}}=\textsf{SNR}_{\textsf{T2}}=+\infty$ dB}}
				\put(66,57){\small{$\propto -20\log \textsf{SNR}$}}
				\end{overpic}
				\label{Fig:OperatingCurve:Failure}}
			}
	\caption{(a) A typical curve of $\textsf{G}(\textsf{SNR})$ has
	failure area, threshold area and operating area, separated by 
	$\textsf{SNR}_{\textsf{T1}}$ and $\textsf{SNR}_{\textsf{T2}}$. 
	(b) Without matrix mismatch, the curve has only operating area. 
	(c) A curve has only failure area for some cases in PAPC.}
	\label{Fig:SpecialOperatingCurve}
\end{figure*}

Though we will give expression for $G_L$ in Sec. 
		\ref{Sec:Performance_Analysis:Deriving_G_SNR}, its specific values are of no interest
		to us.	
Fig. \ref{Fig:OperatingCurve:Regular} shows a typical curve of $\textsf{G}(\textsf{SNR})$.
We plot the curve in failure area and operating area given in 
\eqref{Equ:Performance:G_Eb}, and connect their ends by a dashed line.	 
We can see that the performance of beamformer degrades 
rapidly when the input SNR is below $\textsf{SNR}_{\textsf{T2}}$.
And it fails completely after reaching $\textsf{SNR}_{\textsf{T1}}$.
Therefore, matrix mismatch causes a threshold effect in MPB,
and $\textsf{SNR}_{\textsf{T2}}$ is a critical parameter to be optimized.

There are two special cases of $\textsf{G}(\textsf{SNR})$. 
The first one is $\textsf{SNR}_{\textsf{T1}}\!\!=\!\!\textsf{SNR}_{\textsf{T2}}\!\!=\!\!0$ ($-\infty$dB).
This happens when there is no matrix mismatch, i.e. $\mathbf{Q}_\mathcal{S}\!\!=\!\!\mathbf{Q}_\mathcal{I}$,
so that $\gamma_1 \!\!=\!\! 0$ and $\textsf{SNR}_{\textsf{T}i}$ given by \eqref{Equ:Performance:SNR_T0}
and \eqref{Equ:Performance:SNR_T1_AND_SNR_T2} are zero.
Then, \eqref{Equ:Performance:G_Eb} implies 
the operating curve is a horizontal line as shown in 
Fig. \ref{Fig:OperatingCurve:Operating}. Another interesting case happens in PAPC schemes
mentioned in \ref{Sec:ProblemFormulation:MPBeamformer}, whose PLR is $\beta\!\!=\!\!1\!\!>\!\!0$.
If it further satisfies $(N-\beta)/\gamma_1\!\!<\!\!\beta$, then by
\eqref{Equ:Performance:SNR_T0} and \eqref{Equ:Performance:SNR_T1_AND_SNR_T2},
$\textsf{SNR}_{\textsf{T}1}\!\!=\!\!\textsf{SNR}_{\textsf{T}2}=+\infty$
and $\textsf{G}(\textsf{SNR})$ only has the failure area, which
decreases in the order of  $\mathcal{O}(\textsf{SNR}^{-2})$
(c.f. Fig. \ref{Fig:OperatingCurve:Failure}). 
In the rest of the section, we will derive $\textsf{G}(\textsf{SNR})$
and reveal how MPB works blindly under matrix mismatch.
The discussion of $\textsf{SNR}_{\textsf{T0}}$ is left to Sec. \ref{Sec:Threshold}.

\subsection{Derivation of the Weight Vector for MPB}
As a step to prove Theorem \ref{thm:operating_curve}, we first derive the
expression of the weight vector for MPB. By the arguments in 
Sec. \ref{Sec:ProblemFormulation:MPBeamformer}, it is the solution to 
\eqref{Equ:MPBeamformer:EigenEquation}.
Define $\bm{\Phi}_\Delta \triangleq \bm{\Phi}_\mathcal{S}-\bm{\Phi}_\mathcal{I}$.
Then, by \eqref{Equ:MPBeamformer:Q_S}, \eqref{Equ:MPBeamformer:Q_I}
and \eqref{Equ:MPBeamformer:R_S_minus_R_I}, 
the problem becomes solving the following generalized eigenequation
	\begin{align}
		\label{Equ:Performance:EigenEquation1}
		\left[
			(\sigma_{\mathcal{S}_0}^2-\sigma_{\mathcal{I}_0}^2) 
			\mathbf{a}_0\mathbf{a}_0^H 
			+ 
			\mathbf{A}_I \bm{\Phi}_\Delta \mathbf{A}_I^H
		\right]
		\mathbf{w}
		=
		(\lambda_\mathrm{max}-1)
		\mathbf{R}_\mathcal{I}
		\mathbf{w}, 
	\end{align}
where $\lambda_\mathrm{max}$ is the largest eigenvalue of 
$(\mathbf{R}_\mathcal{S},\mathbf{R}_\mathcal{I})$.
Sec. \ref{Sec:ProblemFormulation:MPBeamformer} 
already gave the result for $\mathbf{w}$ when there is no matrix mismatch, i.e.
$\bm{\Phi}_\Delta \!\!=\!\! \mathbf{O}$ (or $\mathbf{Q}_\mathcal{S}\!\!=\!\!\mathbf{Q}_\mathcal{I}$). 
We concluded that it is optimal in the sense of maximizing the output SINR.
However, in the presence of matrix mismatch, we have $\bm{\Phi}_\Delta \!\!\neq\!\! \mathbf{O}$, 
which is the key challenge for
solving \eqref{Equ:Performance:EigenEquation1}.
To deal with this problem, we first have the following two observations:
	\begin{itemize}
		\item
			By left-multiplying $\frac{1}{\lambda_\mathrm{max}-1}\mathbf{R}_\mathcal{I}^{-1}$ to 
			both sides of \eqref{Equ:Performance:EigenEquation1}, we can see that $\mathbf{w}$
			can be expressed as
				\begin{align}
					\label{Equ:Performance:w1}
					\mathbf{w}	=	\eta_0 \mathbf{R}_\mathcal{I}^{-1}\mathbf{a}_0
									+
									\sum_{i=1}^D
									\eta_i \mathbf{R}_\mathcal{I}^{-1}\mathbf{a}_{i}
				\end{align}
			i.e. it is a linear combination of $\mathbf{R}_\mathcal{I}^{-1}\mathbf{a}_0,
			\mathbf{R}_\mathcal{I}^{-1}\mathbf{a}_1,\ldots,\mathbf{R}_\mathcal{I}^{-1}\mathbf{a}_D$,
			where $\eta_0 = \frac{\sigma_{\mathcal{S}_0}^2-\sigma_{\mathcal{I}_0}^2}
			{\lambda_\mathrm{max}-1}\mathbf{a}_0^H\mathbf{w}$, $\mathbf{a}_i$ is
			the $i$th column of $\mathbf{A}_I$, and
			$\eta_i$ is the $i$th component of $\bm{\eta} = \bm{\Phi}_\Delta\mathbf{A}_I^H\mathbf{w}$.
			To determine $\eta_0, \eta_1,\ldots, \eta_D$, we only need to 
			substitute \eqref{Equ:Performance:w1} back into 
			\eqref{Equ:Performance:EigenEquation1} and solve a
			linear equation. However, we will immediately
			discover that the solution is intractable for further analysis
			because $\mathbf{A}_I^H\mathbf{R}_\mathcal{I}^{-1}\mathbf{A}_I$
			and $\bm{\Phi}_\Delta$ are not diagonal. To overcome this, we need the
			next observation.
		\item
			Suppose we can factorize $\bm{\Phi}_\Delta$ into diagonal form such that
			$\mathbf{A}_I\bm{\Phi}_\Delta\mathbf{A}_I^H=
			\mathbf{A}_\epsilon\bm{\Gamma}\mathbf{A}_\epsilon$ and 
			$\mathbf{A}_\epsilon^H\mathbf{R}_\mathcal{I}^{-1}\mathbf{A}_\epsilon=\mathbf{I}$,
			where $\bm{\Gamma}$ is a diagonal matrix. Then, by repeating the above procedure
			except for replacing $\mathbf{a}_1,\ldots, \mathbf{a}_D$ by 
			$\mathbf{a}_{\epsilon_1},\ldots, \mathbf{a}_{\epsilon_D}$, 
			with $\mathbf{a}_{\epsilon_i}$ being the $i$th column of $\mathbf{A}_\epsilon$,
			we can solve $\mathbf{w}$ as
				\begin{align}
					\label{Equ:Performance:w2}
					\mathbf{w}	=	\eta_0
									\left[
										\mathbf{R}_\mathcal{I}^{-1}\mathbf{a}_0
										+
										\sum_{i=1}^D
										\frac{\gamma_i\tilde{\psi}_{T_i}}
											{\lambda_\mathrm{max}-1-\gamma_i}
										\mathbf{R}_\mathcal{I}^{-1}\mathbf{a}_{\epsilon_i}
									\right],
				\end{align}
			where $\gamma_i$ is the $i$th diagonal component of $\bm{\Gamma}$, $\eta_0$ is an
			arbitrary constant,  and 
			$\tilde{\psi}_{T_i} = \mathbf{a}_{\epsilon_i}^H\mathbf{R}_\mathcal{I}^{-1}\mathbf{a}_0$.
	\end{itemize}
	
So far, the only thing left is to figure out a way to factorize $\bm{\Phi}_\Delta$ so that it meets the above
requirement. Consider the generalized eigen-decomposition of the matrix pair 
$(\bm{\Phi}_\Delta, (\mathbf{A}_I^H\mathbf{R}_\mathcal{I}^{-1}\mathbf{A}_I)^{-1})$. By the simultaneous
diagonalization theorem (c.f. \cite{laub2004matrix}, pp.133), there exists a nonsingular matrix $\mathbf{T}$
such that
	\begin{align}
		\label{Equ:Performance:JointDiag}
		&\mathbf{T}^H\bm{\Phi}_\Delta\mathbf{T} = \bm{\Gamma},
		\qquad
		\mathbf{T}^H (\mathbf{A}_I^H\mathbf{R}_\mathcal{I}^{-1}\mathbf{A}_I)^{-1}\mathbf{T} = \mathbf{I},
	\end{align}
where $\bm{\Gamma} = \mathrm{diag}(\gamma_1,\gamma_2,\ldots,\gamma_D)$ is a diagonal matrix
whose diagonal terms are the generalized eigenvalues of 
$(\bm{\Phi}_\Delta, (\mathbf{A}_I^H\mathbf{R}_\mathcal{I}^{-1}\mathbf{A}_I)^{-1})$.
Define $\mathbf{A}_\epsilon = \mathbf{A}_I(\mathbf{T}^{-1})^H$. Then, by 
\eqref{Equ:Performance:JointDiag},
we can verify that
	\begin{align}
		&\mathbf{A}_I\bm{\Phi}_\Delta\mathbf{A}_I^H 
			= 	\mathbf{A}_I(\mathbf{T}^{-1})^H\bm{\Gamma}\mathbf{T}^{-1}\mathbf{A}_I^H
			=	\mathbf{A}_\epsilon \bm{\Gamma} \mathbf{A}_\epsilon^H
				\nonumber\\
		&\mathbf{A}_\epsilon\mathbf{R}_\mathcal{I}^{-1}\mathbf{A}_\epsilon
			=	\mathbf{T}^{-1}\mathbf{A}_I^H\mathbf{R}_\mathcal{I}^{-1}\mathbf{A}_I(\mathbf{T}^{-1})^H
			=	\mathbf{I}.
				\nonumber
	\end{align}
Therefore, \eqref{Equ:Performance:JointDiag} is
the exact decomposition of $\bm{\Phi}_\Delta$ we are looking for. Furthermore, by
$\mathbf{A}_\epsilon = \mathbf{A}_I(\mathbf{T}^{-1})^H$, we notice that each column of
$\mathbf{A}_\epsilon$ is in fact a linear combination of $\mathbf{a}_1,\ldots,\mathbf{a}_D$.
This, together with \eqref{Equ:Performance:w2}, implies that $\mathbf{w}$ 
is still a a linear combination of $\mathbf{R}_\mathcal{I}^{-1}\mathbf{a}_0,
\mathbf{R}_\mathcal{I}^{-1}\mathbf{a}_1,\ldots,\mathbf{R}_\mathcal{I}^{-1}\mathbf{a}_D$, just
as that in \eqref{Equ:Performance:w1}.

\subsection{How MPB Works Blindly}
\label{Sec:Performance:How_MPB_works}
To fully understand the behavior of MPB given by \eqref{Equ:Performance:w2}
and how it works blindly, 
we still need the expressions of $\lambda_\mathrm{max}$ and $\gamma_i$. 
However, $\lambda_\mathrm{max}$ and $\gamma_i$ are the solutions to
	\begin{align}
		\label{Equ:Performance:EigenEquation_lambdamax1}
		&\mathrm{det}
		\left\{
			(\lambda-1) \mathbf{R}_\mathcal{I}
			-
			(\sigma_{\mathcal{S}_0}^2-\sigma_{\mathcal{I}_0}^2) 
			\mathbf{a}_0\mathbf{a}_0^H 
			-
			\mathbf{A}_I \bm{\Phi}_\Delta \mathbf{A}_I^H
		\right\}
		=
		0
		\\
		\label{Equ:Performance:EigenEquation_lambda_i_1}
		&\mathrm{det}
		\left\{
			\lambda (\mathbf{A}_I^H\mathbf{R}_\mathcal{I}^{-1}\mathbf{A}_I)^{-1}
			-
			\bm{\Phi}_\Delta
		\right\}
		=0
	\end{align}
which are polynomial equations of degree $D+1$ and $D$, respectively.  
It is known that there are no general 
closed-form solutions if their degrees are higher than four. 
Thus, our approach here is to derive  approximate expressions
for $\lambda_\mathrm{max}$ and $\gamma_i$.
The main idea is to transform \eqref{Equ:Performance:EigenEquation_lambdamax1}
and \eqref{Equ:Performance:EigenEquation_lambda_i_1} into eigenvalue problems of
a diagonal matrix perturbed by a small term. 
Then we can approximate the eigenvalues by the diagonal entries, and bound the error
using matrix perturbation theory.

We first discuss $\lambda_\mathrm{max}$. Before this, let's introduce the following
identity which will be used repeatedly in the analysis and can be derived from 
Properties 16 and 17 in \cite[pp.5]{laub2004matrix}.
	\begin{align}
		\label{Equ:Performance:Identity_det}
		\det( \lambda \mathbf{I} - \mathbf{X}\mathbf{Y} )	
				=	\lambda^{m-n} \cdot \det( \lambda \mathbf{I} - \mathbf{Y}\mathbf{X} )												
	\end{align}
where $\mathbf{X} \in \mathbb{C}^{m\times n}$ and $\mathbf{Y} \in \mathbb{C}^{n \times m}$.
Substituting the factorization of $\bm{\Phi}_\Delta$ in \eqref{Equ:Performance:JointDiag}
into \eqref{Equ:Performance:EigenEquation_lambdamax1} and using
\eqref{Equ:Performance:Identity_det} as well as $\mathbf{A}_\epsilon = \mathbf{A}_I(\mathbf{T}^{-1})^H$, 
we can have the equivalent form of \eqref{Equ:Performance:EigenEquation_lambdamax1} as below
	\begin{align}
		\label{Equ:Performance:EigenEquation_lambdamax2}
		(\lambda-1)^{L-D-1}
		\cdot
		\det \mathbf{R}_\mathcal{I}
		\cdot
		\det
			\left\{
				(\lambda-1) \mathbf{I}
				-
				\begin{bmatrix}
					\sigma_{\mathcal{S}_0}^2 - \sigma_{\mathcal{I}_0}^2		&	\mathbf{0}^T	\\
					\mathbf{0}										&	\bm{\Gamma}
				\end{bmatrix}
				\begin{bmatrix}
					\mathbf{a}_0^H\mathbf{R}_\mathcal{I}^{-1}\mathbf{a}_0
						&		\mathbf{a}_0^H\mathbf{R}_\mathcal{I}^{-1}\mathbf{A}_\epsilon	\\
					\mathbf{A}_\epsilon^H\mathbf{R}_\mathcal{I}^{-1}\mathbf{a}_0
						&		\mathbf{A}_\epsilon^H\mathbf{R}_\mathcal{I}^{-1}\mathbf{A}_\epsilon
				\end{bmatrix}
			\right\}
		=
		0.
	\end{align}
Then, \eqref{Equ:Performance:EigenEquation_lambdamax2} implies that 
the solution of \eqref{Equ:Performance:EigenEquation_lambdamax1} is the same as the eigenvalues of
the following matrix
	\begin{align}
		\label{Equ:Performance:M}
		\mathbf{M}	=	\begin{bmatrix}
							\gamma_0+1		&	\gamma_0
												\sqrt{\frac{\sigma^2}{L}}
												\bm{\psi}_T^H
												\\
							\frac{\sqrt{\frac{\sigma^2}{L}}}
								{\frac{L\beta}{N}\textsf{SNR}+1}\bm{\Gamma}\bm{\psi}_T	
											&	\bm{\Gamma}+\mathbf{I}
						\end{bmatrix}
	\end{align}
except for the multiplicity of ones, where we have used the second equality of \eqref{Equ:Performance:JointDiag}.
And $\gamma_0$ and $\bm{\psi}_T$ are 
	\begin{align}
		\label{Equ:Performance:lambda_0_AND_psi_T}
		&\gamma_0	\triangleq	
						(\sigma_{\mathcal{S}_0}^2-\sigma_{\mathcal{I}_0}^2)
						\mathbf{a}_0^H\mathbf{R}_\mathcal{I}^{-1}\mathbf{a}_0,
						\qquad
		\bm{\psi}_T	\triangleq
						\left( \frac{L\beta}{N}\textsf{SNR}+1 \right)
						\mathbf{A}_\epsilon^H\mathbf{R}_\mathcal{I}^{-1}\mathbf{a}_0
	\end{align}	
%
Before proceeding on, we cite the following two lemmas. The first one
is summarized from Lemma 1 and Lemma 2 in \cite{chen2010unpublished}, and the second one
can be found in \cite[pp.344]{horn1990ma}.
    	\begin{lemma}
        	\label{Lem:aRa_psi_T}
       		The quantities $\mathbf{a}_0^H\mathbf{R}_\mathcal{I}^{-1}\mathbf{a}_0$
		and $\bm{\psi}_T$ have the following results
            	\begin{align}
	            	\mathbf{a}_0^H\mathbf{R}_\mathcal{I}^{-1}\mathbf{a}_0 
						\approx			\frac{L}{\sigma^2}
	                                                                            \frac{
	                                                                                    N
	                                                                                 }
	                                                                                 {
	                                                                                    {L\beta}
	                                                                                    \textsf{SNR}
	                                                                                    +N
	                                                                                 },
	           				\qquad
	                	\|\pmb{\psi}_T\|	\approx			\sqrt{\frac{L}{\sigma^2} \kappa_0}
						\ll				\sqrt{\frac{L}{\sigma^2}}.
	                                            \nonumber
            \end{align}
           where $\beta$ is the PLR defined in Theorem \ref{thm:operating_curve},
           and $\kappa_0$ is a very small positive number independent of $\textsf{SNR}$.
    \end{lemma}
	\begin{lemma}[Gerschgorin]
	        \label{Lem:Gerschgorin}
	        Let matrix $\mathbf{A} \in \mathbb{C}^{n \times n}$ and $a_{ij}$ be its $i,j$th element. 
	        Define
	            \begin{align}
	                R_i(\mathbf{A})\triangleq \sum_{j=1,j \neq i}^n | a_{i,j}|, \quad 1\leq i\leq n
	                \nonumber
	            \end{align}
	        as the \emph{deleted absolute row sums} of $\mathbf{A}$ . 
	        Then all the eigenvalues of $\mathbf{A}$ are
	        in the union of $n$ disks, i.e.
	            \begin{align}
	                \lambda(\mathbf{A})\in\bigcup_{i=1}^n G_i(\mathbf{A}),
	                \nonumber
	            \end{align}
	        where $G_i(\mathbf{A})\!\!\triangleq\!\!\{z\!\!\in\!\!\mathbb{C}\!\!:\!\! |z\!\!-\!\!a_{ii}|\!\!\leq\!\!
	        R_i(\mathbf{A})\}$ is the $i$th Gerschgorin disk.
	        Moreover, if a union of  $k$ disks forms a connected 
	        region disjoint from all the remaining disks, then there 
	        are $k$ eigenvalues of $\mathbf{A}$ in this region.
	    \end{lemma}
	    
By Lemma \ref{Lem:aRa_psi_T}, the off-diagonal terms of each row in 
\eqref{Equ:Performance:M} are much smaller than the corresponding diagonal 
terms. Then, according to lemma \ref{Lem:Gerschgorin},
its eigenvalues are approximately $\gamma_0\!+\!1,\ldots,\gamma_D\!+\!1$.
We will give a bound for the approximation error in the next subsection. 
Now we directly use this approximation to discuss the behavior of MPB.
Without loss of generality, assume $\gamma_1 > \ldots > \gamma_D$, then
	\begin{align}
		\label{Equ:Performance:lambda_max1}
		\lambda_\mathrm{max} 	\approx 	\max\{\gamma_0+1, \gamma_1+1, \ldots,
										\gamma_D+1, 1\}
							=		\max\{\gamma_0+1, \gamma_1+1\}.
	\end{align}
Substituting the expression of $\mathbf{a}_0^H\mathbf{R}_\mathcal{I}^{-1}\mathbf{a}_0$
in Lemma \ref{Lem:aRa_psi_T} into its definition in
\eqref{Equ:Performance:lambda_0_AND_psi_T}, we can have
	\begin{align}
		\label{Equ:Performance:lambda_0}
		\gamma_0		=	(N-\beta)\frac{L\textsf{SNR}}{L\beta \textsf{SNR}+N}
	\end{align}
which is a monotonically increasing function of $\textsf{SNR}$.
Furthermore, we will show later that $\gamma_1\!\!+\!\!1$ is approximate to
one of the generalized eigenvalue of $(\mathbf{Q}_\mathcal{S},\mathbf{Q}_\mathcal{I})$
and is almost independent of $\textsf{SNR}$.
Hence, there is a threshold $\textsf{SNR}_{\textsf{T0}}$
such that $\lambda_\mathrm{max}$ switches from $\gamma_1+1$ to $\gamma_0+1$
when $\textsf{SNR}$ exceeds $\textsf{SNR}_{\textsf{T0}}$. Its expression is
in \eqref{Equ:Performance:SNR_T0} and can be derived by 
setting \eqref{Equ:Performance:lambda_0}
to $\gamma_1$ and solving $\textsf{SNR}$. As a result,
	\begin{itemize}
		\item
            		When $\textsf{SNR}>\textsf{SNR}_\textsf{T0}$, 
			$\lambda_\mathrm{max} \approx \gamma_0+1$.
			If $\gamma_0+1$ can be much greater than $\gamma_1+1$ as
			$\textsf{SNR}$ increases, then \eqref{Equ:Performance:w2} implies 
		        $\mathbf{w} \approx \mu_1 \mathbf{R}_\mathcal{I}^{-1} \mathbf{a}_0$.
		        Hence, $\mathbf{a}_0$ is the dominant steering vector, and
		        the beamformer can operate properly by steering the mainlobe
		        to the direction of the desired signal.
		\item		        
            		When $\textsf{SNR}<\textsf{SNR}_\textsf{T0}$, 
			$\lambda_\mathrm{max} \approx \gamma_1+1$. Thus
                		$\mathbf{w} \approx \mu_1 \mathbf{R}_\mathcal{I}^{-1} \mathbf{a}_{\epsilon_1}$, 
			and $\mathbf{a}_{\epsilon_1}$ is the dominant steering vector in
			\eqref{Equ:Performance:w2}. By $\mathbf{A}_\epsilon = \mathbf{A}_I(\mathbf{T}^{-1})^H$,
			$\mathbf{a}_{\epsilon_1}$ is a linear combination of all the interference steering vectors.
            		Therefore, the beamformer fails for its mainlobe 
pointing to the directions of the interferers.
			If $\beta \!\!\neq\!\! 0$, then there would be SOI in $\mathbf{R}_\mathcal{I}$,
			which makes MPB treat it as interference and null it out.
			This explains the reason why $\textsf{G}(\textsf{SNR})$ decreases in the order
			of $\mathcal{O}(\textsf{SNR}^{-2})$ in Fig. \ref{Fig:OperatingCurve:Failure}.
	\end{itemize}

Now, we further analyze how this threshold effect happens. By the factorizations
\eqref{Equ:Performance:JointDiag}, we
can rewrite $\mathbf{R}_\mathcal{S}-\mathbf{R}_\mathcal{I}$ on the left
hand side of \eqref{Equ:Performance:EigenEquation1} as
    \begin{align}
        \label{Equ:Performance:Rs_minus_RI_expr2}
        \mathbf{R}_\mathcal{S}-\mathbf{R}_\mathcal{I}		=	\gamma_0
        												\cdot 
												\mathbf{a}_{S_0}\mathbf{a}_{S_0}^H
                                                                    				+
                                                                    				\sum_{i=1}^D
                                                                    				\gamma_i
												\cdot 
												\mathbf{a}_{\epsilon_i}
												\mathbf{a}_{\epsilon_i}^H,
    \end{align}
where $\mathbf{a}_{S_0} \triangleq \mathbf{a}_0/
[\mathbf{a}_0^H\mathbf{R}_\mathcal{I}^{-1}\mathbf{a}_0]^{\frac{1}{2}}$. 
It is obvious that $\gamma_0$ is the generalized eigenvalue of the 
matrix pair
$\left(  (\sigma_{\mathcal{S}_0}^2\!\!-\!\sigma_{\mathcal{I}_0}^2)\mathbf{a}_0\mathbf{a}_0^H,
\mathbf{R}_\mathcal{I}  \right)$, with $\mathbf{R}_\mathcal{I}^{-1}\mathbf{a}_{S_0}$ being its
``normalized'' eigenvector. 
By applying \eqref{Equ:Performance:Identity_det} to
\eqref{Equ:Performance:EigenEquation_lambda_i_1}, we can also verify that
$\gamma_1,\gamma_2,\ldots,\gamma_D$ are the generalized eigenvalues of
$\left(\mathbf{A}_I \bm{\Phi}_\Delta \mathbf{A}_I^H, \mathbf{R}_\mathcal{I}\right)$,
with the eigenvectors being
$\{\mathbf{R}_\mathcal{I}^{-1}\mathbf{a}_{\epsilon_i}\}_{i=1}^D$.
Comparing \eqref{Equ:Performance:Rs_minus_RI_expr2} with
the left hand side of \eqref{Equ:Performance:EigenEquation1},
we can find that $\gamma_0$ actually measures the
mismatch of the desired signal between the signal channel and the interference channel
in Fig. \ref{Fig:MatrixPairBeamformer_lthAntenna}, and
$\{\gamma_i\}_{i=1}^D$ measures that of the interferers.
From the previous discussion, we know that  
$\lambda_\mathrm{max} = \max\{\gamma_0+1,\gamma_1+1\}$,
and $\mathbf{w}$ takes $\mathbf{R}_\mathcal{I}^{-1}\mathbf{a}_0$ or 
$\mathbf{R}_\mathcal{I}^{-1}\mathbf{a}_{\epsilon_1}$ depending on
which one of $\gamma_0+1$ and $\gamma_1+1$ is larger.
Therefore, MPB \emph{blindly} chooses the 
one with the largest mismatch metrix in \eqref{Equ:Performance:Rs_minus_RI_expr2} 
as its steering vector. We term this as ``\emph{steering vector competition}''.
The threshold happens when the $\textsf{SNR}$ is large enough 
so that $\gamma_0$ exceeds $\gamma_1$.
If there is no mismatch, i.e. $\bm{\Phi}_\Delta = \mathbf{O}$ so that $\gamma_i=0$,
then the desired signal always wins out in the competition
($\gamma_0 > \gamma_1$), and MPB can work properly for all $\textsf{SNR}$.

\subsection{Proof of Theorem \ref{thm:operating_curve}}
\label{Sec:Performance_Analysis:Deriving_G_SNR}
Let us return to our original problem of operating curve for MPB and prove
Theorem \ref{thm:operating_curve}. We first use the expression of $\mathbf{w}$ in \eqref{Equ:Performance:w2} to derive $\textsf{G}(\textsf{SNR})$ in term of $\lambda_\mathrm{max}$.
The result is given by the following lemma, with its proof in 
Appendix \ref{Sec:Appendix:Proof_of_Thm_G_lambda_max}.
    \begin{lemma}
        \label{lemma:G_lambda_max}
        The normalized output SINR in \eqref{Equ:Matrix_Mismatch:G} 
        can be expressed as the following function of $\lambda_\mathrm{max}$
        		\begin{align}
        			\label{Equ:Performance:G_lambda_max_expr3}
                   	\textsf{G}(\textsf{SNR})   
					= 	\frac{
                                                                	\displaystyle
                                                                	\left[
                                                                        1 
                                                                        + 
                                                                        \frac{N}{L\beta\textsf{SNR}+N}  
                                                                        \psi_S(\lambda_{\mathrm{max}})
                                                                	\right]^2
                                                                  }
                                                                  {
                                                                	\displaystyle
                                                                	\left[
                                                                        	\psi_I(\lambda_{\mathrm{max}})
                                                                        	-
                                                                        	\frac{L\beta\textsf{SNR}}{L\beta\textsf{SNR}+N}
                                                                        	\psi_S^2(\lambda_{\mathrm{max}})
                                                                	\right]
                                                                	+
                                                                	\left[
                                                                        	1 
									+ 
									\frac{N}{L\beta\textsf{SNR}+N}  
									\psi_S(\lambda_{\mathrm{max}})
                                                                	\right]^2
                                                             	}.
           	\end{align}
	 where $\psi_S(\lambda_\mathrm{max})$ and $\psi_I(\lambda_\mathrm{max})$ are 
	 in the following forms with $\psi_{T_i}$ being the $i$th component of $\bm{\psi}_T$.
            	\begin{align}
                		\label{Equ:Performance:psi_S_lambda_max}
                		&\psi_S(\lambda_\mathrm{max})
                			\!\!\triangleq\!\!
                						\frac{\sigma^2}{L}
                						\sum_{i=1}^D
                						\!\!
					                	\frac{
					                        	\lambda_\mathrm{max}\!\!-\!\!1
					                        }
					                        {
					                        	\lambda_\mathrm{max}
					                        	\!\!-\!\!
					                        	(\gamma_i\!\!+\!\!1)
					                    	}
					                	|\psi_{T_i}|^2,
			\qquad
		             \psi_I(\lambda_\mathrm{max})
					                	\!\!
					                	\triangleq
					                	\!\!
					                	\frac{\sigma^2}{L}
					                	\!
					                	\sum_{i=1}^D
					                	\!
					                	(\gamma_i \!\!+\!\! 1)
					                	\!
					                	\left[
					                        \frac{
					                                	\lambda_\mathrm{max}\!\!-\!\!1
					                               }
					                               {
					                                	\lambda_\mathrm{max}
					                                	\!\!-\!\!
					                                	(\gamma_i\!\!+\!\!1)
					                               }
					                	\right]^2
					                	\!\!\!
					                	|\psi_{T_i}|^2.
	            \end{align}
    \end{lemma}
    
To further simplify \eqref{Equ:Performance:G_lambda_max_expr3}, we need the
expression of $\lambda_\mathrm{max}$.
In the previous subsection, we used the approximation 
\eqref{Equ:Performance:lambda_max1} to analyze MPB. 
The following lemma quantifies how precise it is by 
bounding the approximation error. Besides checking the validity
of \eqref{Equ:Performance:lambda_max1}, 
this is also critical in deriving $\textsf{G}(\textsf{SNR})$.
	\begin{lemma}
		\label{Lem:lambda_max}
		The largest generalized eigenvalue of
		$(\mathbf{R}_\mathcal{S},\mathbf{R}_\mathcal{I})$ defined in
		\eqref{Equ:MPBeamformer:R_S} and \eqref{Equ:MPBeamformer:R_I} 
		satisfies
			\begin{align}
				\label{Equ:Performance:Thm_lambda_max:lambda_max}
				|\lambda_\mathrm{max}-(\lambda_a+1)|	\le	\lambda_a
													\cdot
													f\left(
														\frac{\lambda_b}{\lambda_a}
													\right)
			\end{align}
		where $\lambda_a \triangleq \max\{\gamma_0,\gamma_1\}$ and
		$\lambda_b \triangleq \min\{\gamma_0,\gamma_1\}$. $f(x)$ and $\delta$
		are defined as
			\begin{align}
				f(x)	&\triangleq	\frac{1}{2}
								\left[1\!\!-\!\!x\!\!-\!\!\sqrt{(1\!\!-\!\!x)^2\!\!-\!\!4\delta |x|}\right],
								\quad							
				\delta	\triangleq	\frac{
									\frac{\sigma^2}{L}|\psi_{T_1}|^2
								       }
								       {
								       	\frac{L\beta}{N}\textsf{SNR}+1
								       } \ll 1,
								\quad
				x		\in		(-\infty,1\!-\!2\gamma_{-}]
								\!\cup\! [1\!+\!2\gamma_{+},+\infty)
								\nonumber
			\end{align}
		with $\gamma_{\pm}\!\triangleq\!\sqrt{\delta^2\!+\!\delta}\! \pm \!\delta$.
		Furthermore,
		$0 \!\le\! f(x) \!\le\! \max\{ \delta, \gamma_{-} \} \!\ll\! 1$
		when $x \in(-\infty,1\!-\!2\gamma_{-}]$.
	\end{lemma}
	\begin{IEEEproof}
		The key idea here is to design the appropriate similarity transform for $\mathbf{M}$ in
		\eqref{Equ:Performance:M} and apply
		Lemma \ref{Lem:Gerschgorin}. Our detailed proof can be found in Appendix 
		\ref{Sec:Appendix:Proof_of_Thm_lambda_max}.
	\end{IEEEproof}
	
Now we are ready to derive the expression of $\textsf{G}(\textsf{SNR})$. 
We first consider the trivial case
of $-1 \!\!\le\!\! \gamma_1 \!\!\le\!\! 0$.
\footnote{Since $\gamma_i$ is the eigenvalue of 
$\left(\mathbf{A}_I \bm{\Phi}_\Delta \mathbf{A}_I^H, \mathbf{R}_\mathcal{I}\right)$,
$\gamma_i\!+\!1$ is for 
$\left(\sigma_{\mathcal{I}_0}^2\mathbf{a}_0\mathbf{a}_0^H\!\!+\!\!\mathbf{Q}_\mathcal{S}, 
\mathbf{R}_\mathcal{I}\right)$, a positive definite pair. 
Hence $\gamma_i\!+\!1\!\!>\!\!0$.}
Since $\gamma_D \!\!<\!\! \ldots \!\!<\!\! \gamma_1$, we can immediately have from
\eqref{Equ:Performance:psi_S_lambda_max} that 
$0 \!\!<\!\! \psi_S(\lambda_\mathrm{max}) \!\!<\!\! \frac{\sigma^2}{L}\|\bm{\psi}_T\|^2 \!\!\ll\!\! 1$ 
and $0 \!\!<\!\! \psi_I(\lambda_\mathrm{max}) \!\!<\!\! \frac{\sigma^2}{L}\|\bm{\psi}_T\|^2 \!\!\ll\!\! 1$.
Therefore, \eqref{Equ:Performance:G_lambda_max_expr3} becomes 
$\textsf{G}(\textsf{SNR}) \!\!\approx\!\! 1$ for all $\textsf{SNR}$. 

Next, we are going to discuss the case of $\gamma_1\!\!>\!\!0$ in two separate cases.

\subsubsection{$\gamma_0>\gamma_1$}
Then $\lambda_a \!=\! \gamma_0$, 
$\lambda_b\!=\!\gamma_1$ and $\lambda_\mathrm{max} \!\approx\! \gamma_0\!+\!1$.
So long as $\textsf{SNR}$ is slightly larger than $\textsf{SNR}_{\textsf{T}0}$ such that
$\gamma_0 \!\!>\!\! \gamma_1/(1\!\!-\!\!\sqrt{{\sigma^2}/{L}}\|\bm{\psi}_T\|) \!\!\approx\!\! \gamma_1$,
we have the following approximation for $\psi_S$:
	\begin{align}
		\psi_S(\lambda_\mathrm{max})	\approx	\frac{\sigma^2}{L}
			                						\sum_{i=1}^D
			                						\!\!
								                	\frac{
								                        	\gamma_0
								                        }
								                        {
								                        	\gamma_0\!\!-\!\!\gamma_i
								                    	}
								                	|\psi_{T_i}|^2
								<		\frac{\gamma_0}{\gamma_0\!\!-\!\!\gamma_1}
										\frac{\sigma^2}{L}
										\sum_{i=1}^D
			                						|\psi_{T_i}|^2
								<		\sqrt{\frac{\sigma^2}{L}}
										\|\bm{\psi}_T\|
								\ll 		1.
										\nonumber
	\end{align}
Moreover, we approximate $\psi_I$ by the following bound,
which is asymptotically tight with respect to $\frac{\gamma_0}{\gamma_1}$.
\footnote{In fact, this approximation is precise enough when $\gamma_0$ is
reasonably larger than $\gamma_1$, say, $\gamma_0>2\gamma_1$.}
	\begin{align}
		\psi_I(\lambda_\mathrm{max})	\approx	\frac{\sigma^2}{L}
								                	\!
								                	\sum_{i=1}^D
								                	\!
								                	(\gamma_i \!\!+\!\! 1)
								                	\!
								                	\left(
								                        \frac{
								                                	\gamma_0
								                               }
								                               {
								                                	\gamma_0
								                                	\!\!-\!\!
								                                	\gamma_i
								                               }
								                	\right)^2
								                	\!\!\!
								                	|\psi_{T_i}|^2
								\le		\left(
								                        \frac{
								                                	\gamma_0
								                               }
								                               {
								                                	\gamma_0
								                                	\!\!-\!\!
								                                	\gamma_1
								                               }
								                	\right)^2
										\frac{\sigma^2}{L}
								                	\!
								                	\sum_{i=1}^D
								                	(\gamma_i \!\!+\!\! 1)
								                	|\psi_{T_i}|^2.
										\nonumber
	\end{align}
Substituting the above two approximations together with \eqref{Equ:Performance:lambda_0}
and \eqref{Equ:Performance:SNR_T0} into \eqref{Equ:Performance:G_lambda_max_expr3},
we can finally have
	\begin{align}
		\textsf{G}(\textsf{SNR})	=	\frac{
										P_I + 1
								       }
								       {
								       		\displaystyle
								       		P_I
										/
										\left[
											1-{\textsf{SNR}_{\textsf{T}0}}/{\textsf{SNR}}
										\right]^2+1
								       }
								       G_U
								       \nonumber
	\end{align}
where $G_U = 1/(P_I+1)$ and $P_I$ is defined as
	\begin{align}
		P_I	\triangleq	\left[
						\frac{
								(N-\beta)/\beta
						       }
						       {
						       		(N-\beta)/\beta-\gamma_1
						       }
					\right]^2
					\frac{\sigma^2}{L}
					\sum_{i=1}^D
					(\gamma_i+1)|\psi_{T_i}|^2
					\nonumber
	\end{align}
However, this expression is difficult to evaluate. Instead, we can 
compute $G_U$ first and have $P_I \!\!=\!\! 1/G_U\!\!-\!\!1$.
Noticing that $G_U$ is the output SINR as $\textsf{SNR}\!\!=\!\!\infty$, 
we can ignore the term $\mathbf{A}_I\bm{\Phi}_\Delta\mathbf{A}_I^H$ on the left hand side of
\eqref{Equ:Performance:EigenEquation1}. Hence, 
$\mathbf{w}_\infty\!\!=\!\!\mu \mathbf{R}_\mathcal{I}^{-1}\mathbf{a}_0
\!\!=\!\!\mu'\mathbf{Q}_\mathcal{I}^{-1}\mathbf{a}_0$, and by 
\eqref{Equ:Matrix_Mismatch:G} and \eqref{Equ:MPBeamformer:SINR_max}, we have its
expression in \eqref{Equ:Performance:G_U}.

\subsubsection{$\gamma_0<\gamma_1$}
Then $\lambda_a \!=\! \gamma_1$, 
$\lambda_b\!=\!\gamma_0$ and $\lambda_\mathrm{max} \!\approx\! \gamma_1\!+\!1$.
As a result,
the term corresponding to $\gamma_1$ would dominate $\psi_S$ and $\psi_I$
in \eqref{Equ:Performance:psi_S_lambda_max}.
To further evaluate these two terms, we need the bound in 
\eqref{Equ:Performance:Thm_lambda_max:lambda_max} to
measure how close $\lambda_\mathrm{max}$ is to $\gamma_1+1$,
namely, we use it to evaluate $\lambda_\mathrm{max}-(\gamma_1+1)$ 
in $\psi_S$ and $\psi_I$:
	\begin{align}
		\psi_S(\lambda_\mathrm{max})
                			&\!\approx\!	\frac{\sigma^2}{L}
					                	\frac{
					                        	\gamma_1
					                        }
					                        {
					                        	\lambda_\mathrm{max}
					                        	\!\!-\!\!
					                        	(\gamma_i\!\!+\!\!1)
					                    	}
					                	|\psi_{T_1}|^2
				\!\approx\!		\frac{\sigma^2}{L}
					                	\frac{
					                        	1
					                        }
					                        {
					                        	f\left(\frac{\gamma_0}{\gamma_1}\right)
					                    	}
					                	|\psi_{T_1}|^2	
				\!=\!			\left(
								\frac{L\beta}{N}\textsf{SNR} \!\!+\!\! 1
							\right)
							\frac{\gamma_1}{\gamma_0}
							\cdot
							g\left(\frac{\gamma_0}{\gamma_1}\right)
							\nonumber\\
		\psi_I(\lambda_\mathrm{max})
				&\!\approx\!	\frac{\sigma^2}{L}							
					                        \frac{
					                        		(\gamma_1 \!\!+\!\! 1)	
					                                	\gamma_1^2
					                               }
					                               {
					                               	[
					                                	\lambda_\mathrm{max}
					                                	\!\!-\!\!
					                                	(\gamma_1\!\!+\!\!1)
									]^2
					                               }
					                	|\psi_{T_1}|^2
				\!\approx\!		\frac{\sigma^2}{L}
					                	\frac{
					                        	\gamma_1+1
					                        }
					                        {
					                        	\left[f\left(\frac{\gamma_0}{\gamma_1}\right)\right]^2
					                    	}
					                	|\psi_{T_1}|^2
				\!=\!			\frac{\gamma_1+1}{\frac{\sigma^2}{L}|\psi_{T_1}|^2}
							\left[\psi_S(\lambda_\mathrm{max})\right]^2
							\nonumber
	\end{align}
where $g(x) = \frac{1}{2}[1\!\!-\!\!x\!\!+\!\!\sqrt{(1\!\!-\!\!x)^2\!\!-\!\!4\delta x}]$.
In fact, we can have the approximation that $g(x)\approx 1-x$, because
	\begin{align}	
		\left| (1-x) - g(x) \right|	=	|f(x)|		\ll	1		\nonumber
	\end{align}	
Substituting the above expressions of $\psi_S$ and $\psi_I$ as well as the 
approximation of $g(x)$ into \eqref{lemma:G_lambda_max}, we get
	\begin{align}
		\textsf{G}(\textsf{SNR})	&\!\!=\!\!	\left[
									\frac{N}
		                                                                  {L\beta\textsf{SNR}\!\!+\!\!N}
		                                                      	\right]^2
									\!\!
									\left[
										\frac{\gamma_1}{\gamma_1\!\!-\!\!\gamma_0}
									\right]^2
									\!\!
									\left[
										\frac{\gamma_1\!\!+\!\!1}
										       {\frac{\sigma^2}{L}|\psi_{T_1}|^2}
										\!\!-\!\!
										\frac{\frac{L\beta}{N}\textsf{SNR}}
										       {\frac{L\beta}{N}\textsf{SNR} \!\!+\!\!1}
										\!\!+\!\!
										\left(\!
											\frac{\gamma_1}{\gamma_1\!\!-\!\!\gamma_0}
											\!
										\right)^2
									\right]^{-1}
							\!\!\!\!\!\!\!\approx\!\!	
									\left[
									\frac{N}
		                                                                  {L\beta\textsf{SNR}\!\!+\!\!N}
		                                                      	\right]^2
									\!\!
									\left[
										\frac{\gamma_1}{\gamma_1\!\!-\!\!\gamma_0}
									\right]^2
									\!\!
									G_L,
									\nonumber
	\end{align}
where $G_L \triangleq \frac{\sigma^2}{L}|\psi_{T_1}|^2/(\gamma_1\!\!+\!\!1)$ is
the output SINR when $\textsf{SNR}\!\!=\!\!0$ ($-\infty$dB). The last approximation holds when 
$\gamma_0 \!\!<\!\! (1\!\!-\!\!\sqrt{\frac{\sigma^2}{L}}|\psi_{T_1}|)\gamma_1$, 
i.e. $\gamma_0$ is slightly smaller than $\gamma_1$, which implies 
$\gamma_1^2/(\gamma_1\!\!-\!\!\gamma_0)^2 \!\!<\!\! 
1/\frac{\sigma^2}{L}|\psi_{T_1}|^2 < (\gamma_1+1)/\frac{\sigma^2}{L}|\psi_{T_1}|^2$.
Furthermore, because $\frac{\sigma^2}{L}|\psi_{T_1}|^2 \!\!\ll\!\! 1$, 
${\frac{L\beta}{N}\textsf{SNR}}/{(\frac{L\beta}{N}\textsf{SNR} \!\!+\!\!1)} \!\!<\!\! 1 \!\!\ll\!\!
(\gamma_1+1)/\frac{\sigma^2}{L}|\psi_{T_1}|^2$. There is, however, no easier way to
evaluate $G_L$ than its definition. Fortunately, we are not interested in its specific
values but the threshold $\textsf{SNR}_{\textsf{T}1}$ that $\textsf{G}$ reaches this value,
as discussed in Sec. \ref{Sec:Performance_Analysis:Operating_Curve}.
Noticing that $\textsf{SNR}_{\textsf{T}0}=\infty$ when $(N-\beta)/\beta<\gamma_1$
(c.f. \eqref{Equ:Performance:SNR_T0}),  we have following fact regarding $\gamma_1$
and $\textsf{SNR}_{\textsf{T}0}$, 
	\begin{align}
		\gamma_1		=	\frac{L(N-\beta)\textsf{SNR}_{\textsf{T}0}}
						       {L\beta\textsf{SNR}_{\textsf{T}0}+N}
						+
						\left(\gamma_1-\frac{N-\beta}{\beta}\right)^{+}.
						\nonumber
	\end{align}
Combining the above expression and \eqref{Equ:Performance:lambda_0}, 
$\textsf{G}(\textsf{SNR})$ can be reduced to
	\begin{align}
		\textsf{G}(\textsf{SNR})	=	\left[
									\frac{1+K_0}
									       {
									       	1-{\textsf{SNR}}/{\textsf{SNR}_{\textsf{T}0}}
										+
										K_0
										\left(
											\frac{L\beta}{N}
											\textsf{SNR}
											+
											1
										\right)
									       }
								\right]^2	
								G_L	
	\end{align}
with $K_0$ defined in \eqref{Equ:Performance:K0}. Finally, the following lemma provides 
an easier way to compute $\textsf{SNR}_{\textsf{T}0}$ from $\gamma_1$.
The proof can also be found in Appendix \ref{Sec:Appendix:Proof_of_Thm_lambda_max}.
	\begin{lemma}
		\label{Lem:gamma_i_approx}
		$\gamma_1,\ldots,\gamma_D$ are approximate to all the nonzero generalized eigenvalues of  
		$(\mathbf{Q}_\mathcal{S}\!\!-\!\!\mathbf{Q}_\mathcal{I},\mathbf{Q}_\mathcal{I})$
		(padded up to $D$ with zeros if not enough), and is almost independent of $\textsf{SNR}$.
	\end{lemma}

\section{Discussion of the Threshold}
\label{Sec:Threshold}

From Theorem \ref{thm:operating_curve} in Sec. \ref{Sec:Performance_Analysis:Operating_Curve},
we know that the empirical threshold $\textsf{SNR}_{\textsf{T}0}$ is a key parameter for MPB.
By the discussion in Sec. \ref{Sec:Performance:How_MPB_works}, it is the intersection
of the curves $\gamma_0\!\!+\!\!1$ and $\gamma_1\!\!+\!\!1$, as shown in 
Fig. \ref{Fig:Performance:lambda_max_gamma0vsgamma1}. Therefore, we need to investigate
these two parameters to gain deeper insight of $\textsf{SNR}_{\textsf{T}0}$.

\subsection{General Results of $\gamma_0$ and $\gamma_1$}
\begin{figure*}[t]
        \centering{
        			\subfigure[The curves of $\gamma_0\!+\!1$, $\gamma_1\!\!+\!\!1$ and 
					$\lambda_\mathrm{max}$ vs. \textsf{SNR}. ($\beta=0$)]
					{
						\includegraphics[width=0.45\textwidth]{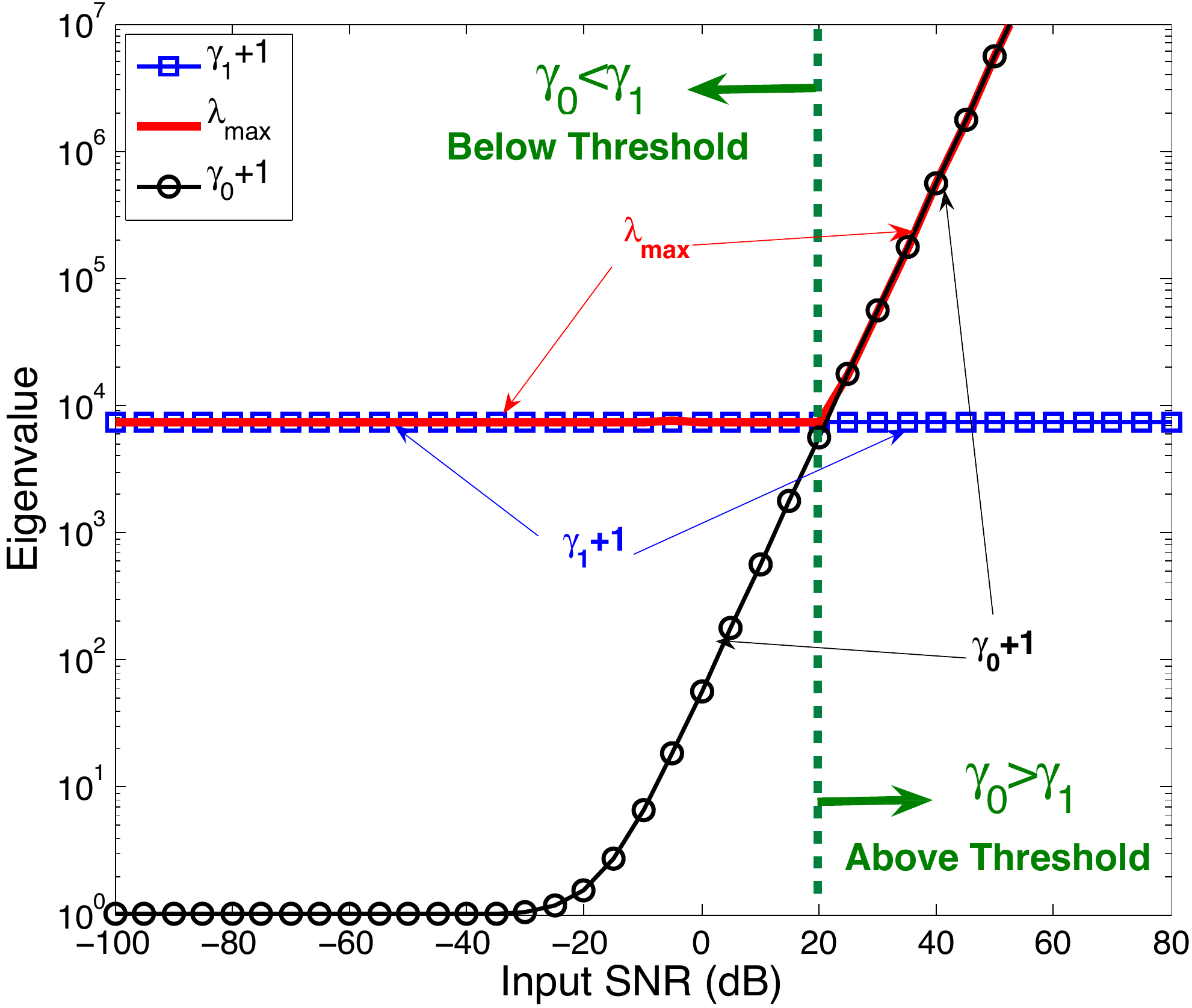}
        						\label{Fig:Performance:lambda_max_gamma0vsgamma1}
					}
        			\subfigure[$\gamma_0\!+\!1$ against $\textsf{SNR}$ with different $\beta$.]
					{
        						\includegraphics[width=0.45\textwidth]{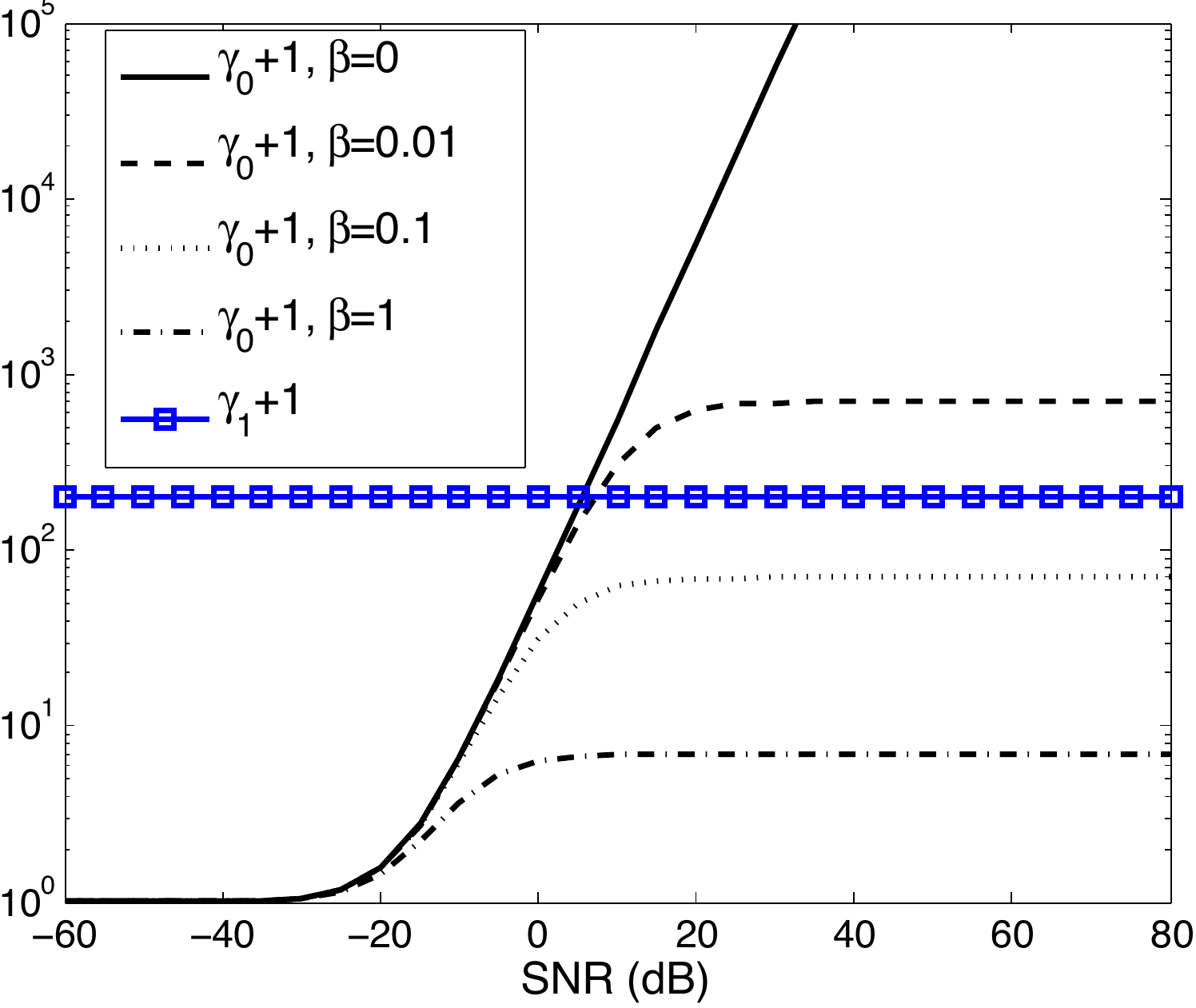}
        						\label{Fig:Performance:gamma_0_DifferentBeta}
					}			
        }
        \caption{The curve of $\gamma_0+1$ and $\gamma_1+1$ against $\textsf{SNR}$. The intersection
        			of them is the empirical threshold $\textsf{SNR}_{\textsf{T}0}$.}
	\label{Fig:Threshold}
    \end{figure*}

$\gamma_0$ has a simple expression of \eqref{Equ:Performance:lambda_0}, from which we know
that the parameter $\beta$ is critical. 
Fig. \ref{Fig:Performance:gamma_0_DifferentBeta} shows the curve of $\gamma_0\!+\!1$
with different $\beta$. We see that, if $\beta \!=\! 0$, i.e.
$\mathbf{H}_\mathcal{I}$ and $\mathbf{h}_\mathcal{S}$ are orthogonal, then 
$\gamma_0 \!\!=\!\! L\textsf{SNR}$ is unbounded as $\textsf{SNR}$ goes to infinity. 
Otherwise, there would be a limiting value of $(N \!-\! \beta)/\beta$
so that $\gamma_0$ might never
exceed $\gamma_1$ and $\textsf{SNR}_{\textsf{T}0}\!\!=\!\!+\infty$.
Therefore, $\beta\!\!=\!\!0$ is the best choice for $\gamma_0$.

$\gamma_1$ is another critical parameter that determines $\textsf{SNR}_{\textsf{T}0}$. 
Fig. \ref{Fig:Threshold} shows that, as $\gamma_1$ increases, its intersection with $\gamma_0$
moves rightward and $\textsf{SNR}_{\textsf{T}0}$ increases. Therefore, knowing how to
control $\gamma_1$ is important in designing MPB. By Lemma \ref{Lem:gamma_i_approx}, 
$\gamma_1$ is the largest nonzero generalized eigenvalue of
$(\mathbf{Q}_\mathcal{S} \!-\! \mathbf{Q}_\mathcal{I},\mathbf{Q}_\mathcal{I})$.
(It is zero if there are less than $D$ nonzero eigenvalues and all of them are negative.)
To solve it directly, we need the roots of a polynomial eigen-equation of order $L$, which has no general 
formula when $L\!\!>\!\!4$. Instead, we resort to matrix perturbation theory again
to derive a bound for it. We first notice that the eigen-decomposition of 
$(\mathbf{Q}_\mathcal{S} \!-\! \mathbf{Q}_\mathcal{I},\mathbf{Q}_\mathcal{I})$ is 
equivalent to $(\mathbf{Q}_\mathcal{S},\mathbf{Q}_\mathcal{I})$ and their eigenvalues only differ
by one. By \eqref{Equ:MPBeamformer:Q_S} and \eqref{Equ:MPBeamformer:Q_I}, 
the eigenvalue of $(\mathbf{Q}_\mathcal{S},\mathbf{Q}_\mathcal{I})$ is further
equivalent to that of $(\mathbf{Y}_\mathcal{S}\!+\!\mathbf{I},\mathbf{Y}_\mathcal{I}\!+\!\mathbf{I})$, where
	\begin{align}
		\label{Equ:Threshold:YS_YI}
		\mathbf{Y}_\mathcal{S} 	= 	\frac{1}{\sigma^2} \mathbf{A}_I \bm{\Phi}_\mathcal{S} \mathbf{A}_I^H
							=	\textsf{INR} 
								\cdot 
								\mathbf{A}_I \bm{\Phi}_{\mathcal{S}_0} \mathbf{A}_I^H,
		\qquad
		\mathbf{Y}_\mathcal{I} 	= 	\frac{1}{\sigma^2} \mathbf{A}_I \bm{\Phi}_\mathcal{I} \mathbf{A}_I^H
							=	\textsf{INR} 
								\cdot 
								\mathbf{A}_I \bm{\Phi}_{\mathcal{I}_0} \mathbf{A}_I^H.
	\end{align}
We term $(\mathbf{Y}_{\mathcal{S}},\mathbf{Y}_{\mathcal{I}})$ as \emph{noise free pair},
since it can be viewed as the covariance matrix pair of MPB without noise.
Our strategy here is to view $(\mathbf{Y}_\mathcal{S}\!+\!\mathbf{I},\mathbf{Y}_\mathcal{I}\!+\!\mathbf{I})$
as a perturbed version of $(\mathbf{Y}_\mathcal{S},\mathbf{Y}_\mathcal{I})$ and
apply the results in matrix perturbation theory to derive a bound for $\gamma_1\!\!+\!\!1$.
Before we proceed on, we cite a more general definition of the generalized eigenvalue
of a matrix pair\cite{stewart1990mpt}.
	\begin{definition}[Generalized eigenvalue]
		\label{Def:Threshold:Generalized_Eigenvalue}
		The generalized eigenvalue of a matrix pair $\left(\mathbf{A},\mathbf{B}\right)$ 
		is a one dimensional space, denoted as 
		$\langle\nu,\mu\rangle \triangleq \left\{\tau \cdot [\nu \; \mu]:\tau\in\mathbb{C}\right\}$, 
		where $[\nu \; \mu]$ is a $1 \times 2$ vector satisfying
			\begin{align}
				\mu\cdot\mathbf{A}\mathbf{x}=\nu \cdot \mathbf{B}\mathbf{x},        \nonumber
			\end{align}
		with $\mathbf{x}$ being its eigenvector. If $\mathbf{B}$ is nonsingular, 
		then $\lambda \!\!=\!\! \nu/\mu$ becomes the conventional definition.
	\end{definition}
	
Comparing to the conventional definition of generalized eigenvalues, this one includes the
special case of $\mathbf{B}$ being singular so that $\mu \!\!=\!\! 0$ and $\nu \!\!\neq\!\! 0$,
namely, $\lambda \!\!=\!\! +\infty$. We will see its importance later. Besides, we
also need the following definition and lemma from \cite[pp.315--316]{stewart1990mpt}.

    \begin{definition}[Definite pair]
        If a matrix pair $(\mathbf{A},\mathbf{B})$ consists of two Hermitian matrices and
            \begin{align}
                C(\mathbf{A},\mathbf{B})     \triangleq      \min_{\|\mathbf{x}\|=1} \sqrt{
                                                                                                        \left(\mathbf{x}^H\mathbf{A}\mathbf{x}\right)^2
                                                                                                        +
                                                                                                        \left(\mathbf{x}^H\mathbf{B}\mathbf{x}\right)^2
                                                                                                     }
                                                                        >0,
                                                                        \nonumber
            \end{align}
        then $(\mathbf{A},\mathbf{B})$ is called definite pair, and $C(\mathbf{A},\mathbf{B})$
        is its Crawford number.
    \end{definition}
	\begin{theorem}[Weyl-Lidskii type]
	\label{Thm:Weyl_Lidskii}
		Assume $\left(\mathbf{A},\mathbf{B}\right)$ is a definite pair and
		$\left(\mathbf{A} \!+\! \mathbf{E},\mathbf{B} \!+\! \mathbf{F}\right)$ is its perturbed version. 
		Let $\langle\alpha_i,\beta_i\rangle$ and $\langle\widetilde{\alpha}_i,\widetilde{\beta}_i\rangle$,
		($i\!\!=\!\!1,\ldots,n$), be their ordered generalized eigenvalues, respectively. If
			\begin{align}
				\label{Equ:Threshold:Thm2_Condition}
				{\sqrt{\|\mathbf{E}\|_S^2+\|\mathbf{F}\|_S^2}}<{C(\mathbf{A},\mathbf{B})},    
			\end{align}
		where $\|\cdot\|_S$ denotes spectral norm of a matrix,
		then $(\widetilde{\mathbf{A}},\widetilde{\mathbf{B}})$ is a definite pair, and
			\begin{align}
				\frac{
					|\alpha_1\widetilde{\beta}_2-\beta_1\widetilde{\alpha}_2|
					}
					{
					\sqrt{|\alpha_1|^2+|\beta_1|^2}
					\sqrt{|\widetilde{\alpha}_2|^2+|\widetilde{\beta}_2|^2}
					}
				\leq
				\frac{
					\sqrt{\|\mathbf{E}\|_S^2+\|\mathbf{F}\|_S^2}
					}
					{
					C(\mathbf{A},\mathbf{B})
					},
				\nonumber
			\end{align}		
	\end{theorem}

We want to apply Theorem \ref{Thm:Weyl_Lidskii} to 
$(\mathbf{Y}_{\mathcal{S}} \!+\! \mathbf{I},\mathbf{Y}_{\mathcal{I}} \!+\! \mathbf{I})$
and $(\mathbf{Y}_{\mathcal{S}},\mathbf{Y}_{\mathcal{I}})$
to derive a bound for its eigenvalues.
However, $(\mathbf{Y}_{\mathcal{S}},\mathbf{Y}_{\mathcal{I}})$ is not a definite pair, 
for the null spaces of $\mathbf{Y}_{\mathcal{S}}$ and $\mathbf{Y}_{\mathcal{I}}$ 
may have nontrivial interset (larger than $\{0\}$). Therefore, 
we need to transform $(\mathbf{Y}_{\mathcal{S}},\mathbf{Y}_{\mathcal{I}})$
into a definite pair. Let $\mathcal{N}_{\mathcal{S}}$ and $\mathcal{N}_{\mathcal{I}}$ be the 
null spaces of $\mathbf{Y}_{\mathcal{S}}$ and $\mathbf{Y}_{\mathcal{I}}$, 
respectively, and 
$\mathcal{N}_{0} \!\triangleq\! \mathcal{N}_{\mathcal{S}} \bigcap \mathcal{N}_{\mathcal{I}}$.
Let $\mathbf{E}_0$ be a matrix whose columns are the orthonormal basis of $\mathcal{N}_{0}^\perp$, 
with $\mathbf{E}_0^H\mathbf{E}_0=\mathbf{I}$. 
Then, by the determinant identity in \eqref{Equ:Performance:Identity_det},
$\left(\mathbf{Y}_{\mathcal{S}} \!+\! \mathbf{I},\mathbf{Y}_{\mathcal{I}} \!+\! \mathbf{I}\right)$ has the same generalized eigenvalues as
$(\mathbf{E}_0^H\mathbf{Y}_{\mathcal{S}}\mathbf{E}_0 \!+\! \mathbf{I},
\mathbf{E}_0^H\mathbf{Y}_{\mathcal{I}}\mathbf{E}_0 \!+\! \mathbf{I})$,
except for the multiplicity of ones. Similarly,  
$(\mathbf{E}_0^H\mathbf{Y}_{\mathcal{S}}\mathbf{E}_0,
\mathbf{E}_0^H\mathbf{Y}_{\mathcal{I}}\mathbf{E}_0)$
has the same eigenvalue as $(\mathbf{Y}_{\mathcal{S}}, \mathbf{Y}_{\mathcal{I}})$ does 
except for the multiplicity of zero. Therefore, instead, we can apply Theorem \ref{Thm:Weyl_Lidskii}
to $(\mathbf{E}_0^H\mathbf{Y}_{\mathcal{S}}\mathbf{E}_0,
\mathbf{E}_0^H\mathbf{Y}_{\mathcal{I}}\mathbf{E}_0)$
and $(\mathbf{E}_0^H\mathbf{Y}_{\mathcal{S}}\mathbf{E}_0 \!+\! \mathbf{I},
\mathbf{E}_0^H\mathbf{Y}_{\mathcal{S}}\mathbf{E}_0 \!+\! \mathbf{I})$
to derive the bound. The Crawford number of
$(\mathbf{E}_0^H\mathbf{Y}_{\mathcal{S}}\mathbf{E}_0, 
\mathbf{E}_0^H\mathbf{Y}_{\mathcal{I}}\mathbf{E}_0)$ is
	\begin{align}
	\label{Equ:Performance:Crawford_YsYi_Eperp}
		C_{Y} 	&=	\min_{\substack{\mathbf{w}\in\mathcal{N}_{0}^\perp, \;  \|\mathbf{w}\|=1}}
					\sqrt{
						\left(\mathbf{w}^H\mathbf{Y}_{\mathcal{S}}\mathbf{w}\right)^2
						\!\!+\!\!
						\left(\mathbf{w}^H\mathbf{Y}_{\mathcal{I}}\mathbf{w}\right)^2
						}
						>0,
	\end{align}
which means $(\mathbf{E}_0^H\mathbf{Y}_{\mathcal{S}}\mathbf{E}_0, 
\mathbf{E}_0^H\mathbf{Y}_{\mathcal{I}}\mathbf{E}_0)$ is a positive pair.
Before applying Theorem \ref{Thm:Weyl_Lidskii}, we first analyze the dependency 
of $C_Y$ on the interference power. Substituting \eqref{Equ:Threshold:YS_YI}
into \eqref{Equ:Performance:Crawford_YsYi_Eperp}, we have
$C_{Y} \!=\! \textsf{INR} \cdot C_{Y_0}$, where
	\begin{align}
		C_{Y_0} 	&=	\min_{{\mathbf{w}\in\mathcal{N}_{0}^\perp, \; \|\mathbf{w}\|=1}}
					\sqrt{
						\left(
							\mathbf{w}^H
							\mathbf{A}_I \bm{\Phi}_{\mathcal{S}_0} \mathbf{A}_I^H
							\mathbf{w}
						\right)^2
						\!\!+\!\!
						\left(
							\mathbf{w}^H
							\mathbf{A}_I \bm{\Phi}_{\mathcal{I}_0} \mathbf{A}_I^H
							\mathbf{w}
						\right)^2
						}.
					\nonumber
	\end{align}
Since $\bm{\Phi}_{\mathcal{S}_0}$ and $\bm{\Phi}_{\mathcal{I}_0}$ 
are independent of $\textsf{INR}$ (c.f. Sec. \ref{Sec:ProblemFormulation:MPBeamformer}),
$C_{Y_0}$ is also independent of $\textsf{INR}$. Therefore, $C_Y$ is proportional to $\textsf{INR}$.
Now, we are ready to use Theorem \ref{Thm:Weyl_Lidskii}, and we only consider the 
case of $\textsf{INR}$ being large. Let $\langle \nu_0,\mu_0 \rangle$ 
and $\langle \nu, \mu \rangle$ be the corresponding generalized eigenvalues of
$(\mathbf{E}_0^H\mathbf{Y}_{\mathcal{S}}\mathbf{E}_0, 
\mathbf{E}_0^H\mathbf{Y}_{\mathcal{I}}\mathbf{E}_0)$ and
$(\mathbf{E}_0^H\mathbf{Y}_{\mathcal{S}}\mathbf{E}_0 \!+\! \mathbf{I},
\mathbf{E}_0^H\mathbf{Y}_{\mathcal{I}}\mathbf{E}_0 \!+\! \mathbf{I})$, respectively.
Then, for large $\textsf{INR}$, \eqref{Equ:Threshold:Thm2_Condition} is satisfied and
	\begin{align}
	\label{Equ:Performance:rho_c_Y}
		\frac{
			|\nu\mu_0-\mu\nu_0|
			}
			{
			\sqrt{|\nu|^2 \!\!+\!\! |\mu|^2}
			\!
			\sqrt{|\nu_0|^2 \!\!+\!\! |\mu_0|^2}
			}
		\!\!<\!\!
		\frac{\sqrt{2}}{C_{Y_0}}
		\cdot 
		\frac{1}{\textsf{INR}}.
	\end{align}
	
We now derive the bound for $\gamma_1\!+\!1$ in two separate cases.
	
	\subsubsection{$\mathcal{N}_{\mathcal{I}} \nsubseteq \mathcal{N}_{\mathcal{S}}$}		        
			There is a nonzero $\mathbf{w}_0$ such that
			$\mathbf{Y}_{\mathcal{S}}\mathbf{w}_0 \!\neq\! \mathbf{0}$
			and $\mathbf{Y}_{\mathcal{I}}\mathbf{w}_0 \!=\! \mathbf{0}$.
			By Definition \ref{Def:Threshold:Generalized_Eigenvalue}, 
			$\langle \nu_0, 0 \rangle$ ($\forall \nu_0 \!\neq\! 0$) is a generalized eigenvalue of
            		$(\mathbf{Y}_\mathcal{S}, \mathbf{Y}_\mathcal{I})$ and 
			$(\mathbf{E}_0^H\mathbf{Y}_\mathcal{S}\mathbf{E}_0, 
			\mathbf{E}_0^H\mathbf{Y}_\mathcal{I}\mathbf{E}_0)$.
			This means the noise free pair 
			$(\mathbf{Y}_{\mathcal{S}},\mathbf{Y}_{\mathcal{I}})$ has an infinite
            		generalized eigenvalue. As a result, \eqref{Equ:Performance:rho_c_Y}
			becomes
				\begin{align}
					\frac{
						|\mu|
						}
						{
						\sqrt{|\nu|^2 \!\!+\!\! |\mu|^2}
						}
					\!\!<\!\!
					\frac{\sqrt{2}}{C_{Y_0}} \cdot \frac{1}{\textsf{INR}}
					\quad
					\Leftrightarrow
					\quad
					\lambda
					\triangleq
					\frac{\nu}{\mu}         
					>       
					\sqrt{
					\frac{C_{Y_0}^2}{2}
					\textsf{INR}^2
					\!\!-\!\!
					1
					}
					\approx
					\frac{C_{Y_0}}{\sqrt{2}}
					\textsf{INR}.
					\nonumber
				\end{align}
            		This means 
			$(\mathbf{E}_0^H\mathbf{Y}_{\mathcal{S}}\mathbf{E}_0 \!+\! \mathbf{I},
			\mathbf{E}_0^H\mathbf{Y}_{\mathcal{S}}\mathbf{E}_0 \!+\! \mathbf{I})$
			would always have an eigenvalue that satisfies the above inequality.
			Since $(\mathbf{Q}_\mathcal{S},\mathbf{Q}_\mathcal{I})$ has the
			same eigenvalue except for the multiplicity of ones, $\gamma_1\!+\!1$ satisfies
				\begin{align}
					\label{Equ:Threshold:gamma1_plus_1_LowerBound}
					\gamma_{1}+1		>	\frac{C_{Y_0}}{\sqrt{2}}
										\textsf{INR},
				\end{align}
			which gives a lower bound for $\gamma_1\!+\!1$. We can see that
			it goes up unboundedly as $\textsf{INR}$ increases. 
			Hence, by \eqref{Equ:Performance:SNR_T0} and 
			\eqref{Equ:Performance:SNR_T1_AND_SNR_T2}, the threshold SNR
			also increases unboundedly with $\textsf{INR}$.
	\subsubsection{$\mathcal{N}_{\mathcal{I}} \subseteq \mathcal{N}_{\mathcal{S}}$}
			Then, $\mathcal{N}_0\!=\!\mathcal{N}_\mathcal{I}$ and 
			$\mathbf{E}_0^H\mathbf{Y}_\mathcal{I}\mathbf{E}_0$ is nonsingular. As a 
			result, all eigenvalues of 
			$(\mathbf{E}_0^H\mathbf{Y}_\mathcal{S}\mathbf{E}_0, 
			\mathbf{E}_0^H\mathbf{Y}_\mathcal{I}\mathbf{E}_0)$ and 
			$(\mathbf{Y}_\mathcal{S}, \mathbf{Y}_\mathcal{I})$
			are bounded.
			When $\textsf{INR}$ is large, \eqref{Equ:Performance:rho_c_Y} becomes
				\begin{align}
					\frac{|\lambda-\lambda_0|}{\sqrt{1+\lambda^2}\sqrt{1+\lambda_0^2}}
					<
					\frac{C_{Y_0}}{\sqrt{2}} \cdot \frac{1}{\textsf{INR}}
					\quad
					\Longrightarrow
					\quad
					|\lambda-\lambda_0|		<	\frac{C_{Y_0}}{\sqrt{2}} (1+\lambda_0^2)
											\cdot \frac{1}{\textsf{INR}}
					\nonumber
				\end{align}
			where $\lambda_0 \!=\! \nu_0/\mu_0$ is the largest eigenvalue of 
			$(\mathbf{E}_0^H\mathbf{Y}_{\mathcal{S}}\mathbf{E}_0,
			\mathbf{E}_0^H\mathbf{Y}_{\mathcal{S}}\mathbf{E}_0)$
			and we used $\lambda \!\approx \lambda_0$ in the second inequality.	
			Since $(\mathbf{Q}_\mathcal{S},\mathbf{Q}_\mathcal{I})$ has the
			same eigenvalue as 
			$(\mathbf{E}_0^H\mathbf{Y}_{\mathcal{S}}\mathbf{E}_0 \!+\! \mathbf{I},
			\mathbf{E}_0^H\mathbf{Y}_{\mathcal{S}}\mathbf{E}_0 \!+\! \mathbf{I})$
			except for the multiplicity of ones, $\gamma_1+1$ is bounded
			around the largest eigenvalue of $(\mathbf{Y}_\mathcal{S},\mathbf{Y}_\mathcal{I})$ 
			or one. Furthermore, from 
			\eqref{Equ:Threshold:YS_YI}, we know that $\lambda_0$ is independent of $\textsf{INR}$.
			Therefore, as all eigenvalues of noise free pair is finite, $\gamma_1\!+\!1$ is 
			bounded and independent of $\textsf{INR}$.

\subsection{Two Typical Scenarios of $\gamma_1$}
\label{Sec:TwoTypicalScenarios}

From the previous part, we know that whether the threshold of MPB is unbounded is determined by
the existence of infinite eigenvalue of the noise free pair. Now, we discuss two typical classes
of interferences. In the first case, all eigenvalues are finite, and in the second case, there may be
infinite eigenvalues.

\subsubsection{Directional White Noise}
\label{Sec:TwoTypicalScenarios:DWN}

By uncorrelated directional white noise, we mean an interference that arrives from a specific direction, 
with its samples in time domain being uncorrelated. This means the entries of 
$\mathbf{S}_{I}(k)$ are independent identically distributed random variables with zero mean
and unit variance. (In section \ref{Sec:ProblemFormulation:SignalModel}, we 
have already normalized the interference power in $\mathbf{s}_i(k)$.)
Then, by \eqref{Equ:MPBeamformer:Phi_S0_AND_Phi_I0}, we can immediately
have $\bm{\Phi}_{\mathcal{S}_0} \!=\! \bm{\Phi}_{\mathcal{I}_0}$, and
all the eigenvalue of the noise free pair $(\mathbf{Y}_\mathcal{S},\mathbf{Y}_\mathcal{I})$
are ones. Thus, in this case, $\gamma_1 \!+\! 1$ is bounded and is independent of $\textsf{INR}$.
In fact, we can further have $\mathbf{Q}_\mathcal{S}\!=\!\mathbf{Q}_\mathcal{I}$ according to
\eqref{Equ:MPBeamformer:Q_S} and \eqref{Equ:MPBeamformer:Q_I}. Therefore, $\gamma_1\!=\!0$,
$\textsf{SNR}_{\textsf{T}0}\!=\!0$ and $\textsf{G}(\textsf{SNR}) \!=\! 1$.

\subsubsection{Directional Periodical Interference}
\label{Sec:TwoTypicalScenarios:DPI}

If an interference has periodical structure in time domain and arrives from certain direction, 
then we term it as directional periodical interference. By periodical, we mean the interfer is 
periodic with respect to the projection basis, i.e.
	\begin{align}
		\label{Equ:Performance:Periodical_PI_AND_PS}
		&\mathbf{H}_\mathcal{I}^H\mathbf{S}_I(k)		=		e^{j\phi_k}
													\mathbf{H}_{\mathcal{I}}^H\mathbf{S}_I,
		\qquad
		\mathbf{h}_{\mathcal{S}}^H\mathbf{S}_I(k)	=		e^{j\phi_k'}
													\mathbf{h}_{\mathcal{S}}^H\mathbf{S}_I.
	\end{align}
A stronger condition would be $\mathbf{S}_I(k) \!=\! \mathbf{S}_I$.
However, \eqref{Equ:Performance:Periodical_PI_AND_PS} is good enough for our analysis. 
Now, we will discuss the existence of infinite generalized eigenvalue in the noise free pair 
$(\mathbf{Y}_\mathcal{S},\mathbf{Y}_\mathcal{I})$. To do this, we need to consider the 
relationship between their null spaces.

In practice, $\mathbf{R}_\mathcal{S}$ and $\mathbf{R}_\mathcal{I}$ are estimated from sample average.
Therefore, replacing the expectation in \eqref{Equ:MPBeamformer:Phi_S0_AND_Phi_I0}
by sample average and using \eqref{Equ:Performance:Periodical_PI_AND_PS}, we can get
	\begin{align}
		\bm{\Phi}_{\mathcal{S}_0}		=	\bm{\Omega}_I^{1/2} 
									\mathbf{S}_I^T
									\mathbf{h}_\mathcal{S}^\ast
									\mathbf{h}_\mathcal{S}^T
									\mathbf{S}_I^\ast
									\bm{\Omega}_I^{1/2},
		\qquad
		\bm{\Phi}_{\mathcal{I}_0}		=	\frac{1}{r_\mathcal{I}}
									\bm{\Omega}_I^{1/2} 
									\mathbf{S}_I^T
									\mathbf{H}_\mathcal{I}^\ast
									\mathbf{H}_\mathcal{I}^T
									\mathbf{S}_I^\ast
									\bm{\Omega}_I^{1/2}.
									\nonumber
	\end{align}
Let $\bm{\Pi} \!\triangleq\! \mathbf{S}_I\bm{\Omega}_I^{1/2}\mathbf{A}_I^T$.
Then, by \eqref{Equ:Threshold:YS_YI}, $\mathbf{Y}_\mathcal{S}$ and $\mathbf{Y}_\mathcal{I}$ 
in this case can be expressed as
	\begin{align}
		\mathbf{Y}_\mathcal{S}	=	\textsf{INR} 
								\cdot 
								\left(
									\bm{\Pi}^H
									\mathbf{h}_\mathcal{S}
									\mathbf{h}_\mathcal{S}^H
									\bm{\Pi}
								\right)^\ast,
		\qquad
		\mathbf{Y}_\mathcal{I}	=	\textsf{INR} 
								\cdot 
								\frac{1}{r_\mathcal{I}}
								\left(
									\bm{\Pi}^H
									\mathbf{H}_\mathcal{I}
									\mathbf{H}_\mathcal{I}^H
									\bm{\Pi}
								\right)^\ast.
	\end{align}
Therefore, the noise free pair $(\mathbf{Y}_\mathcal{S}, \mathbf{Y}_\mathcal{I})$
has the same eigenvalue as 
$(\bm{\Pi}^H\mathbf{h}_\mathcal{S}\mathbf{h}_\mathcal{S}^H\bm{\Pi},
\frac{1}{r_\mathcal{I}}\bm{\Pi}^H\mathbf{H}_\mathcal{I}\mathbf{H}_\mathcal{I}^H\bm{\Pi})$ does,
and we only have to check the null spaces of later matrix pair. 
Let $\mathcal{V}_I \!\triangleq\! \mathcal{R}(\bm{\Pi})$  be the range space of $\bm{\Pi}$.
By definition, it is also the range space of $\mathbf{S}_I$, namely the space spanned by
all interference waveforms in one period. Then, we can express the null spaces of 
$\bm{\Pi}^H\mathbf{h}_\mathcal{S}\mathbf{h}_\mathcal{S}^H\bm{\Pi}$
and $\bm{\Pi}^H\mathbf{H}_\mathcal{I}\mathbf{H}_\mathcal{I}^H\bm{\Pi}$ as
	\begin{align}
		&\mathcal{N}_\mathcal{S}'		
						\triangleq		\mathcal{N}
									\left(
										\bm{\Pi}^H
										\mathbf{h}_\mathcal{S}\mathbf{h}_\mathcal{S}^H
										\bm{\Pi}
									\right)
						=			\left\{
										\bm{\Pi}^\dagger \mathbf{x}: \; 
										\mathbf{x} \in \mathcal{S}^\perp \cap \mathcal{V}_I
									\right\}		
									\oplus
									\mathcal{N}(\bm{\Pi})
									\nonumber\\
		&\mathcal{N}_\mathcal{I}'
						\triangleq		\mathcal{N}
									\left(
										\bm{\Pi}^H
										\mathbf{H}_\mathcal{I}\mathbf{H}_\mathcal{I}^H
										\bm{\Pi}
									\right)
						=			\left\{
										\bm{\Pi}^\dagger \mathbf{x}: \; 
										\mathbf{x} \in \mathcal{I}^\perp \cap \mathcal{V}_I
									\right\}		
									\oplus
									\mathcal{N}(\bm{\Pi}),
									\nonumber				
	\end{align}
where $\mathcal{S} \!=\! \mathcal{R}(\mathbf{h}_\mathcal{S})$,
$\mathcal{I} \!=\! \mathcal{R}(\mathbf{H}_\mathcal{I})$ as defined in 
Sec. \ref{Sec:ProblemFormulation:MPBeamformer}, and $\dagger$ denotes 
Moore-Penrose pseudoinverse \cite{laub2004matrix}. To see the relationship
between $\mathcal{N}_\mathcal{S}'$ and $\mathcal{N}_\mathcal{I}'$, we only
need to check the relationship between $\mathcal{S}^\perp \cap \mathcal{V}_I$
and $\mathcal{I}^\perp \cap \mathcal{V}_I$ which can be expressed as
	\begin{align}
		\label{Equ:Threshold:SVI_AND_IVI}
		\mathcal{S}^\perp \cap \mathcal{V}_I		=	\{
												\mathbf{x}: \; 
												\mathbf{B}_\mathcal{S}^H\mathbf{x}=\mathbf{0}
											\},
		\qquad
		\mathcal{I}^\perp \cap \mathcal{V}_I		=	\{
												\mathbf{x}: \; 
												\mathbf{B}_\mathcal{I}^H\mathbf{x}=\mathbf{0}
											\},
	\end{align}
where
	\begin{align}
		\mathbf{B}_{\mathcal{I}}   		\triangleq      	\begin{bmatrix}
												\mathbf{H}_{\mathcal{V}_I^\perp} 
														& \mathbf{H}_\mathcal{I} \\
											\end{bmatrix},
		\qquad
		\mathbf{B}_{\mathcal{S}}   	\triangleq      	\begin{bmatrix}
												\mathbf{H}_{\mathcal{V}_I^\perp} 
														& \mathbf{h}_\mathcal{S} \\
											\end{bmatrix}.
											\nonumber
	\end{align}
with $\mathbf{H}_{\mathcal{V}_I^\perp}$ being the matrix whose columns are the orthonormal basis
of $\mathcal{V}_I^\perp$. Then, by \eqref{Equ:Threshold:SVI_AND_IVI}, whether
the eigenvalues of $(\mathbf{Y}_\mathcal{S}, \mathbf{Y}_\mathcal{I})$ are finite is equivalent 
to the validity of
$\mathcal{R}\{ \mathbf{B}_{\mathcal{I}} \} \!\!\supseteq\!\!  \mathcal{R}\{ \mathbf{B}_{\mathcal{S}} \}$.
Define
	\begin{align}
		&\mathbf{T}_{ \mathbf{B}_{\mathcal{I}} }	
				\triangleq  		\begin{bmatrix}
								\mathbf{I} 		& 		-\mathbf{H}_{\mathcal{V}_I^\perp}^H 
													\mathbf{H}_\mathcal{I}\\
								0 			& 		\mathbf{I} 
							\end{bmatrix},
		\qquad
		\mathbf{T}_{ \mathbf{B}_{\mathcal{S}} }    
				\triangleq  		\begin{bmatrix}
								\mathbf{I} 		& 		-\mathbf{H}_{\mathcal{V}_I^\perp}^H 
													\mathbf{h}_\mathcal{S} \\
								0	 		& 		\mathbf{I} 
							\end{bmatrix}.
							\nonumber
	\end{align}
Right-multiplying $\mathbf{B}_{\mathcal{I}}$ and $\mathbf{B}_{\mathcal{S}}$ by $\mathbf{T}_{ \mathbf{B}_{\mathcal{I}} }$ and $\mathbf{T}_{ \mathbf{B}_{\mathcal{S}} } $,
respectively, we can have
	\begin{align}
		\mathbf{B}_{\mathcal{I}} \cdot \mathbf{T}_{ \mathbf{B}_{\mathcal{I}} }
					=		\begin{bmatrix}
								\mathbf{H}_{\mathcal{V}_c^\perp}      
									&       \mathbf{P}_{\mathcal{V}_I} \mathbf{H}_\mathcal{I}
							\end{bmatrix},
		\qquad
		\mathbf{B}_{\mathcal{S}} \cdot \mathbf{T}_{ \mathbf{B}_{\mathcal{S}} }
					=       	\begin{bmatrix}
								\mathbf{H}_{\mathcal{V}_c^\perp}      
									&       \mathbf{P}_{\mathcal{V}_I} \mathbf{h}_\mathcal{S}
							\end{bmatrix},
		\nonumber
	\end{align}
where $\mathbf{P}_{\mathcal{V}_I} \!\triangleq\! \mathbf{H}_{\mathcal{V}_I} \mathbf{H}_{\mathcal{V}_I}^H$ 
is the projection matrix onto $\mathcal{V}_I$. Since $\mathbf{T}_{ \mathbf{B}_{\mathcal{I}} }$ and 
$\mathbf{T}_{ \mathbf{B}_{\mathcal{S}} }$ are nonsingular, the above 
two matrices have the same range spaces
as $\mathbf{B}_\mathcal{S}$ and $\mathbf{B}_\mathcal{I}$ do, namely
we can have
    \begin{align}
        \mathcal{R}\{ \mathbf{B}_{\mathcal{I}} \}      
                                                        =       \mathcal{V}_I^\perp
                                                                \oplus
                                                                \mathcal{R}\{\mathbf{P}_{\mathcal{V}_I} \mathbf{H}_\mathcal{I}\},
        \qquad
        \mathcal{R}\{ \mathbf{B}_{\mathcal{S}} \}
                                                        =       \mathcal{V}_I^\perp
                                                                \oplus
                                                                \mathcal{R}\{\mathbf{P}_{\mathcal{V}_I} \mathbf{h}_\mathcal{S}\}
                                                                \nonumber
    \end{align}
Therefore, whether the generalized eigenvalues of the noise free pair are bounded is finally equivalent to whether 
    \begin{align}
        \label{Equ:Performance:JudgeCriteria_PeriodicalInterference}
        \mathcal{R}\big\{\mathbf{P}_{\mathcal{V}_I} \mathbf{H}_\mathcal{I}\big\} 
        \!\!\supseteq\!\! 
        \mathcal{R}\big\{\mathbf{P}_{\mathcal{V}_c} \mathbf{h}_\mathcal{S}\big\}.
    \end{align}
    
    \begin{figure}[t]
        \centering
        \includegraphics[width=0.5\textwidth]{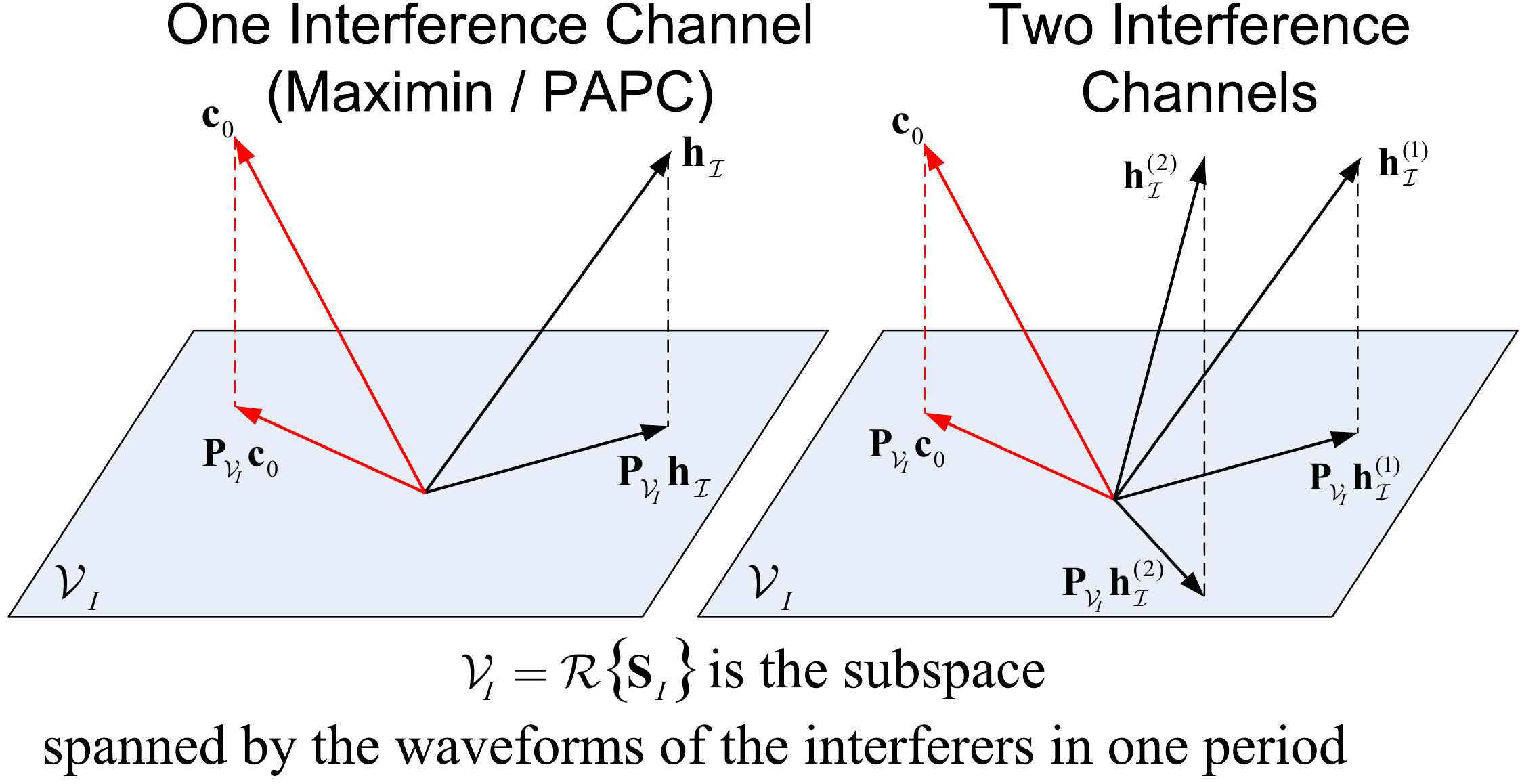}
        \caption{
        Geometrical interpretation of 
        (\ref{Equ:Performance:JudgeCriteria_PeriodicalInterference}). 
        $\mathbf{h}_\mathcal{I}^{(1)}$ and $\mathbf{h}_\mathcal{I}^{(2)}$ 
        denote the columns of $\mathbf{H}_\mathcal{I}$.
        For one interference channel, $\mathbf{H}_\mathcal{I}\!=\!\mathbf{h}_\mathcal{I}$. }
        \label{Fig:Performance:Geometric_gamma1_boundedness}
    \end{figure}

Fig. \ref{Fig:Performance:Geometric_gamma1_boundedness} shows the geometrical interpretation 
of (\ref{Equ:Performance:JudgeCriteria_PeriodicalInterference}).
To see if $\gamma_1$ is bounded, we can project $\mathbf{h}_\mathcal{S}$ 
and all the columns of $\mathbf{H}_\mathcal{I}$ onto $\mathcal{V}_I$.
If the former projection lies in the space spanned by the later ones,
then $\gamma_1$ is bounded. Otherwise, there is an infinite eigenvalue in the noise free pair
and $\gamma_1$ will increase unboundedly with $\textsf{INR}$. Since $\mathbf{H}_\mathcal{I}$ of the Maximin 
scheme and PAPC scheme has only one column, this means the projections of $\mathbf{h}_\mathcal{S}$
and $\mathbf{H}_\mathcal{I}$ should be on the same line, which is hardly valid. 
Thus, these two approaches are sensitive to
directional periodical interferences, as we will see in later simulation results. Furthermore, 
Fig. \ref{Fig:Performance:Geometric_gamma1_boundedness}
also shows the case of multiple interference channels, i.e. the dimension 
of $\mathcal{I}$ is larger than one, then $\gamma_1$ is more likely to be bounded.  This issue is out of the scope of this paper and will be discussed in 
in \cite{wang2009micmpb}.

\section{Simulation Results}

In this section, we simulate various scenarios and compare them to our theoretical results.
The matrix pair beamformers implemented include the  Maximin algorithm and PAPC algorithm, 
which have many kinds of adaptive algorithms. However, we are only interested in their steady 
state performance. The reason for this is that, if the methods suffer from poor steady state 
performance, then it is meaningless to investigate their adaptive algorithms.
Therefore, we directly calculate their weight vectors by performing generalized 
eigen-decompositions on the estimated matrix pairs. The interference signals encountered in
the simulation include BPSK signal, tones, periodical noise and multiuser interference. 
The first one is the directional white noise and the others are directional periodical interferences.

In all cases, we consider a uniform linear array (ULA) with eight isotropic antennas ($L=8$) spaced half a wavelength apart. For each user, a $100$ kbps DPSK signal is randomly generated and spreaded 
by a distinct 31-chip Gold sequence in each simulation trial. Then it is modulated onto a $1$ 
GHz carrier to form a RF signal with bandwidth $3.1$ MHz. In all the simulations, we assume  
the SOI arrives from $0^{\circ}$ and the interferers have equal power.

\begin{figure*}[t]
	\centering{
			\subfigure[$\textsf{G}(\textsf{SNR})$ of Maximin and PAPC under BPSK jammers.]
				{
					\includegraphics[width=0.45\textwidth]{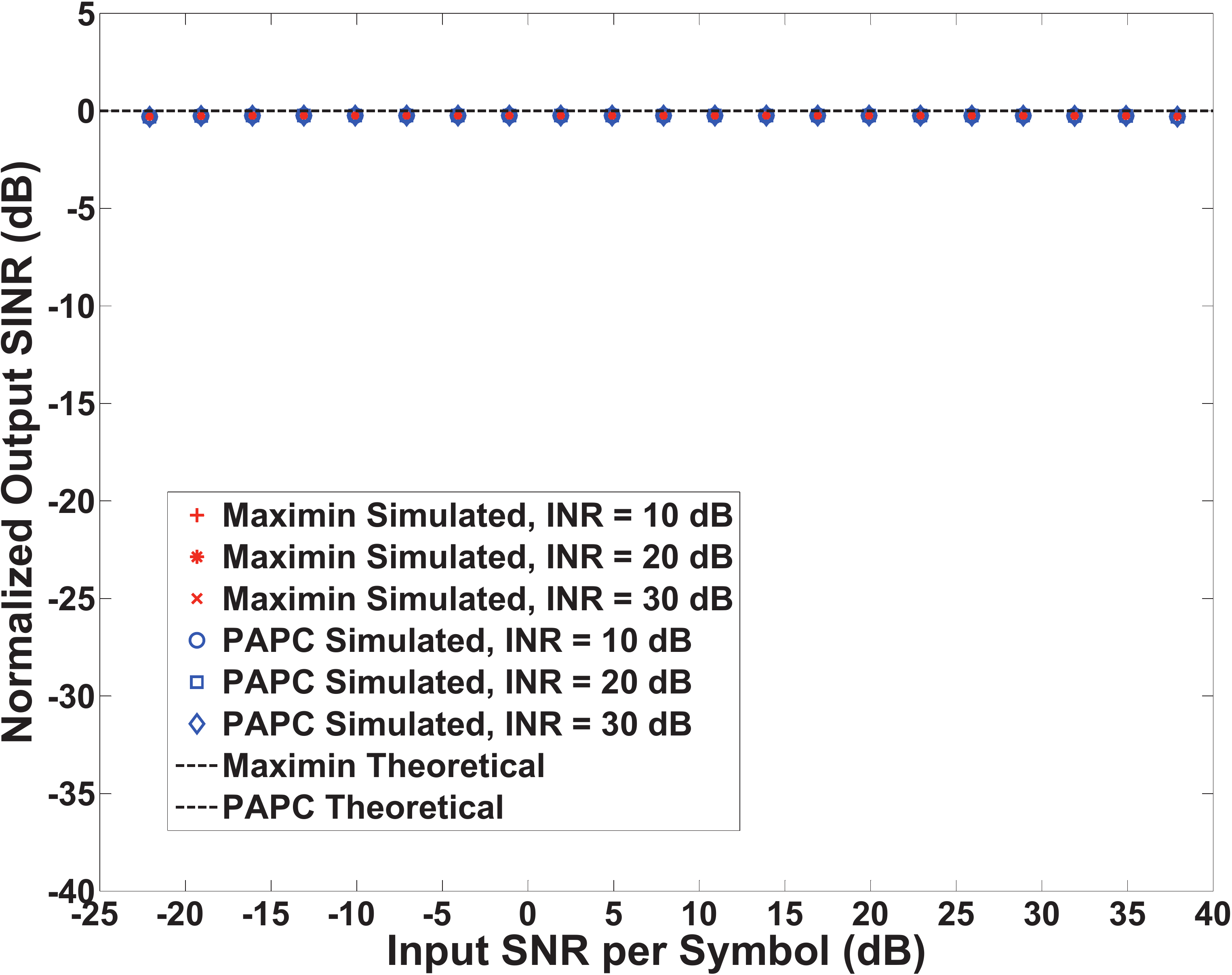}
					\label{Fig:Simulation:GvsEb:TB_GvsSNR_noMIC_3Inf}
				}
			\subfigure[$\textsf{G}(\textsf{SNR})$ of  Maximin and PAPC under periodical noise.]
				{
					\includegraphics[width=0.45\textwidth]{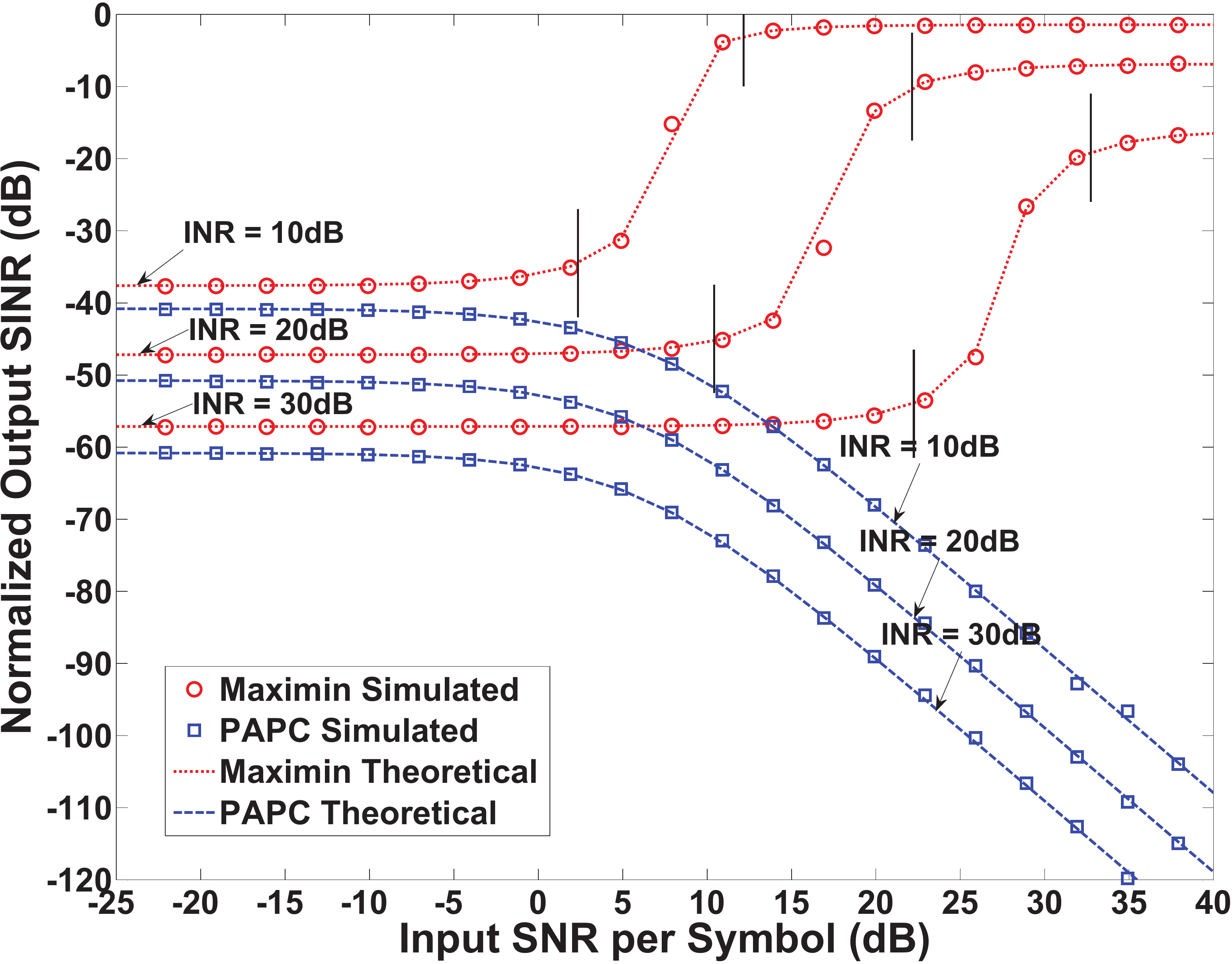}
					\label{Fig:Simulation:GvsEb:PN_GvsSNR_noMIC_3Inf}
				}
	}
	\centering{
			\subfigure[$\textsf{G}(\textsf{SNR})$ of  Maximin and PAPC under tone jammers.]
				{
					\includegraphics[width=0.45\textwidth]{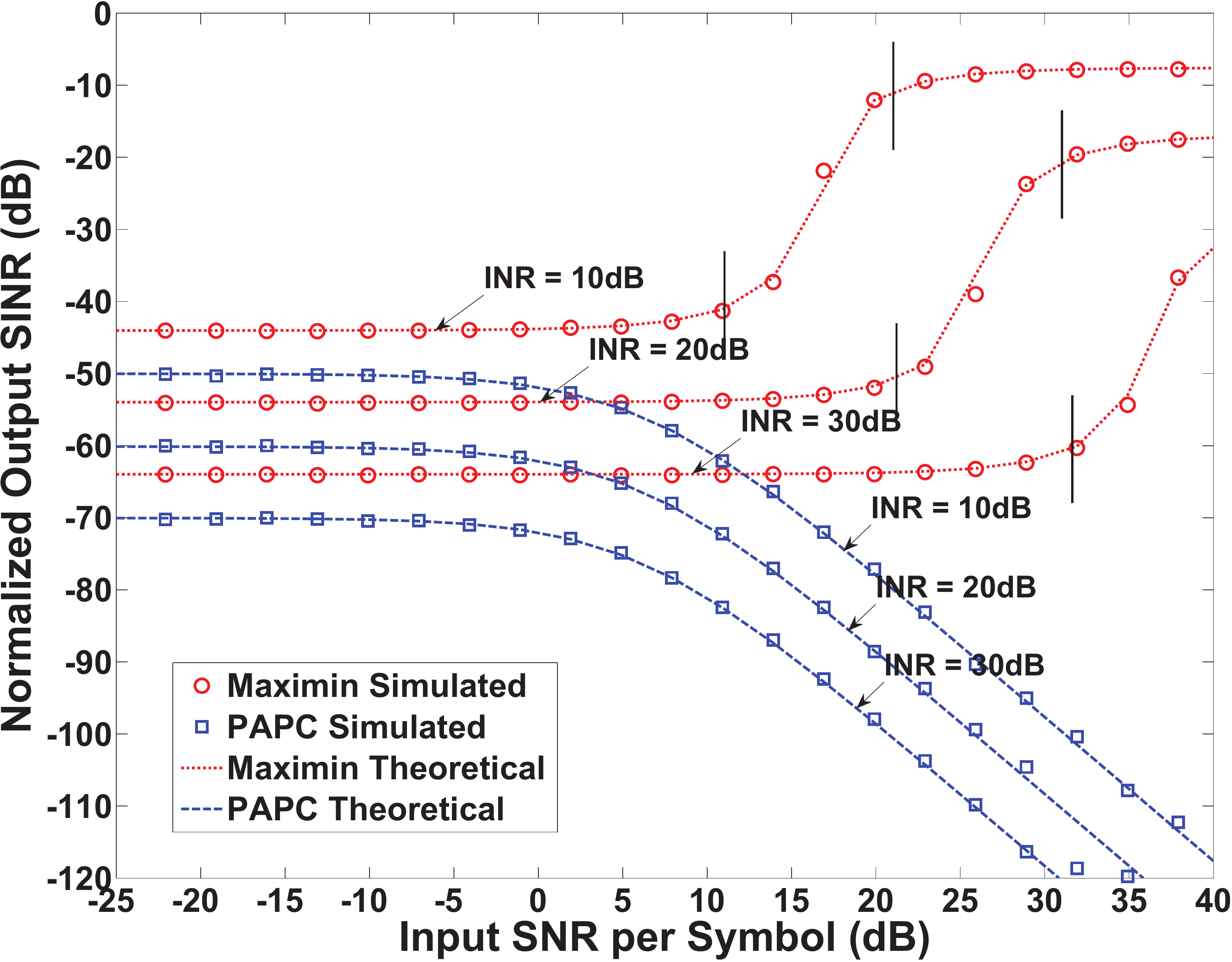}
					\label{Fig:Simulation:GvsEb:TT_GvsSNR_noMIC_5Inf}
				}
			\subfigure[$\textsf{G}(\textsf{SNR})$ of  Maximin and PAPC 
					under multipath MAI.]
				{
					\includegraphics[width=0.45\textwidth]{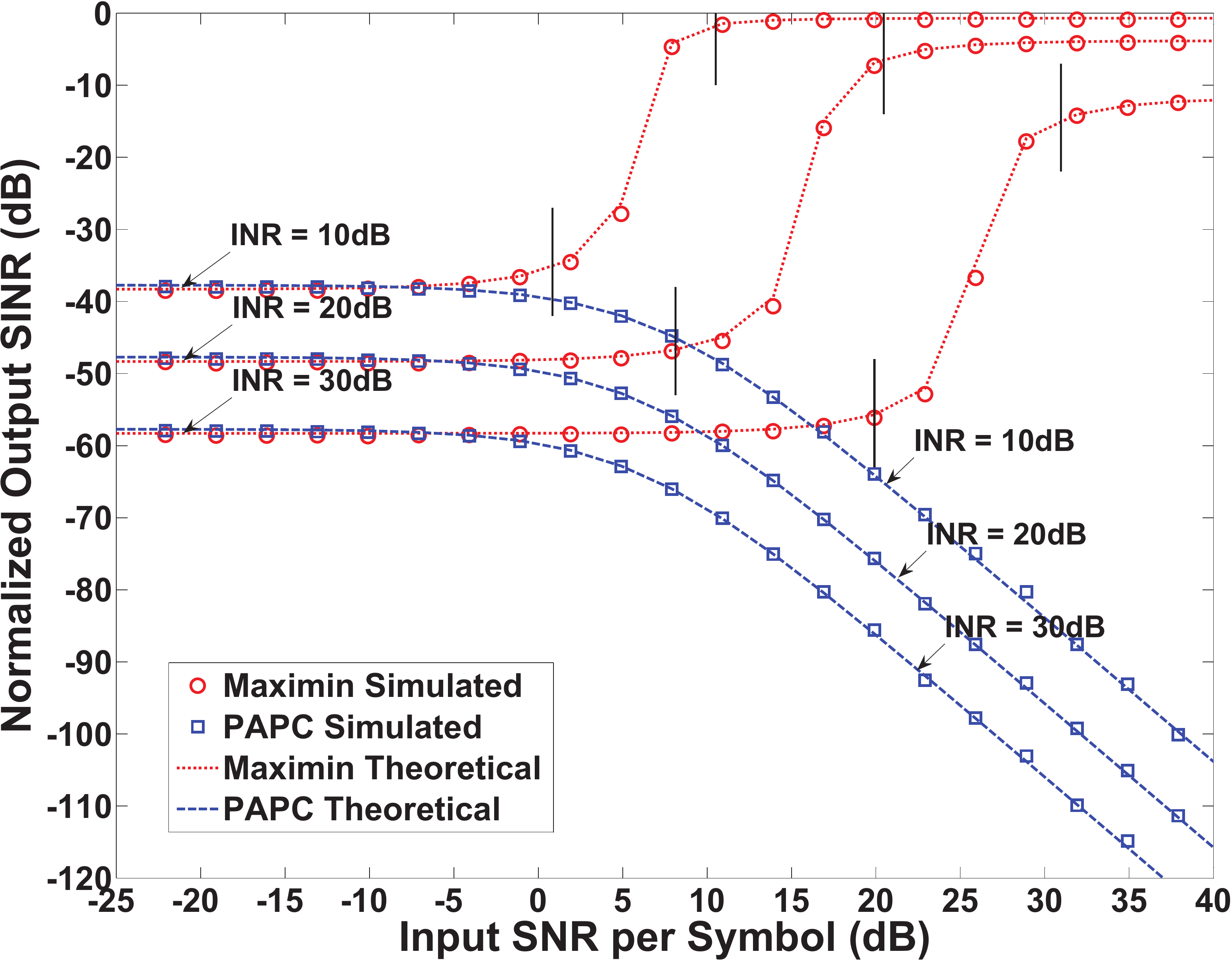}
					\label{Fig:Simulation:GvsEb:TM_GvsSNR_noMIC_3Inf}
				}
	}
	\caption{Simulated and theoretical operating curve $\textsf{G}(\textsf{SNR})$ under
			different kinds of interference for Maximin and PAPC.}
	\label{Fig:Simulation:G_SNR}
\end{figure*}

Fig. \ref{Fig:Simulation:GvsEb:TB_GvsSNR_noMIC_3Inf}--Fig. \ref{Fig:Simulation:GvsEb:TM_GvsSNR_noMIC_3Inf} show the operating curve
$\textsf{G}(\textsf{SNR})$ of the Maximin and PAPC beamformers
in the four scenarios. The simulated values are obtained by using 
the simulated data and the theoretical ones are computed by the piecewise function
$\textsf{G}(\textsf{SNR})$ in (\ref{Equ:Performance:G_Eb}). 
To eliminate the randomness caused by finite sample effects, $K=10^{6}$ symbols are simulated for
each $\textsf{SNR}$ and INR in every experiment. 


In Fig. \ref{Fig:Simulation:GvsEb:TB_GvsSNR_noMIC_3Inf}, the interferers are three uncorrelated BPSK modulated signals, i.e. random sequences of $\pm1$.
The rates of the interferers are all 3.1Mbps, which is the same as the chip rate of the SOI and covers its whole bandwidth.
Therefore, they belong to the type of directional white noise. We can see that $\textsf{G}(\textsf{SNR}) \approx 1$ and there is no threshold effect,
which is consistent with the analysis in Section \ref{Sec:TwoTypicalScenarios:DWN}.

Fig. \ref{Fig:Simulation:GvsEb:PN_GvsSNR_noMIC_3Inf}--Fig. 
\ref{Fig:Simulation:GvsEb:TM_GvsSNR_noMIC_3Inf} show the results for three types of
directional periodical interferences: periodical noise, tones and multiple access 
interference (MAI) with multipath. In Fig. \ref{Fig:Simulation:GvsEb:PN_GvsSNR_noMIC_3Inf},
we consider two periodical noises, which arrives from $30^\circ$ and $-40^\circ$, respectively,
and each of them is generated by repeating a segment of Gaussian 
white noise over times, with the repeating frequency being $100$ kHz 
(same as that of SOI's symbol rate). The frequency offsets of the 
tones with respect to the carrier in 
Fig. \ref{Fig:Simulation:GvsEb:TT_GvsSNR_noMIC_5Inf} are $100$kHz,
$-300$kHz, $0$Hz, $400$kHz and $-100$kHz, respectively, and their DOAs are 
$30^\circ$, $-50^\circ$, $-20^\circ$, $19^\circ$ and $45^\circ$. In Fig. 
\ref{Fig:Simulation:GvsEb:TM_GvsSNR_noMIC_3Inf},
there is one incident MAI signal with three-ray multipath delays of 
$3$ chips, $5$ chips and $4$ chips arriving from $30^\circ$, $-20^\circ$ and $-50^\circ$, respectively.
We can see that the theoretical values of $\textsf{G}(\textsf{SNR})$ 
match with the simulated ones very well.

\begin{figure*}[t]
	\centering{
			\subfigure[Maximin scheme]
				{
					\includegraphics[width=0.45\textwidth]{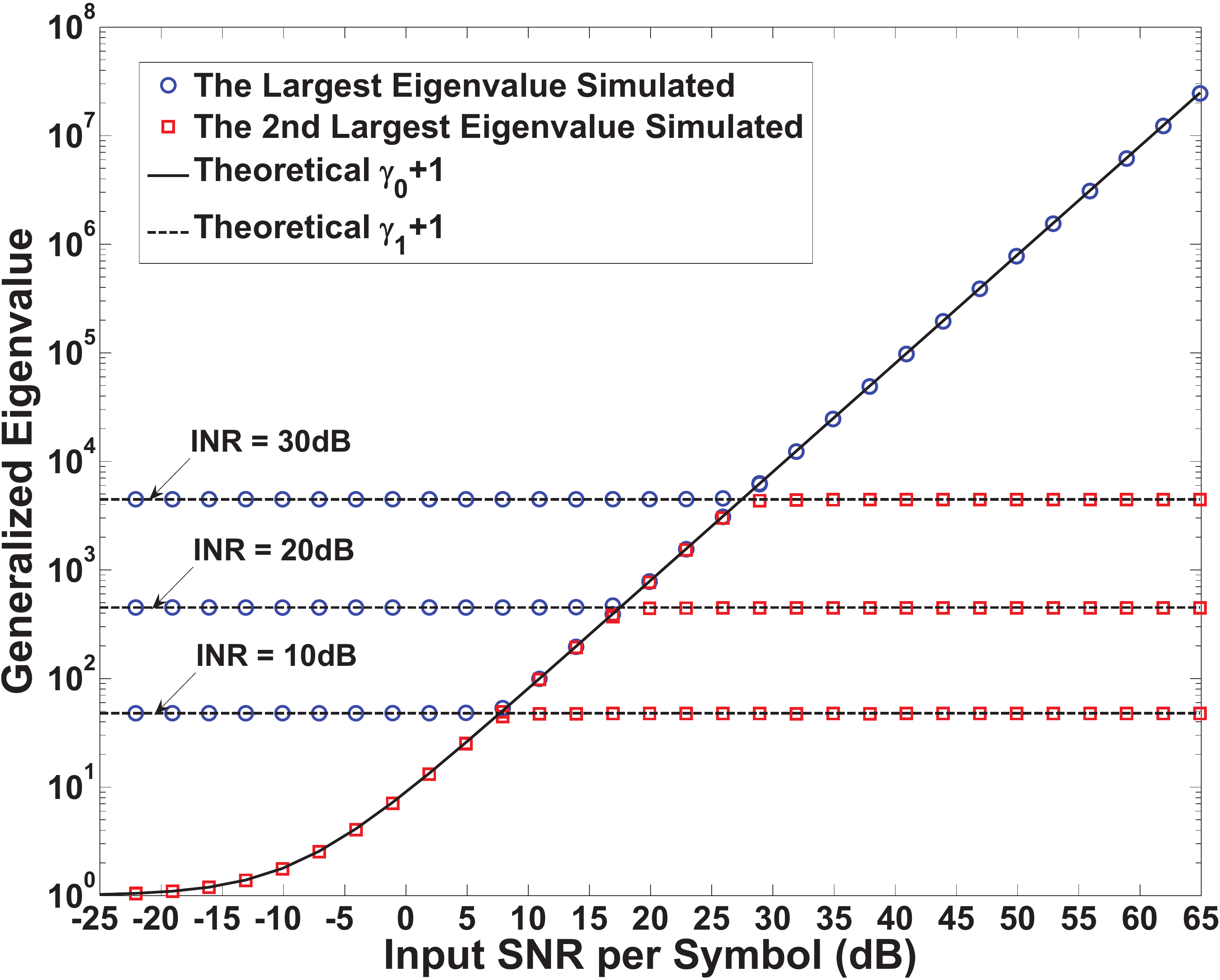}
					\label{Fig:Simulation:PN:PN_EigvsSNR_Max_3Inf}
				}
	}
	\centering{
			\subfigure[PAPC scheme]
				{
					\includegraphics[width=0.45\textwidth]{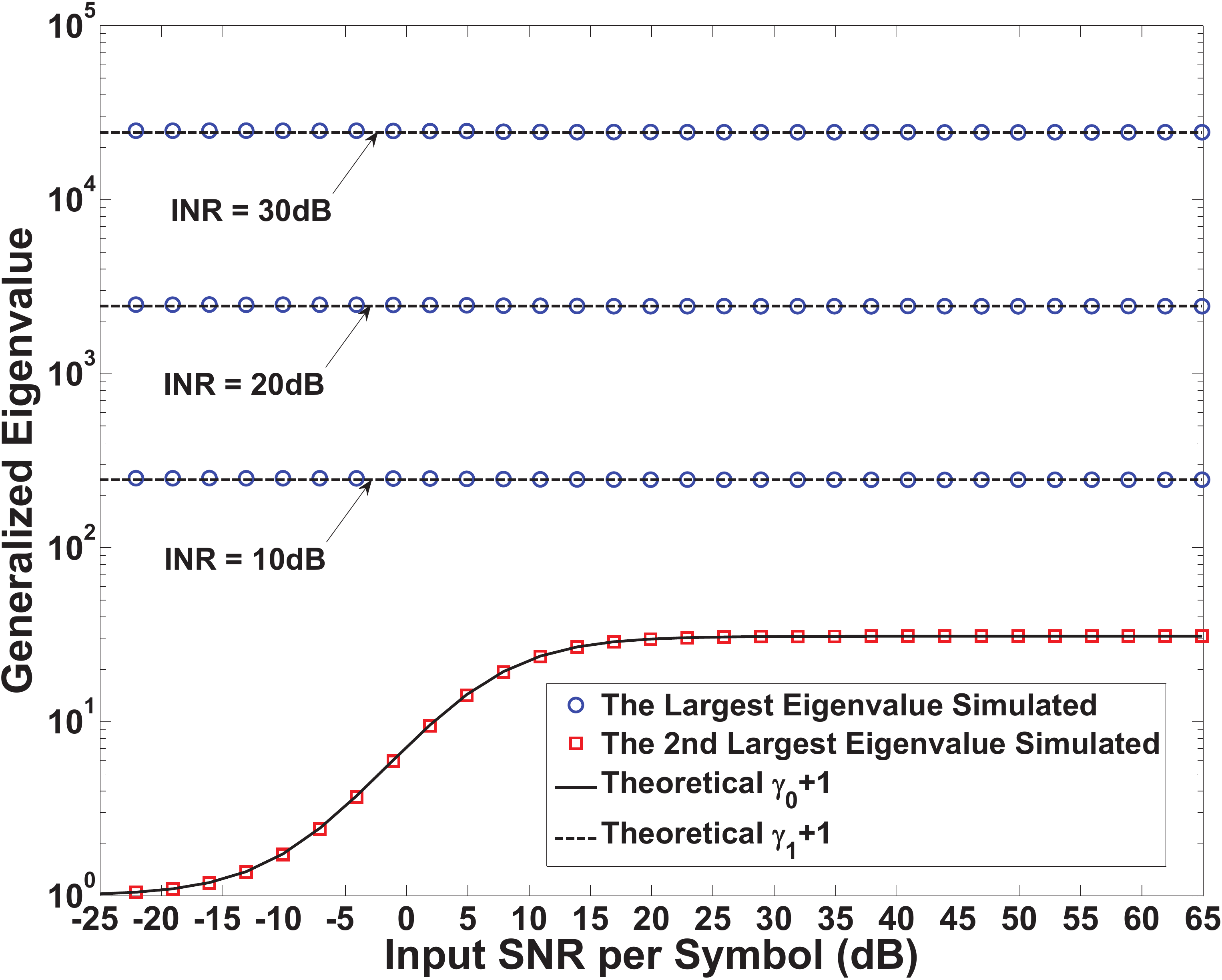}
		        			\label{Fig:Simulation:PN:PN_EigvsSNR_PAPC_3Inf}
				}
	}
	\caption{The largest and second largest generalized eigenvalues of the 
			Maximin scheme vs $\textsf{SNR}$ under two periodical noises.}
	\label{Fig:Simulation:Eigenvalue}
\end{figure*}

\begin{figure*}[t]
	\centering{
			\subfigure[Case I: below threshold ($\textsf{SNR}\!\!=\!\!-10.1$dB, $\textrm{INR}\!\!=\!\!30$dB)]
				{
					\includegraphics[width=0.45\textwidth]{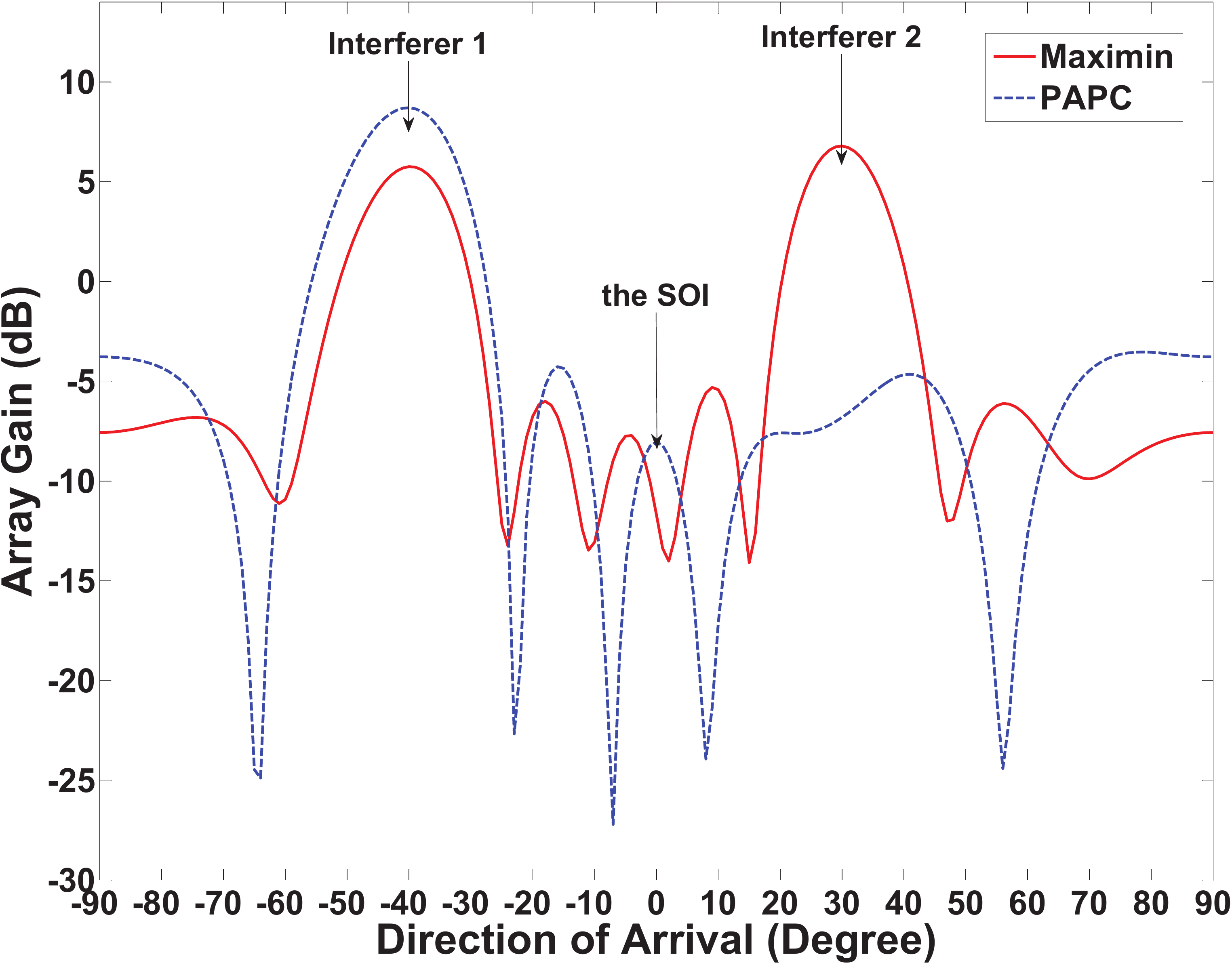}
					\label{Fig:Simulation:PN:PN_AP_below_3Inf}
				}
	}
	\centering{
			\subfigure[Case II: above threshold ($\textsf{SNR}\!\!=\!\!40.9$dB, $\textrm{INR}\!\!=\!\!30$dB)]
				{
					\includegraphics[width=0.45\textwidth]{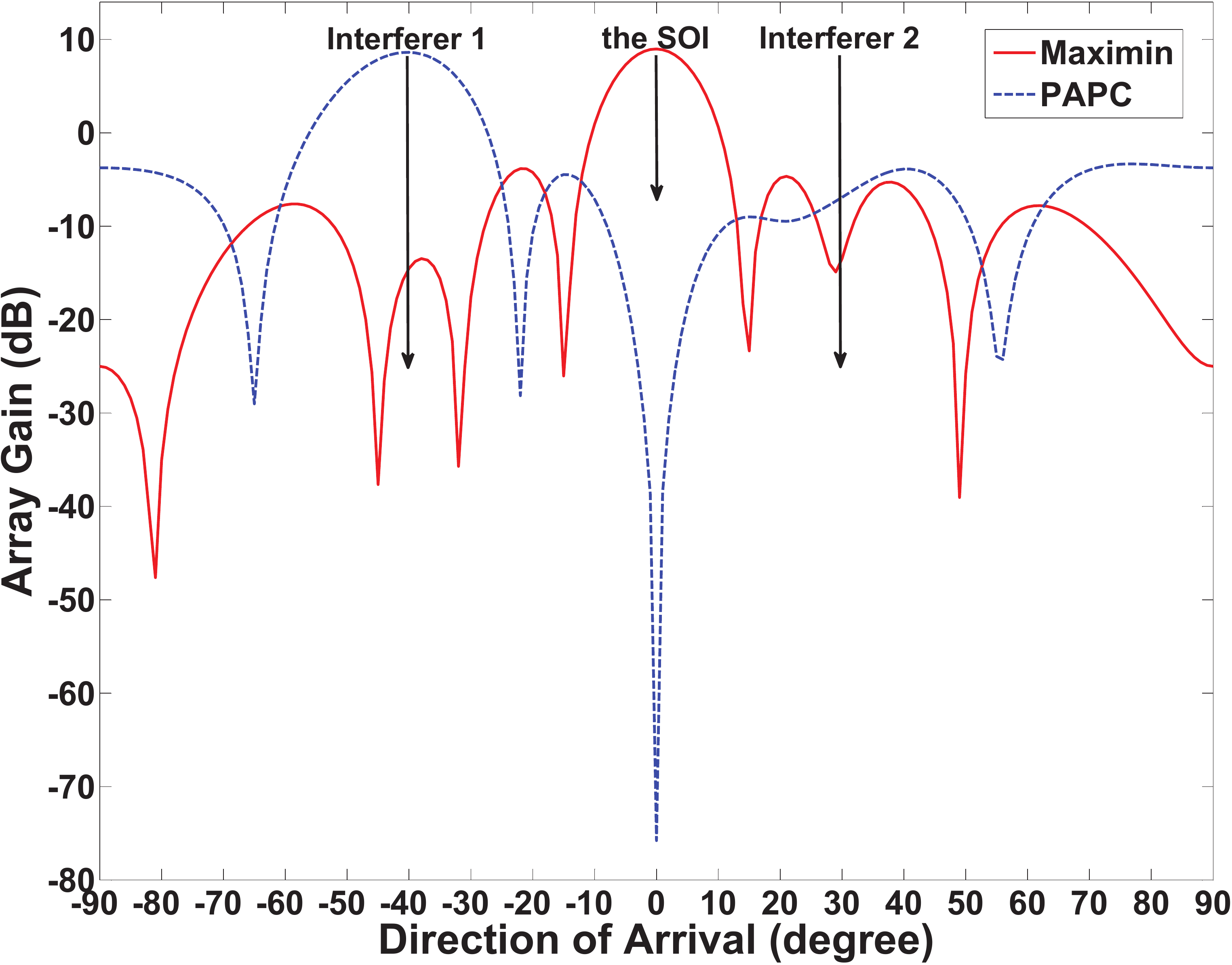}
					\label{Fig:Simulation:PN:PN_AP_above_3Inf}
				}
	}
	\caption{The array patterns corresponding to the Maximin and PAPC under two periodical noises.}
	\label{Fig:Simulation:ArrayPattern}
\end{figure*}

To verify the validity of the approximation given by Lemma \ref{Lem:lambda_max}, 
in Fig. \ref{Fig:Simulation:PN:PN_EigvsSNR_Max_3Inf} and Fig. \ref{Fig:Simulation:PN:PN_EigvsSNR_PAPC_3Inf},
we also show the curves of $\gamma_0\!+\!1$, $\gamma_1\!+\!1$,
and $\lambda_\mathrm{max}$. 
We can see that $\max\{\gamma_0+1,\gamma_1+1\}$ can be an 
excellent approximation for $\lambda_\mathrm{max}$. 
In Fig. \ref{Fig:Simulation:PN:PN_AP_below_3Inf}
and Fig. \ref{Fig:Simulation:PN:PN_AP_above_3Inf}, 
we show the array patterns that are below and above the threshold. 
The results in these four figures are simulated under two periodical noises, 
with the same parameter as Fig. \ref{Fig:Simulation:GvsEb:PN_GvsSNR_noMIC_3Inf}.
The case for the tone jammers and the multiple access interference cases are quite similar, 
and is thus omitted.

For the Maximin algorithm, all the curves of $\textsf{G}(\textsf{SNR})$ 
have failure area, threshold area and operating area, 
which are consistent with the typical curve in 
Fig. \ref{Fig:OperatingCurve:Regular}. We have also marked the predicted
thresholds of $\textsf{SNR}_{\textsf{T}1}$ and $\textsf{SNR}_{\textsf{T}2}$, 
computed by \eqref{Equ:Performance:SNR_T1_AND_SNR_T2}, 
in the figures as well, which confirm our theoretical calculations.
Furthermore, they also show that the threshold SNR would increase with the interference INR. This is because $\gamma_1+1$ (c.f.
Fig. \ref{Fig:Simulation:PN:PN_EigvsSNR_Max_3Inf}) would moving upward as INR increases, making the intersecting point of $\gamma_0+1$
and $\gamma_1+1$ move rightwards. This is consistent with our claim in Section \ref{Sec:TwoTypicalScenarios:DPI} that $\gamma_1$
would increase unboundedly with the INR. Fig. \ref{Fig:Simulation:PN:PN_AP_below_3Inf} and Fig. \ref{Fig:Simulation:PN:PN_AP_above_3Inf}
show its array pattern below and above threshold, respectively. We can see that the mainlobe of the beamformer would mistakenly point to the interferers
once the SNR is below the threshold, as we have predicted in Section 
\ref{Sec:Performance:How_MPB_works}.

For the PAPC algorithm, Fig. \ref{Fig:Simulation:GvsEb:PN_GvsSNR_noMIC_3Inf}--Fig. 
\ref{Fig:Simulation:GvsEb:TM_GvsSNR_noMIC_3Inf} show
that there are only failure areas. This is because, no matter how large $\textsf{SNR}$ is, 
$\gamma_0\!\!+\!\!1$ will never exceeds $\gamma_1\!\!+\!\!1$,
as shown by Fig. \ref{Fig:Simulation:PN:PN_EigvsSNR_PAPC_3Inf}. 
Therefore, the threshold of PAPC in this case is infinity. The array pattern
shown in Fig. \ref{Fig:Simulation:PN:PN_AP_below_3Inf} also indicates that 
its mainlobe has pointed to the interferers. We have stated in section 
\ref{Sec:Performance:How_MPB_works} that the presence of SOI in $\mathbf{R}_\mathcal{I}$ 
will make the beamformer mistakenly null the SOI. 
The curves of PAPC in Fig. \ref{Fig:Simulation:GvsEb:PN_GvsSNR_noMIC_3Inf}--Fig. 
\ref{Fig:Simulation:GvsEb:TM_GvsSNR_noMIC_3Inf} confirms this observation,
and Fig. \ref{Fig:Simulation:PN:PN_AP_above_3Inf} shows that PAPC beamformer 
has a deep null in the direction of the SOI.

In summary, the conventional MPB like Maximin and PAPC work well 
in the presence of directional white noise, even
when the INR is large. However, they are very vulnerable to multiple 
directional periodical interferers with repeating structures in the time domain and
arrive from different directions.

\section{Conclusions}

Matrix pair beamformer (MPB) is a general framework we proposed to model a class of blind beamformers that exploit
the temporal signature of the signal of interest (SOI). It has the advantages 
 only relying on the second order
statistics to achieve blind processing. 
In this paper, we have analyzed the mechanism of MPB with matrix mismatch, and showed how it worked ``blindly''.
We have  discovered that there is a threshold effect in MPB, i.e. the beamformer would fail completely if the SOI's input SNR
is below that threshold.
Meanwhile, its normalized output SINR has been derived as the performance measure, and the threshold SNR has also been predicted.
We have also observed that the existence of infinite generalized eigenvalue in what is called
\emph{noise free pair}  makes the threshold increase unboundedly with the interference power. This
is highly probable when there are multiple periodical interferers. All our theoretical analysis matches
with the simulation results very well.

Our analysis indicates that the conventional MPB is very vulnerable to multiple periodical interferers.
Moreover, it also implies the importance of choosing the appropriate projection space for the interference channel.
And we will address this issue in another paper\cite{wang2009micmpb}.

\appendices

\section{Approximation of $\lambda_\mathrm{max}$ and $\gamma_i$}
\label{Sec:Appendix:Proof_of_Thm_lambda_max}
\subsection{Approximation of $\lambda_\mathrm{max}$}
Since similarity transform does not change the eigenvalues,
we can apply it to $\mathbf{M}$ before using Lemma \ref{Lem:Gerschgorin}. 
As we will see later, the bounds derived in this way can be surprisingly tight. 
Though any transform matrix can be used, we prefer the following diagonal 
matrix that can preserve the diagonal terms of $\mathbf{M}$
	\begin{align}
		\mathbf{F}		=	\mathrm{diag}\{1, f_1,f_2,\ldots,f_D\},
						\quad (f_i>0,\; i=1,2,\ldots,D)	\nonumber
	\end{align}
Applying it to $\mathbf{M}$ in \eqref{Equ:Performance:M} and using 
Lemma \ref{Lem:Gerschgorin},  we can have the following Gerschgorin disks for 
$\mathbf{F}\mathbf{M}\mathbf{F}^{-1}$:
	\begin{align}
		G_i	&=	\{
					\lambda: \; 
					|\lambda-(\gamma_i+1)|	<	R_i(\mathbf{F})
				\},		
				\qquad
				i=0,1,2,\ldots,D,
				\nonumber
	\end{align}
where the radii of the disks are
	\begin{align}
		\label{Equ:Performance:GerschgorinDisk_Radius_0}
		R_0(\mathbf{F})	
				&=		\gamma_0	\sqrt{\frac{\sigma^2}{L}}
						\sum_{i=1}^D |\psi_{T_i}|
						\frac{1}{f_i}
						\\
		\label{Equ:Performance:GerschgorinDisk_Radius_i}
		R_i(\mathbf{F})
				&=		\gamma_i	
						\frac{
							\sqrt{\frac{\sigma^2}{L}}
							|\psi_{T_i}|
							\cdot
							\mathrm{sgn}(\gamma_i)
							}
							{
							\frac{L\beta}{N}
							\textsf{SNR}
							+
							1
							}
						f_i,
						\quad		1 \le i \le D,
	\end{align}
with  $\psi_{T_i}$ being the $i$th component of $\bm{\psi}_T$. Now, we are going to optimize
$f_1,\ldots,f_D$.  To derive an effective bound for $\lambda_\mathrm{max}$,
we should ensure the rightmost disk is separated from the others (c.f. Lemma \ref{Lem:Gerschgorin}).
Therefore, our criterion for finding the optimal $\mathbf{F}$ is
\emph{minimizing the radius of the rightmost Gerschgorin disk subject to the
constraint that it is separated from all the remaining ones}. Depending on
$\gamma_0+1>\gamma_1+1$ or $\gamma_0+1<\gamma_1+1$,
the rightmost disk might be $G_0$ or $G_1$, and we now discuss them separately.

If $\gamma_0+1>\gamma_1+1$, then $G_0$ is the rightmost disk,
and we can formulate the optimization of $\mathbf{F}$ as
	\begin{align}
		\label{Equ:Performance:OptimizeG0_Objective}
		\min				~&~		R_0(\mathbf{F})	=	\gamma_0 \sqrt{\frac{\sigma^2}{L}}
											\sum_{i=1}^D
											|\psi_{T_i}| \frac{1}{f_i}
											\\
		\label{Equ:Performance:OptimizeG0_Constraint}
		\mathrm{s.t.}		~&~		\gamma_i+1+R_i(F) \le \gamma_0+1-R_0(\mathbf{F}),
								\quad	1 \le i \le D.
	\end{align}
To minimize $R_0(\mathbf{F})$, we want $f_1,\ldots,f_D$ to be as large as possible.
But this will increase the radius of $G_1,\ldots,G_D$, and to avoid
connecting $G_0$, they cannot be too large. In fact, 
$f_1,\ldots, f_D$ have different importance in this tradeoff.
Since $\gamma_0>\ldots>\gamma_D$, $G_2,\ldots, G_D$ are
farther away from $G_0$ than $G_1$. Therefore, $f_2,\ldots,f_D$ can be
reasonably larger than $f_1$ when keeping separate from $G_0$.
Thus, to simplify analysis, we can ignore all the terms except $|\psi_{T_1}|/f_1$
in $R_0(\mathbf{F})$, i.e.
$R_0(\mathbf{F}) \approx \gamma_0 \sqrt{\frac{\sigma^2}{L}}\frac{|\psi_{T_1}|}{f_1}$
in \eqref{Equ:Performance:OptimizeG0_Objective} and \eqref{Equ:Performance:OptimizeG0_Constraint},
and only the constraint for $i=1$ is effective in \eqref{Equ:Performance:OptimizeG0_Constraint}.
Then, the problem can be reduced to
	\begin{align}
		\min				~&~		R_0(\mathbf{F})	\approx	\gamma_0
														\sqrt{\frac{\sigma^2}{L}}
														|\psi_{T_1}|
														\frac{1}{f_1}
								\nonumber\\
		\mathrm{s.t.}		~&~		\gamma_1	
								\frac{
									\sqrt{\frac{\sigma^2}{L}}
									|\psi_{T_1}|
									\cdot
									\mathrm{sgn}(\gamma_1)
									}
									{
									\frac{L\beta}{N}
									\textsf{SNR}
									+
									1
									}
								f_1^2
								-
								(\gamma_0-\gamma_1)f_1
								+
								\gamma_0
								\sqrt{\frac{\sigma^2}{L}}
								|\psi_{T_1}|
								\le 0.
								\nonumber
	\end{align}
This is a simple convex optimization problem, feasible when
$\frac{\gamma_1}{\gamma_0} \in (-\infty, 1+2\delta-\sqrt{(1+2\delta)^2-1}]$,
where $\delta	\triangleq		\frac{\sigma^2}{L}
						|\psi_{T_1}|^2
						/(\frac{L\beta}{N}\textsf{SNR}+1)
						\ll 1$.
Thus, we can easily solve the above optimization and get
	\begin{align}
		|\lambda_\mathrm{max}-(\gamma_0+1)| 
						< 	\gamma_0 \cdot f\left(\frac{\gamma_1}{\gamma_0}\right)
										\nonumber
	\end{align}
where $f(x) \triangleq \frac{1}{2} [1\!-\!x\!-\!\sqrt{(1\!-\!x)^2\!-\!4\delta |x|}]$ 
with $x \in(-\infty,1\!-\!2\sqrt{\delta\!+\!\delta^2}\!+\!2\delta]
		\!\cup\! [1\!+\!2\sqrt{\delta\!+\!\delta^2}\!+\!2\delta,+\infty)$.
Furthermore, by taking the derivative of $f(x)$, we can easily get 
$0 \!\!\le\!\! f(x) \!\!\le\!\! \max\{ \delta, \sqrt{\delta^2\!\!+\!\!\delta}\!\!-\!\!\delta \} \!\!\ll\!\! 1$
when $x \in(-\infty,1\!-\!2\sqrt{\delta\!+\!\delta^2}\!+\!2\delta]$.
	
If $\gamma_0+1<\gamma_1+1$, 
then $G_1$ is the rightmost disk and $\gamma_1>0$. The optimization problem is
	\begin{align}
		\label{Equ:Performance:OptimizeG1_Objective}
		\min			~&~		R_1(\mathbf{F})	=	\gamma_1 
												\frac{
													\sqrt{\frac{\sigma^2}{L}}
													|\psi_{T_1}|
												       }
												       {
												       	\frac{L\beta}{N}
													\textsf{SNR}
													+
													1
												       }
												f_1
												\\
		\label{Equ:Performance:OptimizeG1_Constraint1}
		\mathrm{s.t.}	~&~		\gamma_0+1+R_0(F) \le \gamma_1+1-R_1(\mathbf{F}),\\
		\label{Equ:Performance:OptimizeG1_Constraint2}
					~&~		\gamma_i+1+R_i(F) \le \gamma_1+1-R_1(\mathbf{F}),
							\quad	2 \le i \le D.
	\end{align}
According to \eqref{Equ:Performance:GerschgorinDisk_Radius_0}
and \eqref{Equ:Performance:GerschgorinDisk_Radius_i}, 
$f_1$ should be as small as possible to minimize $R_1(\mathbf{F})$
in \eqref{Equ:Performance:OptimizeG1_Objective},
\eqref{Equ:Performance:OptimizeG1_Constraint1}
and \eqref{Equ:Performance:OptimizeG1_Constraint2}. 
However, this will increase $R_0(\mathbf{F})$ in 
\eqref{Equ:Performance:OptimizeG1_Constraint1}, making $G_0$ connect
with $G_1$. Therefore, $f_1$ cannot be arbitrarily small.
On the other hand, $f_2,\ldots,f_D$ should be as large as possible to reduce $R_0(\mathbf{F})$
in \eqref{Equ:Performance:OptimizeG1_Constraint1}
while keeping \eqref{Equ:Performance:OptimizeG1_Constraint2} valid.
As a result, $|\psi_{T_1}|/f_1$ is still the dominant term in $R_0(\mathbf{F})$ 
(c.f. \eqref{Equ:Performance:GerschgorinDisk_Radius_0}) and the key point here
remains the tradeoff between $R_1(\mathbf{F})$ and $R_0(\mathbf{F})$. 
In other words, the optimization becomes
	\begin{align}
		\min			~&~		R_1(\mathbf{F})	=	\gamma_1 
												\frac{
													\sqrt{\frac{\sigma^2}{L}}
													|\psi_{T_1}|
												       }
												       {
												       	\frac{L\beta}{N}
													\textsf{SNR}
													+
													1
												       }
												f_1
												\nonumber\\
		\mathrm{s.t.}	~&~		\gamma_1	
								\frac{
									\sqrt{\frac{\sigma^2}{L}}
									|\psi_{T_1}|
									}
									{
									\frac{L\beta}{N}
									\textsf{SNR}
									+
									1
									}
								f_1^2
								-
								(\gamma_1-\gamma_0)f_1
								+
								\gamma_0
								\sqrt{\frac{\sigma^2}{L}}
								|\psi_{T_1}|
								\le 0.
								\nonumber
	\end{align}
The feasible region for this convex optimization is
$\frac{\gamma_1}{\gamma_0} \in [1+2\delta+\sqrt{(1+2\delta)^2-1},+\infty)$,
with $\delta$ defined in the previous case.  
By solving it, we can finally 
get the bound for $\lambda_\mathrm{max}$ as
	\begin{align}
		|\lambda_\mathrm{max}-(\gamma_1+1)| 	\le
										\gamma_1 
										\cdot 
										f\left(\frac{\gamma_0}{\gamma_1}\right)
										\nonumber
	\end{align}
where $f(x)$ is the same as in the previous case.

\subsection{Approximation of $\gamma_i$}
Then, we discuss the approximation of $\gamma_i$. By substituting \eqref{Equ:MPBeamformer:R_I}
into \eqref{Equ:Performance:EigenEquation_lambda_i_1} and using matrix inversion
lemma, we can have the equivalent equation of \eqref{Equ:Performance:EigenEquation_lambda_i_1}
as
	\begin{align}
		\label{Equ:Performance:EigenEquation_lambda_i_2}
		\det(\mathbf{A}_I^H\mathbf{R}_\mathcal{I}^{-1}\mathbf{A}_I)^{-1}
		\cdot
		\det
			\left\{
				\lambda \mathbf{I}
				-
				\left[
					\mathbf{A}_I^H\mathbf{Q}_\mathcal{I}^{-1}\mathbf{A}_I
					-
					\frac{					
						\sigma_\mathcal{I}^2
						\mathbf{A}_I^H\mathbf{Q}_\mathcal{I}^{-1}\mathbf{a}_0
						\cdot
						\mathbf{a}_0^H\mathbf{Q}_\mathcal{I}^{-1}\mathbf{A}_I
					       }
					       {
					       1
					       +
					       \sigma_\mathcal{I}^2
					       \mathbf{a}_0^H\mathbf{Q}_\mathcal{I}^{-1}\mathbf{a}_0
					       }
				\right]
				\bm{\Phi}_\Delta
			\right\}
			=0
	\end{align}
To further reduce the above expression, we need factorizations of 
$\mathbf{A}_I^H\mathbf{Q}_\mathcal{I}^{-1}\mathbf{A}_I$ and $\bm{\Phi}_\Delta$
like \eqref{Equ:Performance:JointDiag}.
Let $\sigma_{\mathcal{I}_0}^2=0$ in \eqref{Equ:Performance:JointDiag}. 
Then, $\mathbf{R}_\mathcal{I} = \mathbf{Q}_\mathcal{I}$,
and \eqref{Equ:Performance:JointDiag}
become
	\begin{align}
		\mathbf{T}^H_0\bm{\Phi}_\Delta\mathbf{T}_0 = \bm{\Gamma}_0,
			\qquad
		\mathbf{T}^H_0 (\mathbf{A}_I^H\mathbf{Q}_\mathcal{I}^{-1}\mathbf{A}_I)^{-1}\mathbf{T}_0 
											= \mathbf{I}.			\nonumber
	\end{align}
where $\mathbf{T}_0$ and $\bm{\Gamma}_0$ are the counterparts of $\mathbf{T}$ and $\bm{\Gamma}$,
respectively.
Substituting them into \eqref{Equ:Performance:EigenEquation_lambda_i_2}, 
we have
	\begin{align}
		\det(\mathbf{A}_I^H\mathbf{R}_\mathcal{I}^{-1}\mathbf{A}_I)^{-1}
		\cdot
		\det
			\left\{
				\lambda \mathbf{I}
				-
				\left[
					\mathbf{I}
					-
					\frac{					
						\sigma_\mathcal{I}^2
						\tilde{\bm{\psi}}_{T_0} \tilde{\bm{\psi}}_{T_0}^H
					       }
					       {
					       1
					       +
					       \sigma_\mathcal{I}^2
					       \mathbf{a}_0^H\mathbf{Q}_\mathcal{I}^{-1}\mathbf{a}_0
					       }
				\right]
				\bm{\Gamma}_0
			\right\}
			=0
			\nonumber
	\end{align}
where $\tilde{\bm{\psi}}_{T_0} = \mathbf{T}_0^{-1}\mathbf{A}_I^H\mathbf{Q}_\mathcal{I}^{-1}\mathbf{a}_0$
is the counterpart of $\tilde{\bm{\psi}}_T$ when $\sigma_{\mathcal{I}_0}^2=0$. 
By letting $\textsf{SNR}=0$ in Lemma \ref{Lem:aRa_psi_T}, $\mathbf{R}_\mathcal{I}$ becomes
$\mathbf{Q}_\mathcal{I}$ and we can have 
$\mathbf{a}_0^H\mathbf{Q}_\mathcal{I}^{-1}\mathbf{a}_0 \approx L/\sigma^2$ and 
$\|\bm{\psi}_{T_0}\|^2 =L\kappa_0/\sigma^2 \ll L/\sigma^2$. Therefore, 
by the similar argument of $\lambda_\mathrm{max}$ using Gerschgorin theorem, 
we can have $\gamma_i \approx \lambda_{i,0}$,
where $\lambda_{i,0}$ is the $i$th diagonal term of $\bm{\Gamma}_0$, namely, the $i$th eigenvalue
of  $(\bm{\Phi}_\Delta, (\mathbf{A}_I^H\mathbf{Q}_\mathcal{I}^{-1}\mathbf{A}_I)^{-1})$.
In fact, there is a correspondence between $\lambda_{i,0}$ and 
the generalized eigenvalue of $(\mathbf{Q}_\mathcal{S}-\mathbf{Q}_\mathcal{I},\mathbf{Q}_\mathcal{I})$. 
By \eqref{Equ:Performance:Identity_det},
	\begin{align}
		\det
		\left\{
			\lambda (\mathbf{A}_I^H\mathbf{Q}_\mathcal{I}^{-1}\mathbf{A}_I)^{-1}
			-
			\bm{\Phi}_\Delta
		\right\}
			&=	\lambda^{D-L} \cdot \det (\mathbf{A}_I^H\mathbf{Q}_\mathcal{I}^{-1}\mathbf{A}_I)^{-1}
				\det \mathbf{Q}_\mathcal{I}^{-1}
				\cdot
				\det \{\lambda \mathbf{Q}_\mathcal{I} - \mathbf{A}_I\bm{\Phi}_\Delta\mathbf{A}_I^H\}.
				\nonumber
	\end{align}
Since 
$\mathbf{Q}_\mathcal{S}-\mathbf{Q}_\mathcal{I} = \mathbf{A}_I\bm{\Phi}_\Delta\mathbf{A}_I^H$,
the above expression implies that 
$(\bm{\Phi}_\Delta, (\mathbf{A}_I^H\mathbf{Q}_\mathcal{I}^{-1}\mathbf{A}_I)^{-1})$ has the
same eigenvalues as $(\mathbf{Q}_\mathcal{S}-\mathbf{Q}_\mathcal{I},\mathbf{Q}_\mathcal{I})$
except for multiplicity of zeros. Thus, we can estimate $\gamma_i$ like this: 1) take out
all nonzero eigenvalues of $(\mathbf{Q}_\mathcal{S}-\mathbf{Q}_\mathcal{I},\mathbf{Q}_\mathcal{I})$, 
2) pad them up to $D$ eigenvalues with zeros, 3) order them decreasingly to get $\lambda_{1,0},\ldots,\lambda_{D,0}$, and 4) let $\gamma_i \approx \lambda_{i,0}$.

\section{Proof of Lemma \ref{lemma:G_lambda_max}}
\label{Sec:Appendix:Proof_of_Thm_G_lambda_max}

Substituting the definitions of $y_S(k)$ and $y_I(k)$ (c.f. \eqref{Equ:MPBeamformer:y_0}),
\eqref{Equ:MPBeamformer:Q_S} and \eqref{Equ:MPBeamformer:SINR_max} 
into \eqref{Equ:Matrix_Mismatch:G} , we can have
    \begin{align}
    	\label{Equ:Appendix:G_SNR1}
        \textsf{G}(\textsf{SNR})     
        					&=    \frac{
                                                        |\mathbf{w}^H\mathbf{a}_0|^2
                                                     }
                                                     {
                                                        \mathbf{w}^H\mathbf{A}_I
                                                        \bm{\Phi}_\mathcal{S}
                                                        \mathbf{A}_I^H\mathbf{w}
                                                        +
                                                        \sigma^2
                                                        \|\mathbf{w}\|^2
                                                     }
                                                \frac{
                                                    	1
					     	     }
                                                       {
                                                		\mathbf{a}_0^H\mathbf{Q}_\mathcal{S}^{-1}\mathbf{a}_0
						     }
					\approx    
						\frac{
                                                        |\mathbf{w}^H\mathbf{a}_0|^2
                                                     }
                                                     {
                                                        \mathbf{w}^H\mathbf{A}_I
                                                        \bm{\Phi}_\mathcal{S}
                                                        \mathbf{A}_I^H\mathbf{w}
                                                        +
                                                        \sigma^2
                                                        \|\mathbf{w}\|^2
                                                     }
                                                \big/
                                                \frac{
                                                    	L
					     	     }
                                                       {
                                                		\sigma^2
						     },
    \end{align}
where we used
$\mathbf{a}_0^H\mathbf{Q}_\mathcal{S}^{-1}\mathbf{a}_0 \!\!\approx\!\! {L}/{\sigma^2}$, 
derived by replacing $\mathbf{R}_\mathcal{I}$ with $\mathbf{R}_\mathcal{S}$
and letting $\sigma_{\mathcal{S}_0}^2\!\!=\!\!0$ in Lemma \ref{Lem:aRa_psi_T}.

To further derive $\textsf{G}$, we first need the expression for $\|\mathbf{w}\|^2$.
The key trick is to recognize that $\mathbf{w} \in \mathcal{R}(\mathbf{A})$ so that
its projection onto $\mathcal{R}(\mathbf{A})$ equals itself, where
$\mathcal{R}(\cdot)$ is the range space of a matrix and 
$\mathbf{A} \triangleq [ \mathbf{a}_0 \; \mathbf{A}_I ]$. This can be proved by
substituting \eqref{Equ:MPBeamformer:Q_I} into \eqref{Equ:MPBeamformer:R_I},
applying matrix inversion lemma and pluging it into \eqref{Equ:Performance:w2}.
Furthermore, $\mathcal{R}(\mathbf{A}_I)$ is a subspace of $\mathcal{R}(\mathbf{A})$
with the dimension lower by one. 
Thus, $\mathcal{R}(\mathbf{A}) \!\!=\!\! \mathcal{R}(\mathbf{A}) \!\oplus\! \mathcal{R}(\hat{\mathbf{b}}_0)$,
where $\oplus$ denotes the direct sum and $\hat{\mathbf{b}}_0$ is the unit vector
in $\mathcal{R}(\mathbf{A})$ that is orthogonal to $\mathcal{R}(\mathbf{A}_I)$.
Then, the projection matrix of $\mathcal{R}(\mathbf{A})$ can be written as
	\begin{align}
		\mathbf{P}_A     =      \mathbf{A}_I(\mathbf{A}_I^H\mathbf{A}_I)^{-1}\mathbf{A}_I^H
                                             	+
                                                   \mathbf{\hat b}_0\mathbf{\hat b}_0^H.
                                                   \nonumber
	\end{align}
This together with $\mathbf{P}_A \mathbf{w} = \mathbf{w}$ and $\mathbf{P}_A^2 = \mathbf{P}_A$
yields
	\begin{align}
		\label{Equ:Appendix:w_norm2_expr1}
		\|\mathbf{w}\|^2        	=      \mathbf{w}^H \mathbf{P}_A\mathbf{w}
						=	\mathbf{w}^H
							\mathbf{A}_I(\mathbf{A}_I^H\mathbf{A}_I)^{-1}\mathbf{A}_I^H
							\mathbf{w}
                                             		+
                                                   	|\mathbf{w}^H\mathbf{\hat b}_0|^2
	\end{align}
Assumption \ref{Asm:Orthogonal} in Sec. \ref{Sec:Performance_Analysis:Operating_Curve} 
implies that $\mathbf{a}_0$  is almost orthogonal to $\mathcal{R}(\mathbf{A}_I)$. 
Thus, it is nearly aligned with $\hat{\mathbf{b}}_0$, and intuitively,
$|\mathbf{w}^H\hat{\mathbf{b}}_0| \!\!\approx\!\! |\mathbf{w}^H\hat{\mathbf{a}}_0|$,
where $\hat{\mathbf{a}}_0 \!\!\triangleq\!\! \mathbf{a}_0/\|\mathbf{a}_0\|$ 
is the unit vector of $\mathbf{a}_0$. To prove it, let 
$\mathcal{R}_0 \!\!\triangleq\!\! \mathrm{span}\{\hat{\mathbf{a}}_0,\hat{\mathbf{b}}_0\}$
and $\{\hat{\mathbf{b}}_0, \hat{\mathbf{b}}_1\}$ be the orthonormal basis of $\mathcal{R}_0$.
Its projection matrix becomes
$\mathbf{P}_{\mathcal{R}_0} \!\!=\!\! \hat{\mathbf{b}}_0\hat{\mathbf{b}}_0^H 
\!\!+\!\!\hat{\mathbf{b}}_1\hat{\mathbf{b}}_1^H$, which satisfies 
$\mathbf{P}_{\mathcal{R}_0}\hat{\mathbf{a}}_0 \!\!=\!\! \hat{\mathbf{a}}_0$
and $\mathbf{P}_{\mathcal{R}_0}\hat{\mathbf{b}}_0 \!\!=\!\! \hat{\mathbf{b}}_0$.  Define
$\mathbf{w}_0 \!\!=\!\! \mathbf{P}_{\mathcal{R}_0}\mathbf{w}/\|\mathbf{P}_{\mathcal{R}_0}\mathbf{w}\|$.
Then,
	\begin{align}
		|\mathbf{w}^H\mathbf{\hat b}_0|^2	&=	|\mathbf{w}^H\mathbf{\hat a}_0|^2
										\!\!+\!\!
										\left[
											|\mathbf{w}^H\mathbf{\hat b}_0|^2     
											\!\!-\!\!  
											|\mathbf{w}^H\mathbf{\hat a}_0|^2
										\right]
									=	|\mathbf{w}^H\mathbf{\hat a}_0|^2
										\!\!+\!\!
										\left[
											|\mathbf{w}^H\mathbf{P}_{\mathcal{R}_0}
											\mathbf{\hat b}_0|^2     
											\!\!-\!\!  
											|\mathbf{w}^H\mathbf{P}_{\mathcal{R}_0}
											\mathbf{P}_{\mathcal{R}_0}
											\mathbf{\hat a}_0|^2
										\right]
										\nonumber\\
									&=	|\mathbf{w}^H\mathbf{\hat a}_0|^2
										\!\!+\!\!
										\|\mathbf{P}_{\mathcal{R}_0}\mathbf{w}\|^2
										\left[
											|\mathbf{w}_0^H\mathbf{\hat b}_0|^2     
											\!\!-\!\!  
											|
												\mathbf{w}_0^H\hat{\mathbf{b}}_0
												\cdot
												\hat{\mathbf{b}}_0^H\mathbf{\hat a}_0
												\!\!+\!\!
												\mathbf{w}_0^H\hat{\mathbf{b}}_1
												\cdot
												\hat{\mathbf{b}}_1^H\mathbf{\hat a}_0
											|^2
										\right]
										\nonumber\\
		\label{Equ:Appendix:wb0}
									&=	|\mathbf{w}^H\mathbf{\hat a}_0|^2
										\!\!+\!\!
										\|\mathbf{P}_{\mathcal{R}_0}\mathbf{w}\|^2
										\cdot
										\cos \phi_I
										\cdot
										\cos (\phi_1+2\phi_I)							
	\end{align}
where $\phi_I$ is the angle between $\hat{\mathbf{a}}_0$ and $\mathcal{R}(\mathbf{A}_I)$,
and $\phi_1$ is the angle between $\hat{\mathbf{a}}_0$ and $\hat{\mathbf{b}}_0$.
And the derivation of the last step in \eqref{Equ:Appendix:wb0} involves 
some simple trigonometry identities like product-to-sum formula.
Let $\pmb{\Psi}_I \!\!\triangleq\!\! \mathbf{A}_I^H\mathbf{A}_I/L$.
Combining \eqref{Equ:Appendix:w_norm2_expr1},
\eqref{Equ:Appendix:wb0}, 
$\|\mathbf{P}_{\mathcal{R}_0}\mathbf{w}\| \!\!\le\!\! \|\mathbf{w}\|$
and $|\cos\phi_I| \!\!\ll\!\! 1$ (c.f. Assumption \ref{Asm:Orthogonal}), we can have
	\begin{align}
        		\|\mathbf{w}\|^2		&=	\frac{
	                                                           \frac{1}{L} \left[
	                                                                             	|\mathbf{a}_0^H\mathbf{w}|^2
	                                                                                	+
	                                                                                	\mathbf{w}^H 
	                                                                                	\mathbf{A}_I 
	                                                                                	\pmb{\Psi}_I^{-1} 
	                                                                                	\mathbf{A}_I^H 
	                                                                                	\mathbf{w}
	                                                                        	\right]
                                                         		}
                                                         		{
	                                                            	1
	                                                            	-
	                                                            	\frac{\|\mathbf{P}_{\mathcal{R}_0}\mathbf{w}\|^2}
									       {\|\mathbf{w}\|^2}
									\cdot
									\cos \phi_I
									\cdot
									\cos (\phi_1+2\phi_I)	
	                                                           }
                                        		\approx    \frac{1}{L} 
								\left[
                                                                        |\mathbf{a}_0^H\mathbf{w}|^2
                                                                        +
                                                                        \mathbf{w}^H 
                                                                        \mathbf{A}_I \pmb{\Psi}_I^{-1} 
                                                                        \mathbf{A}_I^H 
                                                                        \mathbf{w}
                                                                \right],
                                                    \nonumber
    \end{align}
where we also used $\|\mathbf{a}_0\|^2\!\!=\!\!L$ in Assumption \ref{Asm:Norm}.
Substitute the above expression back to \eqref{Equ:Appendix:G_SNR1}, and we get
	\begin{align}
	        \label{Equ:Appendix:G_lambda_max_expr2}
	        \textsf{G}(\textsf{SNR})	\approx\	   \frac{
	                                                            			|\mathbf{w}^H\mathbf{a}_0|^2
	                                                        			  }
	                                                        			  {
	                                                            			 \mathbf{w}^H\mathbf{A}_I   \!
				                                                            \left[
				                                                                    \frac{L}{\sigma^2} \bm{\Phi}_\mathcal{S}   
				                                                                    \!\!+\!\!   
				                                                                    \pmb{\Psi}_I^{-1}
				                                                            \right]
				                                                            \!
				                                                            \mathbf{A}_I^H\mathbf{w}
				                                                            \!\!+\!\!
				                                                            |\mathbf{w}^H\mathbf{a}_0|^2
				                                                     }.
    \end{align}


Next, we are going to evaluate $\mathbf{w}^H\mathbf{a}_0$
and $\mathbf{w}^H\mathbf{A}_I [\frac{L}{\sigma^2} \bm{\Phi}_\mathcal{S}\!\!+\!\!\pmb{\Psi}_I^{-1}]
\mathbf{A}_I^H\mathbf{w}$. By the expression of $\mathbf{w}$
in \eqref{Equ:Performance:w2},
	\begin{align}
        \mathbf{a}_0^H\mathbf{w}        &=          \mathbf{a}_0^H\mathbf{R}_\mathcal{I}^{-1}\mathbf{a}_0
	                                                                +
	                                                                \sum_{i=1}^D
	                                                                \frac{
	                                                                        \gamma_i
	                                                                    }
	                                                                    {
	                                                                        \lambda_\mathrm{max}
	                                                                        \!-\!
	                                                                        (\gamma_i\!+\!1)
	                                                                    }
	                                                                |\widetilde{\psi}_{T_i}|^2
	                                                 =          \mathbf{a}_0^H\mathbf{R}_\mathcal{I}^{-1}\mathbf{a}_0
	                                                                \!\!-\!\!
	                                                                \|\widetilde{\pmb{\psi}}_T\|^2
	                                                                \!\!+\!\!
	                                                                \sum_{i=1}^D
	                                                                \frac{
	                                                                        \lambda_\mathrm{max}\!\!-\!\!1
	                                                                    }
	                                                                    {
	                                                                        \lambda_\mathrm{max}
	                                                                        \!-\!
	                                                                        (\gamma_i+1)
	                                                                    }
	                                                                |\widetilde{\psi}_{T_i}|^2
	                                                                \nonumber\\
        \label{Equ:Appendix:a0H_w_mp}
                                                    &\approx  \frac{L}{\sigma^2}  
                                                                     \left(
                                                                        \frac{L\beta}{N} \textsf{SNR}
                                                                        +
                                                                        1
                                                                     \right)
                                                                \left[
                                                                        1
                                                                        +
                                                                        \left(
                                                                        	\frac{L\beta}{N} \textsf{SNR}
                                                                        	+
                                                                        	1
                                                                        \right)^{-1}
                                                                        \psi_S(\lambda_\mathrm{max})
                                                                \right],
    \end{align}
where we used Lemma \ref{Lem:aRa_psi_T} in the approximation and $\psi_S(\lambda_\mathrm{max})$
is defined as \eqref{Equ:Performance:psi_S_lambda_max} in Lemma 
\ref{lemma:G_lambda_max}. Before the derivation of  
$\mathbf{w}^H\mathbf{A}_I [\frac{L}{\sigma^2} \bm{\Phi}_\mathcal{S}\!\!+\!\!\pmb{\Psi}_I^{-1}]
\mathbf{A}_I^H\mathbf{w}$, we first cite the following identity from Lemma 2 in \cite{chen2010unpublished}.
	\begin{align}
        		\frac{L}{\sigma^2}\bm{\Phi}_\mathcal{I}\!\!+\!\!\pmb{\Psi}_I^{-1}
                                                                        	\!\!=\!\!      	\frac{L}{\sigma^2}
                                                                                        	(\mathbf{T}^{-1})^H
                                                                                        	\!\!
                                                                                        	\left[
                                                                                                	\mathbf{I}
                                                                                                	\!-\!
                                                                                                	\frac{
                                                                                                        	{L\beta}\textsf{SNR}
                                                                                                        	\cdot
                                                                                                        	\frac{\sigma^2}{L}
                                                                                                        	\pmb{\psi}_T\pmb{\psi}_T^H
                                                                                                    		}
                                                                                                    		{
                                                                                                        	{L\beta}
													(1 \!\!-\!\! \rho_0)\textsf{SNR} 
													\!\!+\!\! N
                                                                                                    		}
                                                                                        	\right]
                                                                                        	\!\!
                                                                                        	\mathbf{T}^{-1}.
                                                                                        	\nonumber
    	\end{align}
Then, combining the above expression together with 
$\bm{\Phi}_\Delta \!\!=\!\! \bm{\Phi}_\mathcal{S} \!\!-\!\! \bm{\Phi}_\mathcal{I}$,
\eqref{Equ:Performance:JointDiag} and $\mathbf{A}_\epsilon \!\!\triangleq\!\! \mathbf{A}_I(\mathbf{T}^{-1})^H$
we can have
    \begin{align}
        \mathbf{w}^H
        \mathbf{A}_I
        \left[
        		\frac{L}{\sigma^2}
		\bm{\Phi}_\mathcal{S}
		\!\!+\!\!
		\pmb{\Psi}_I^{-1}
	\right]
	\mathbf{A}_I^H
	\mathbf{w}                                                &=          \frac{L}{\sigma^2}
										    \mathbf{w}^H
										    \mathbf{A}_I
                                                                                        (\mathbf{T}^{-1})^H
                                                                                        \!\!
                                                                                        \left[
                                                                                                \mathbf{I}
                                                                                                \!+\!
                                                                                                \bm{\Gamma}
                                                                                                \!-\!
                                                                                                \frac{
                                                                                                        {L\beta}\textsf{SNR}
                                                                                                        \cdot
                                                                                                        \frac{\sigma^2}{L}
                                                                                                        \pmb{\psi}_T\pmb{\psi}_T^H
                                                                                                    }
                                                                                                    {
                                                                                                        {L\beta}
                                                                                                        (1 \!\!-\!\! \rho_0)
                                                                                                        \textsf{SNR} 
                                                                                                        \!\!+\!\! 
                                                                                                        N
                                                                                                    }
                                                                                        \right]
                                                                                        \!\!
                                                                                        \mathbf{T}^{-1}
                                                                                        \mathbf{A}_I^H
                                                                                        \mathbf{w}
                                                                                        \nonumber\\
             \label{Equ:Appendix:wAIAIw}
                                                                          &\approx
                                                                      		    \left[
										                    \frac{
										                            N{L}/{\sigma^2}
										                         }
										                         {
										                            {L\beta}\textsf{SNR} + N
										                         }
										            \right]^2
										            \left[
										                    \psi_I(\lambda_\mathrm{max})
										                    \!-\!
										                    \frac{
										                            {L\beta}\textsf{SNR}
										                         }
										                         {
										                            {L\beta}\textsf{SNR} + N
										                         }
										                    \psi_S^2(\lambda_\mathrm{max})
								                 \right],
    \end{align}
where $\psi_I(\lambda_\mathrm{max})$ is also given by
\eqref{Equ:Performance:psi_S_lambda_max} in Lemma 
\ref{lemma:G_lambda_max}, and $\rho_0 \!\!\ll\!\! 1$ is a small number independent
of $\textsf{SNR}$.

Finally, substituting \eqref{Equ:Appendix:a0H_w_mp}
and \eqref{Equ:Appendix:wAIAIw} into \eqref{Equ:Appendix:G_lambda_max_expr2}, 
we can derive $\textsf{G}(\textsf{SNR})$ in \eqref{Equ:Performance:G_lambda_max_expr3}
immediately.


\ifCLASSOPTIONcaptionsoff
  \newpage
\fi



\bibliographystyle{IEEEtran}
\bibliography{IEEEabrv,array_signal_processing,beamforming_DSCDMA,communication_general,math,PNCodeAcq_DSCDMA}

\end{document}


%
\title{Some Fundamental Results of Matrix for the Paper ``Performance Analysis
of the Matrix Pair Beamformer with Matrix Mismatch''}

       \author{
        Jianshu~Chen,~\IEEEmembership{Student Member,~IEEE},
        and~Jian~Wang
        \thanks{
                Jianshu Chen was with the Department of Electronic Engineering,
                Tsinghua University, Beijing, P. R. China, 100084. He is currently with the 
                Department of Electrical Engineering, University of
                California, Los Angeles, CA 90095-1594, USA. (e-mail: jshchen@ee.ucla.edu)
               }
        \thanks{
                Jian Wang, is with the Department of Electronic Engineering,
                Tsinghua University, Beijing, P. R. China, 100084. (e-mail: jian-wang@tsinghua.edu.cn)
               }
       }

\maketitle

\begin{abstract}

We give some key results for the paper ``Performance Analysis of the 
Matrix Pair Beamformer with Matrix Mismatch''.
\end{abstract}

\begin{IEEEkeywords}
Adaptive beamforming, Matrix pair beamformer, MPB, Generalized eigen-decomposition, Matrix mismatch
\end{IEEEkeywords}

%
\IEEEpeerreviewmaketitle

\section{Introduction}
We give some key results for the paper ``Performance Analysis of the 
Matrix Pair Beamformer with Matrix Mismatch''.

\section{Problem Formulation}
\label{Sec:Problem_Formulation}
The analysis of the performance of the matrix pair beamformer requires the expression
of $\mathbf{A}^H\mathbf{R}_\mathcal{I}^{-1}\mathbf{A}$,  where
	\begin{align}
	        	\label{Equ:MPBeamformer:R_I}
	        \mathbf{R}_\mathcal{I}          	&\triangleq        \sigma_{\mathcal{I}_0}^2
											\mathbf{a}_0 \mathbf{a}^H_0 
											+ 
											\underbrace{
											\mathbf{A}_I\bm{\Phi}_\mathcal{I}\mathbf{A}_I^H
											+ \sigma^2\mathbf{I}}_{\mathbf{Q}_\mathcal{I}},
											\\
		\mathbf{A} 				&\triangleq	\begin{bmatrix}
												\mathbf{a}_0 & \mathbf{a}_1 
														& \cdots & \mathbf{a}_D
											\end{bmatrix}
        								=			\begin{bmatrix}
												\mathbf{a}_0 & \mathbf{A}_I
											\end{bmatrix}
											\nonumber
    	\end{align}
Furthermore, we base our discussion on the following assumptions.
	\begin{assumption}
		\label{Asm:Orthogonal}
		The spacing of DOA between any two signals is large enough (greater than a mainlobe),
		so that $\{\mathbf{a}_i\}_{i=0}^D$ is linearly independent and
		the projection of $\mathbf{a}_0$ onto 
		$\mathrm{span}\{\!\mathbf{a}_i\}_{i=1}^D$
		is much less than $\|\mathbf{a}_0\|$.
	\end{assumption}
	\begin{assumption}
		\label{Asm:Norm}
		The steering vector of all signals are normalized so that $\|\mathbf{a}_i\|^2=L$, ($i=0,1,\ldots,D$).
	\end{assumption}

\section{Some Fundamental Results}
\label{Sec:Appendix:Proof_of_Lem_ARA}
The main results are summarized in the following two Lemmas.
    \begin{lemma}
        \label{Lem:ARA}
        The expression of  $\mathbf{A}^H\mathbf{R}_\mathcal{I}^{-1}\mathbf{A}$
        is given by
            \begin{align}
                \mathbf{A}^H\mathbf{R}_\mathcal{I}^{-1}\mathbf{A}
                =
                \begin{bmatrix}
                  \mathbf{a}_0^H\mathbf{R}_\mathcal{I}^{-1}\mathbf{a}_0 & \mathbf{a}_0^H\mathbf{R}_\mathcal{I}^{-1}\mathbf{A}_I \\
                  \mathbf{A}_I^H\mathbf{R}_\mathcal{I}^{-1}\mathbf{a}_0 & \mathbf{A}_I^H\mathbf{R}_\mathcal{I}^{-1}\mathbf{A}_I \\
                \end{bmatrix}
                \nonumber
            \end{align}
        where
            \begin{align}
                \label{Equ:Appendix:ARA:a0_R_a0}
                \mathbf{a}_0^H\mathbf{R}_\mathcal{I}^{-1}\mathbf{a}_0 \!&=\!\frac{L}{\sigma^2}
                                                                            \frac{
                                                                                    1-\xi
                                                                                 }
                                                                                 {
                                                                                    \frac{L\beta}{N}
                                                                                    (1\!-\!\xi)
                                                                                    \textsf{SNR}
                                                                                    +1
                                                                                 }
                                                                            \\
                \label{Equ:Appendix:ARA:AI_R_a0}
                \mathbf{A}_I^H\mathbf{R}_\mathcal{I}^{-1}\mathbf{a}_0 \!&=\!\frac{L}{\sigma^2}
                                                                            \frac{
                                                                                    1
                                                                                 }
                                                                                 {
                                                                                    \frac{L\beta}{N}
                                                                                    (1\!-\!\xi)
                                                                                    \textsf{SNR}
                                                                                    +1
                                                                                 }
                                                                                 \pmb{\Xi}
                                                                                 \pmb{\Psi}_I^{-1}
                                                                                 \pmb{\psi}
                                                                            \\
                \label{Equ:Appendix:ARA:AI_R_AI}
                \mathbf{A}_I^H\mathbf{R}_\mathcal{I}^{-1}\mathbf{A}_I \!\!&=\!\!\frac{L}{\sigma^2}
                                                                            \!\!
                                                                            \left[
                                                                                    \pmb{\Xi}
                                                                                    \!-\!
                                                                                    \frac{
                                                                                            \frac{L\beta}{N}
                                                                                            \textsf{SNR}
                                                                                         }
                                                                                         {
                                                                                            \frac{L\beta}{N}
                                                                                            (1\!\!-\!\!\xi)
                                                                                            \textsf{SNR}
                                                                                            \!\!+\!\!
                                                                                            1
                                                                                         }
                                                                                    \pmb{\Xi}
                                                                                    \pmb{\Psi}_I^{-1}
                                                                                    \pmb{\psi}
                                                                                    \pmb{\psi}^H
                                                                                    \pmb{\Psi}_I^{-1}
                                                                                    \pmb{\Xi}
                                                                            \right]
            \end{align}
        with $\pmb{\psi} \triangleq \frac{\mathbf{A}_I^H\mathbf{a}_0}{L}$, 
        $\pmb{\Psi}_I \triangleq \frac{\mathbf{A}_I^H\mathbf{A}_I}{L}$;
        $\beta = N\sigma_{\mathcal{I}_0}/\sigma_{\mathcal{S}_0}$; 
        $\pmb{\Xi}$ and $\xi$ are defined as
            \begin{align}
                \pmb{\Xi}   \triangleq     \left[\frac{L}{\sigma^{2}}\mathbf{S}_{\mathcal{I}}
                						+\mathbf{\Psi}_{\mathcal{I}}^{-1}\right]^{-1}
							,\qquad
                \xi         \triangleq     \rho_{0}-\kappa_0.                          \nonumber
            \end{align}
        $\rho_{0} \triangleq {\pmb\psi}^{H}\pmb\Psi_{\mathcal{I}}^{-1}\pmb\psi$ 
        is the projection power of the unit vector of
        $\mathbf{a_{0}}$ onto $\mathrm{span}\{\mathbf{a}_{0},\cdots,\mathbf{a}_{D}\}$ 
        and is thus much smaller than 1; $\kappa_0 \triangleq
        \pmb\psi^{H}\pmb\Psi_I^{-1}\pmb\Xi\pmb\Psi_I^{-1}\pmb\psi$. Furthermore,
        $\xi$ and $\kappa_0$ satisfy
            \begin{align}
                &0\leq\kappa_0\leq\rho_0 \ll 1,	
                \qquad
                0\leq\xi\leq\rho_0 \ll 1         \nonumber
            \end{align}
    \end{lemma}
    \begin{IEEEproof}
        According to (\ref{Equ:MPBeamformer:R_I}), $\mathbf{R}_\mathcal{I}$ can be written as
            \begin{align}
                \label{Equ:Appendix:ARA:R_I}
                \mathbf{R}_\mathcal{I}  =  \mathbf{A}\mathbf{S}\mathbf{A}^H + \sigma^2\mathbf{I}
            \end{align}
        where $\mathbf{A}$ and $\mathbf{S}$ are defined as
            \begin{align}
                \label{Equ:Appendix:ARA:A}
                \mathbf{A}  &\triangleq     \Big[
                                                \begin{array}{cc}
                                                    \mathbf{a}_0 & \mathbf{A}_I
                                                \end{array}
                                            \Big]
                                            \\
                \label{Equ:Appendix:ARA:S}
                \mathbf{S}  &\triangleq     \left[
                                                \begin{array}{cc}
                                                    \sigma_{\mathcal{I}_0}^2    &   \mathbf{0}^H \\
                                                    \mathbf{0}                  &   \bm{\Phi}_\mathcal{I} \\
                                                \end{array}
                                            \right].
            \end{align}
       According to matrix inversion lemma \cite[pp.124 (3.8.3)]{Meyer2000}, we can derive the $\mathbf{R}_\mathcal{I}^{-1}$ as
            \begin{align}
                \label{Equ:Appendix:ARA:R_inv}
                \mathbf{R}_\mathcal{I}^{-1} =   \frac{L}{\sigma^2}      \left\{
                                                                            \frac{1}{L}\mathbf{I}
                                                                            - \frac{1}{L^2}
                                                                            \mathbf{A} \frac{L}{\sigma^2}\mathbf{S}
                                                                            \left[ \frac{L}{\sigma^2} \cdot \pmb{\Psi}\mathbf{S}+\mathbf{I} \right]^{-1}
                                                                            \!\!\!\!
                                                                            \mathbf{A}^H
                                                                        \right\}
            \end{align}
        where $\pmb{\Psi} \triangleq \mathbf{A}^H\mathbf{A}/L$. Let $\pmb{\psi} \triangleq \mathbf{A}_I^H\mathbf{a}_0/L$
        and $\pmb{\Psi}_I \triangleq \mathbf{A}_I^H\mathbf{A}_I/L$, then $\pmb{\Psi}$ can be written in the following block form
            \begin{align}
                \label{Equ:Appendix:ARA:Psi}
                \pmb{\Psi}      =   \begin{bmatrix}
                                    1                   &       \pmb{\psi}^H \\
                                    \pmb{\psi}          &       \pmb{\Psi}_I \\
                                    \end{bmatrix}.
            \end{align}
        By (\ref{Equ:Appendix:ARA:R_inv}), we can write $\mathbf{A}^H\mathbf{R}_\mathcal{I}^{-1}\mathbf{A}$ as
            \begin{align}
                \label{Equ:Appendix:ARA:ARA1}
                \mathbf{A}^H\mathbf{R}_\mathcal{I}^{-1}\mathbf{A}   &=  \frac{L}{\sigma^2}
                                                                        \left\{
                                                                            \pmb{\Psi}
                                                                            -
                                                                            \frac{L}{\sigma^2}\pmb{\Psi}\mathbf{S}
                                                                            \left[ \frac{L}{\sigma^2}  \pmb{\Psi}\mathbf{S}+\mathbf{I} \right]^{-1}
                                                                            \pmb{\Psi}
                                                                        \right\}
                                                                    \nonumber\\
                                                                    &=  \frac{L}{\sigma^2}
                                                                        \left[ \frac{L}{\sigma^2} \cdot \pmb{\Psi}\mathbf{S}+\mathbf{I} \right]^{-1}
                                                                        \pmb{\Psi}
            \end{align}
        In order to further derive the expression for $\mathbf{A}^H\mathbf{R}_\mathcal{I}^{-1}\mathbf{A}$,
        we need to write $\left[ \pmb{\Psi} \cdot L\mathbf{S}/\sigma^2+\mathbf{I} \right]^{-1}$ in the following block form
            \begin{align}
                \label{Equ:Appendix:ARA:Kernal_inv}
                \left[ \frac{L}{\sigma^2} \cdot \pmb{\Psi}  \mathbf{S}+\mathbf{I} \right]^{-1} =   \left[
                                                                                                    \begin{array}{cc}
                                                                                                        x_{11}          &   \mathbf{x}_{12} \\
                                                                                                        \mathbf{x}_{21} &   \mathbf{X}_{22} \\
                                                                                                    \end{array}
                                                                                                    \right],
            \end{align}
        where $x_{11}$, $\mathbf{x}_{12}$, $\mathbf{x}_{21}$ and $\mathbf{X}_{22}$
        are $1\times1$, $1\times D$, $D\times1$ and $D\times D$ matrix, respectively.
        In order to derive their expression, we need the following equality\cite[pp123]{Meyer2000}
            \begin{align}
                \label{Equ:Appendix:ARA:Inverse_of_BlockMatrix}
                \begin{bmatrix}
                \mathbf{A} & \mathbf{C} \\
                \mathbf{R} & \mathbf{B} \\
                \end{bmatrix}^{-1}
                \!\!\!\!\!\!=\!\!
                \begin{bmatrix}
                \mathbf{A}^{-1}\!+\!\mathbf{A}^{-1}\mathbf{C}\mathbf{S}^{-1}\mathbf{R}\mathbf{A}^{-1}     &       -\mathbf{A}^{-1}\mathbf{C}\mathbf{S}^{-1} \\
                -\mathbf{S}^{-1}\mathbf{R}\mathbf{A}^{-1}                                             &       \mathbf{S}^{-1} \\
                \end{bmatrix},
            \end{align}
        where $\mathbf{S}=\mathbf{B}-\mathbf{R}\mathbf{A}^{-1}\mathbf{C}$ is the Schur complement.
        Substituting (\ref{Equ:Appendix:ARA:S}) and (\ref{Equ:Appendix:ARA:Psi}) into (\ref{Equ:Appendix:ARA:Kernal_inv})
        and using (\ref{Equ:Appendix:ARA:Inverse_of_BlockMatrix}) as well as matrix inversion lemma,
        we can derive
            \begin{align}
                \label{Equ:Appendix:ARA:x11}
                &x_{11}                   =             (1-\alpha)\frac{\alpha_1}{\alpha}       \\
                \label{Equ:Appendix:ARA:x12}
                &\mathbf{x}_{12}          =             -(1-\alpha)\frac{\alpha_1}{\alpha}\frac{L}{\sigma^2} \cdot \pmb{\psi}^H\bm{\Phi}_\mathcal{I}\mathbf{P}   \\
                \label{Equ:Appendix:ARA:x21}
                &\mathbf{x}_{21}          =             -\alpha_1 \cdot \mathbf{P}\pmb{\psi}       \\
                \label{Equ:Appendix:ARA:X22}
                &\mathbf{X}_{22}          =             \mathbf{P}+\alpha_1\frac{L}{\sigma^2} \cdot \mathbf{P}\pmb{\psi}\pmb{\psi}^H\bm{\Phi}_\mathcal{I}\mathbf{P}
            \end{align}
        where
            \begin{align}
                \label{Equ:Appendix:ARA:P}
                &\mathbf{P}               \triangleq    \left[ \frac{L}{\sigma^2} \cdot \pmb{\Psi}_I \bm{\Phi}_\mathcal{I}+\mathbf{I} \right]^{-1}       \\
                \label{Equ:Appendix:ARA:alpha}
                &\alpha                   \triangleq    \frac{L\sigma_{\mathcal{I}_0}^2}{\sigma^2}\left/ \left[ \frac{L\sigma_{\mathcal{I}_0}^2}{\sigma^2}+1 \right]\right. \\
                \label{Equ:Appendix:ARA:alpha1}
                &\alpha_1                 \triangleq    \frac{L\sigma_{\mathcal{I}_0}^2}{\sigma^2}\left/ \left[ \frac{L\sigma_{\mathcal{I}_0}^2}{\sigma^2}(1-\xi)+1 \right]\right.       \\
                \label{Equ:Appendix:ARA:xi}
                &\xi                      \triangleq      \frac{L}{\sigma^2} \cdot \pmb{\psi}^H\bm{\Phi}_\mathcal{I}\mathbf{P}\pmb{\psi}.
            \end{align}
        It is easy to prove that $\alpha$, $\alpha_1$, $\xi$ and $\mathbf{P}$ satisfy the following equalities
            \begin{align}
                &\alpha_1=\alpha(1+\alpha_1\xi)
                \nonumber\\
                &\mathbf{P}^H=\mathbf{I}-\frac{L}{\sigma^2}\bm{\Phi}_\mathcal{I}\mathbf{P}\pmb{\Psi}_I
                \nonumber
            \end{align}
        which have already been used during the derivation of (\ref{Equ:Appendix:ARA:x11})--(\ref{Equ:Appendix:ARA:X22})
        and will be used to simplify the expression of $\mathbf{A}^H\mathbf{R}_\mathcal{I}^{-1}\mathbf{A}$.
        Substituting (\ref{Equ:Appendix:ARA:Psi}), (\ref{Equ:Appendix:ARA:Kernal_inv}),
        (\ref{Equ:Appendix:ARA:x11})--(\ref{Equ:Appendix:ARA:X22}) into (\ref{Equ:Appendix:ARA:ARA1}), after some lengthy but straightforward
        derivations, we can have
            \begin{align}
                \label{Equ:Appendix:ARA:ARA2}
                \mathbf{A}^H \! \mathbf{R}_\mathcal{I}^{-1} \! \mathbf{A}
                \!\!=\!\!
                    \frac{L}{\sigma^2}
                    \!\!
                    \begin{bmatrix}
                        (1\!\!-\!\!\alpha)\frac{\alpha_1}{\alpha}(1\!\!-\!\!\xi)                            &       (1\!\!-\!\!\alpha)\frac{\alpha_1}{\alpha}\pmb{\psi}^H \! \mathbf{P}^H \\
                        (1\!\!-\!\!\alpha)\frac{\alpha_1}{\alpha}\mathbf{P}\pmb{\psi}        &       \mathbf{P}\pmb{\Psi}_I\!\!-\!\!\alpha_1 \mathbf{P}\pmb{\psi}\pmb{\psi}^H \! \mathbf{P}^H \\
                    \end{bmatrix}.
            \end{align}
        Using (\ref{Equ:Appendix:ARA:A}), (\ref{Equ:Appendix:ARA:P})--(\ref{Equ:Appendix:ARA:xi}),
        $\textsf{SNR} = \sigma_{\mathcal{S}_0}^2/\sigma^2$ and $\beta = N \sigma_{\mathcal{I}_0}^2/\sigma_{\mathcal{S}_0}^2$
        we can finally derive (\ref{Equ:Appendix:ARA:a0_R_a0})--(\ref{Equ:Appendix:ARA:AI_R_AI}).

        The remaining problem is to prove $\xi = \rho_0-\kappa_0$ and $\kappa \in [0,\rho_0]$, which can
        also naturally yield $\xi \in [0, \rho_0]$. According to (\ref{Equ:Appendix:ARA:xi}) and the definition of $\pmb{\Xi}$,
        $\xi$ can be written as
            \begin{align}
                \xi         &=      \frac{L}{\sigma^2}
                                    \pmb{\psi}^H
                                    \bm{\Phi}_\mathcal{I}
                                    \pmb{\Xi}\pmb{\Psi}_I^{-1}
                                    \pmb{\psi}
                                    \nonumber\\
                            &=      \pmb{\psi}^H
                                    \pmb{\Xi}^{-1}
                                    \pmb{\Xi}
                                    \pmb{\Psi}_I^{-1}
                                    \pmb{\psi}
                                    -
                                    \pmb{\psi}^H
                                    \pmb{\Psi}_I^{-1}
                                    \pmb{\Xi}
                                    \pmb{\Psi}_I^{-1}
                                    \pmb{\psi}
                                    \nonumber\\
                            &=      \underbrace{
                                                    \pmb{\psi}^H\pmb{\Psi}_I^{-1}\pmb{\psi}
                                               }_{\rho_0}
                                    -
                                    \underbrace{
                                                    \pmb{\psi}^H\pmb{\Psi}_I^{-1}
                                                    \pmb{\Xi}
                                                    \pmb{\Psi}_I^{-1}
                                                    \pmb{\psi}
                                               }_{\kappa_0}.
                                    \nonumber
            \end{align}
        Since $\bm{\Phi}_\mathcal{I}$ is positive semi-definite, $\pmb{\Xi}_I^{-1}-\pmb{\Psi}_I^{-1}=\frac{L}{\sigma^2}\bm{\Phi}_\mathcal{I}$ is also positive semi-definite,
        i.e. $\pmb{\Psi}_I^{-1} \le \pmb{\Xi}_I^{-1}$.
        Then $\pmb{\Psi}_I \ge \pmb{\Xi}_I$, thus
            \begin{align}
                \xi         &=      \pmb{\psi}^H\pmb{\Psi}_I^{-1}
                                    \big[
                                            \pmb{\Psi}_I-\pmb{\Xi}
                                    \big]
                                    \pmb{\Psi}_I^{-1}\pmb{\psi}
                                    \ge
                                    0,
                                    \nonumber
            \end{align}
        which means $\rho_0 \ge \kappa_0 \ge 0$.
        By the definitions of $\pmb{\psi}$ and $\pmb{\Psi}_I$, we have $\rho_0 =
        \frac{1}{\|\mathbf{a}_0\|^2}\mathbf{a}_0^H \mathbf{A}_I (\mathbf{A}_I^H\mathbf{A}_I)^{-1} \mathbf{A}_I\mathbf{a}_0$, i.e.
        the projection power of $\mathbf{a}_0/\|\mathbf{a}_0\|$ onto $\mathrm{span}\{\mathbf{a}_1,\mathbf{a}_2,\ldots,\mathbf{a}_D\}$,
        which we have assumed to be much smaller than 1 in section \ref{Sec:Problem_Formulation}
        Thus $\rho_0 \ll 1$ and this completes the proof.
    \end{IEEEproof}

    \begin{lemma}
        \label{Lem:psi_T}
        Let $\pmb{\psi}_T \triangleq \left[\frac{L\beta}{N}(1-\rho_0)\textsf{SNR}+1\right]\widetilde{\pmb{\psi}}_T$ with
        $\widetilde{\pmb{\psi}}_T \triangleq \mathbf{A}_\epsilon \mathbf{R}_\mathcal{I}^{-1}\mathbf{a}_0$, then
            \begin{align}
                \label{Equ:Appendix:lambda_max:norm_psi_T}
                &\|\pmb{\psi}_T\|    =  \sqrt{
                                                \frac{L}{\sigma^2}
                                                \cdot
                                                \frac{
                                                        \frac{L\beta}{N}(1-\rho_0)\textsf{SNR}+1
                                                    }
                                                    {
                                                        \frac{L\beta}{N}(1-\xi)\textsf{SNR}+1
                                                    }
                                                \kappa_0
                                             }
                                            \ll
                                            \sqrt{\frac{L}{\sigma^2}}
                                            \\
                \label{Equ:Appendix:lambda_max:Xi_inv}
                &\frac{L}{\sigma^2}\bm{\Phi}_\mathcal{I} \!\!+\!\! \pmb{\Psi}_I^{-1}
	                                    =  \frac{L}{\sigma^2}
	                                        \mathbf{T}^{-H}
	                                        \!\!
	                                        \left[
	                                                \mathbf{I}
	                                                \!-\!
	                                                \frac{
	                                                        \frac{L\beta}{N}\textsf{SNR}
	                                                        \cdot
	                                                        \frac{\sigma^2}{L}
	                                                     }
	                                                     {
	                                                        \frac{L\beta}{N}(1 \!\!-\!\! \rho_0)\textsf{SNR}  \!\!+\!\! 1
	                                                     }
	                                                \pmb{\psi}_T
	                                                \pmb{\psi}_T^H
	                                        \right]
	                                        \!\!
	                                        \mathbf{T}^{-1}
            \end{align}
        where $\rho_0$, $\xi$ and $\kappa_0$ are defined in Lemma \ref{Lem:ARA},
        and are independent of $\textsf{SNR}$.
    \end{lemma}
    \begin{IEEEproof}
        Since $\mathbf{A}_\epsilon \!\!=\!\! \mathbf{A}_I\mathbf{T}^{-H}$ and 
        $\mathbf{A}_I^H\mathbf{R}_\mathcal{I}^{-1}\mathbf{A}_I \!\!=\!\! \mathbf{T}\mathbf{T}^H$
        (c.f. (26) in \cite{chen2010dmm}),
        \eqref{Equ:Appendix:ARA:AI_R_AI} is equivalent to
            \begin{align}
                \label{Equ:Appendix:lambda_max:TT}
                \frac{L}{\sigma^2}
                \left[
                        \pmb{\Xi}
                        \!-\!
                        \frac{
                                \frac{L\beta}{N}
                                \textsf{SNR}
                            }
                            {
                                \frac{L\beta}{N}
                                (1\!\!-\!\!\xi)
                                \textsf{SNR}
                                \!\!+\!\!
                                1
                            }
                        \pmb{\Xi}
                        \pmb{\Psi}_I^{-1}
                        \pmb{\psi}
                        \pmb{\psi}^H
                        \pmb{\Psi}_I^{-1}
                        \pmb{\Xi}
                \right]
                =
                \mathbf{T}\mathbf{T}^H.
            \end{align}
        Right-multiplying it by $\pmb{\Psi}_I^{-1}\pmb{\psi}$ and using $\rho_0 = \xi+\kappa_0$, we can derive
            \begin{align}
                \label{Equ:Appendix:lambda_max:Xi_Psi_psi}
                \pmb{\Xi}\pmb{\Psi}_I^{-1}\pmb{\psi}    =   \frac{\sigma^2}{L}
                                                            \cdot
                                                            \frac{
                                                                    \frac{L\beta}{N}(1-\xi)\textsf{SNR}+1
                                                                 }
                                                                 {
                                                                    \frac{L\beta}{N}(1-\rho_0)\textsf{SNR}+1
                                                                 }
                                                            \mathbf{T}
                                                            \mathbf{T}^H
                                                            \pmb{\Psi}_I^{-1}
                                                            \pmb{\psi}.
            \end{align}
        Since $\mathbf{A}_\epsilon = \mathbf{A}_I\mathbf{T}^{-H}$, then by (\ref{Equ:Appendix:ARA:AI_R_a0}) and above expression, we have
            \begin{align}
                \widetilde{\pmb{\psi}}_T    &=  \mathbf{A}_\epsilon^H\mathbf{R}_\mathcal{I}^{-1}\mathbf{a}_0 
                                            =  \frac{L}{\sigma^2}
                                                \cdot
                                                \frac{
                                                        1
                                                     }
                                                     {
                                                        \frac{L\beta}{N}(1-\xi)\textsf{SNR}+1
                                                     }
                                                \mathbf{T}^{-1}
                                                \pmb{\Xi}
                                                \pmb{\Psi}_I^{-1}
                                                \pmb{\psi}
                                            =  \frac{1}{\frac{L\beta}{N}(1-\rho_0)\textsf{SNR}+1}
                                                \mathbf{T}^H
                                                \pmb{\Psi}_I^{-1}
                                                \pmb{\psi}.
                                                \nonumber
            \end{align}
        Let $\pmb{\psi}_T \!\!\triangleq\!\! \left[\frac{L\beta}{N}(1 \!\!-\!\! \rho_0)\textsf{SNR} \!\!+\!\! 1\right]\widetilde{\pmb{\psi}}_T$, then
        $\pmb{\psi}_T \!\!=\!\! \mathbf{T}^H\pmb{\Psi}_I^{-1}\pmb{\psi}$. Using (\ref{Equ:Appendix:lambda_max:TT}), $\rho_0 \!\!=\!\! \kappa_0 \!\!+\!\! \xi$
        and the definition of $\kappa_0$ (Lemma \ref{Lem:ARA}), we have
            \begin{align}
                \|\pmb{\psi}_T\|^2  &=  \pmb{\psi}^H
                                        \pmb{\Psi}_I^{-1}
                                        \mathbf{T}
                                        \mathbf{T}^H
                                        \pmb{\Psi}_I^{-1}
                                        \pmb{\psi}
                                    =  \frac{L}{\sigma^2}
                                        \cdot
                                        \frac{
                                                \frac{L\beta}{N}(1-\rho_0)\textsf{SNR}+1
                                             }
                                             {
                                                \frac{L\beta}{N}(1-\xi)\textsf{SNR}+1
                                             }
                                        \kappa_0.
                                        \nonumber
            \end{align}
        According to Lemma \ref{Lem:ARA}, $0 \!\!\le\!\! \xi \!\!\ll\!\! 1$, $0 \!\!\le\!\! \rho_0 \!\!\ll\!\! 1$ and $0 \!\!\le\!\! \sqrt{\kappa_0} \!\!\le\!\! \sqrt{\rho_0}$.
        Therefore, $\|\pmb{\psi}_T\| \!\!\approx\!\! \sqrt{\frac{L}{\sigma^2}\kappa_0} \!\!\ll\!\! \sqrt{\frac{L}{\sigma^2}}$.
        We then prove (\ref{Equ:Appendix:lambda_max:Xi_inv}). Substituting (\ref{Equ:Appendix:lambda_max:Xi_Psi_psi})
        into (\ref{Equ:Appendix:lambda_max:TT}) and using $\pmb{\psi}_T \!\!=\!\! \mathbf{T}^H\pmb{\Psi}_I^{-1}\pmb{\psi}$,
        we can have
            \begin{align}
                \pmb{\Xi}           &=  \mathbf{T}
                                        \!\!
                                        \left\{
                                                \frac{\sigma^2}{L}
                                                \mathbf{I}
                                                \!+\!
                                                \frac{
                                                        \left[
                                                                \frac{L\beta}{N}(1 \!\!-\!\! \xi)\textsf{SNR}  \!\!+\!\! 1
                                                        \right]
                                                        \frac{L\beta}{N}\textsf{SNR}
                                                     }
                                                     {
                                                        \left[
                                                                \frac{L\beta}{N}(1 \!\!-\!\! \rho_0)\textsf{SNR}  \!\!+\!\! 1
                                                        \right]^2
                                                     }
                                                \pmb{\psi}_T
                                                \pmb{\psi}_T^H
                                        \right\}
                                        \!\!
                                        \mathbf{T}^H.
                                        \nonumber
            \end{align}
        Applying matrix inversion lemma\cite{Meyer2000} to the above equation and using  $\rho_0 \!\!=\!\! \kappa_0 \!\!+\!\! \xi$ (Lemma \ref{Lem:ARA}),
        we can finally derive (\ref{Equ:Appendix:lambda_max:Xi_inv}).
    \end{IEEEproof}


\ifCLASSOPTIONcaptionsoff
  \newpage
\fi



\bibliographystyle{IEEEtran}
\bibliography{IEEEabrv,array_signal_processing,beamforming_DSCDMA,communication_general,math,PNCodeAcq_DSCDMA}